\documentclass[11pt,a4paper]{article}
\pdfoutput=1
\usepackage{amssymb,amsmath,amsfonts, mathtools, mathrsfs}
\usepackage[utf8]{inputenc} 
\usepackage[dvipsnames]{xcolor}
\usepackage{float}

\newif\ifnatbibsort\natbibsorttrue

\relax

\ifnatbibsort\RequirePackage[numbers,sort&compress]{natbib}\else\RequirePackage[numbers,compress]{natbib}\fi
\RequirePackage[colorlinks=true
,urlcolor=blue
,anchorcolor=blue
,citecolor=blue
,filecolor=blue
,linkcolor=blue
,menucolor=blue
,pagecolor=blue
,linktocpage=true
,pdfproducer=medialab
,pdfa=true
]{hyperref}

\usepackage{graphicx}
\usepackage{caption}
\usepackage{tensor}
\usepackage{subfigure}
\usepackage{enumerate}
\usepackage{dsfont}
\usepackage{verbatim}
\usepackage{amsfonts,euscript,amssymb,stmaryrd,braket}
\usepackage{tikz}
\usetikzlibrary{arrows,decorations.markings,patterns}
\usepackage{comment}
\usepackage{slashed}
\usepackage{soul}
\usepackage{cancel}
\usepackage[normalem]{ulem}

\def\clock{{\count0=\time
		\divide\count0 60
		\ifnum\count0<10 0\fi\the\count0
		\multiply\count0 -60 \advance\count0 \time
		:\ifnum\count0<10 0\fi \the\count0
}}
\newcommand{\timestamp}{{\small\vbox{\hbox{\tt\jobname.tex}
			\hbox{\the\day/\the\month/\the\year, \clock}}}}

\newcommand{\be}{\begin{equation}}
\newcommand{\ee}{\end{equation}}
\newcommand{\bea}{\begin{eqnarray}}
\newcommand{\eea}{\end{eqnarray}}
\newcommand{\nn}{\nonumber}

\newcommand{\rst}[1] {\unskip}

\makeatletter
\let\old@startsection=\@startsection
\let\oldl@section=\l@section
\renewcommand{\@startsection}[6]{\old@startsection{#1}{#2}{#3}{#4}{#5}{#6\mathversion{bold}}}
\renewcommand{\l@section}[2]{\oldl@section{\mathversion{bold}#1}{#2}}
\makeatother

\numberwithin{equation}{section}
\def \RR {{\mathbb R}}

\def\ri {{\rm i}}
\def\rd {{\rm d}}
\def\e {{\rm e}}

\begin{document}
	\renewcommand{\thefootnote}{\arabic{footnote}}

	\overfullrule=0pt
	\parskip=2pt
	\parindent=12pt
	\headheight=0in \headsep=0in \topmargin=0in \oddsidemargin=0in

	\vspace{ -3cm} \thispagestyle{empty} \vspace{-1cm}
	\begin{flushright} 
		\footnotesize
		\textcolor{red}{\phantom{print-report}}
	\end{flushright}

	\begin{center}
	\vspace{.0cm}

	{\Large\bf \mathversion{bold}
	Entanglement entropies of an interval
	}
	\\
	\vspace{.25cm}
	\noindent
	{\Large\bf \mathversion{bold}
	in the free Schr\"odinger field theory at finite density}

	\vspace{0.8cm} {
		Mihail Mintchev$^{\,a}$,
		Diego Pontello$^{\,b}$,
		Alberto Sartori$^{\,c,d}$
		and Erik Tonni$^{\,b}$
	}
	\vskip  0.7cm
	
	\small
	{\em
		$^{a}\,$Dipartimento di Fisica, Universit\'a di Pisa and INFN Sezione di Pisa,\\
		largo Bruno Pontecorvo 3, 56127 Pisa, Italy
		\vskip 0.1cm
		$^{b}\,$SISSA and INFN Sezione di Trieste, via Bonomea 265, 34136, Trieste, Italy 
		\vskip 0.1cm
		$^{c}\,$Intelligent Cloud Technologies Laboratory, Huawei Munich Research Center,\\
		Riesstraße 25, 80992 München, Germany		
		\vskip 0.1cm
		$^{d}\,$International Centre for Theoretical Physics (ICTP),\\
		Strada Costiera 11, 34151, Trieste, Italy
	}


\normalsize

\end{center}

\vspace{0.3cm}
\begin{abstract} 

We study the entanglement entropies of an interval on the infinite line
in the free fermionic spinless Schr\"odinger field theory at finite density and zero temperature,
which is a non-relativistic model with Lifshitz exponent $z=2$.
We prove that the entanglement entropies are finite functions of one dimensionless parameter
proportional to the area of a rectangular region in the phase space 
determined by the Fermi momentum and the length of the interval.
The entanglement entropy is a  monotonically increasing function. 
By employing the properties of the prolate spheroidal wave functions of order zero 
or the asymptotic expansions of the tau function of the sine kernel,
we find analytic expressions for the expansions of the entanglement entropies 
in the asymptotic regimes of small and large area of the rectangular region in the phase space.
These expansions lead to prove that 
the analogue of the relativistic entropic $C$ function is not monotonous. 
Extending our analyses
to a class of free fermionic Lifshitz models 
labelled by their integer dynamical exponent $z$,
we find that the parity of this exponent determines 
the properties of the bipartite entanglement for an interval on the line.
\end{abstract}

\vspace{1cm}

\newpage

\newpage

\tableofcontents

\section{Introduction}
\label{sec_intro}

The bipartite entanglement associated to a spatial bipartition has been largely studied during the past three decades in quantum field theories, quantum many-body systems and quantum gravity (see e.g. the reviews \cite{EislerPeschel:2009review, Casini:2009sr, Calabrese-Doyon, Eisert:2008ur,Rangamani:2016dms, Headrick:2019eth, Tonni:2020bjq}).

Consider a quantum system in a state characterised by the density matrix $\rho$
and the bipartition $A \cup B$ of the space given by a region $A$ and its complement $B$.
Assuming that the Hilbert space of the system 
can be factorised as $\mathcal{H} =\mathcal{H}_A \otimes \mathcal{H}_B $,
the reduced density matrix of the subsystem $A$ is $\rho_A \equiv \textrm{Tr}_{\mathcal{H}_B} \rho $,
where the normalisation condition $\textrm{Tr}_{\mathcal{H}_A} \rho_A = 1$ is imposed
(hereafter we enlighten the notation by using $\textrm{Tr} (\cdots)= \textrm{Tr}_{\mathcal{H}_A} (\cdots)$).

The entanglement entropies provide an important set of quantities to study 
in order to understand the bipartite entanglement of the system in the state $\rho$
for the bipartition $A\cup B$.
They are the entanglement entropy, the R\'enyi entropies and the single copy entanglement.
The entanglement entropy $S_A$ is the von Neumann entropy of the reduced density matrix
\be
\label{ee-def-intro}
S_A
\,\equiv\,
  -\,\textrm{Tr}\big(\rho_A \log \rho_A\big) 
  \,=\,
  \lim_{\alpha \to 1} S_A^{(\alpha)} 
\ee
which can be obtained also through the analytic continuation $\alpha \to 1$ 
of the R\'enyi entropies (replica limit), defined in terms of the moments of $\rho_A$ as follows
\be
\label{renyi-def-intro}
S_A^{(\alpha)} 
\equiv \frac{1}{1-\alpha}\,\log\!\big[\textrm{Tr}(\rho_A^\alpha)\big]
\ee
where $\alpha \neq 1$ is a real and positive parameter. 
The replica limit in (\ref{ee-def-intro}) naturally leads to identify $S_A^{(1)} \equiv S_A$.
The single copy entanglement $S_A^{(\infty)}$ 
can be defined as the limit $\alpha \to +\infty$ of the R\'enyi entropies (\ref{renyi-def-intro})
\cite{peschel-05,eisert-05,orus-06}.
Since $\textrm{Tr}(\rho_A^\alpha) = \sum_j \lambda_j^\alpha$, 
with $\lambda_j \in (0,1)$ being the eigenvalues of $\rho_A$,
it is straightforward to realise that 
$S_A^{(\infty)}= - \log (\lambda_{\textrm{\tiny max}})$,
where $\lambda_{\textrm{\tiny max}} $ is the largest eigenvalue of $\rho_A$.
Among these entanglement entropies, $S_A$ is the most important quantity
because it measures the bipartite entanglement when $\rho$ is a pure state.

Many fundamental results have been obtained 
for the entanglement entropies
in relativistic quantum field theories in $d+1$ spacetime dimensions.
In this class of quantum field theories and when $\rho$ is the ground state,
the entanglement entropies are divergent as $\epsilon\to 0$,
where $\epsilon$ is the ultraviolet (UV) cut off.
The leading divergence exhibits the celebrated area law behaviour
$S_A^{(\alpha)} \propto \textrm{Area}(\partial A)/\epsilon^{d-1} +\cdots$ as $\epsilon\to 0$,
where the dots correspond to the subleading terms
\cite{Bombelli:1986rw, Srednicki:1993im,Ryu:2006bv,Ryu:2006ef}.
Our analyses are restricted to $d=1$ translation invariant quantum field theories on the line
and to the bipartition given by an interval $A$;
hence we are allowed to set $A=[-R,R]$ without loss of generality.
A notable exception to the area law behaviour 
are the conformal field theories in $d=1$, where
$S_A^{(\alpha)} = \tfrac{c}{6} (1+\frac{1}{\alpha}) \log(2R/\epsilon) + \cdots$
\cite{Callan:1994py, Holzhey:1994we, Calabrese:2004eu}, 
with $c$ being the central charge of the model,
which occurs in the Virasoro algebra  \cite{Belavin:1984vu}.
Another important result for the relativistic field theories in $d=1$ 
involves the function $C \equiv R\,\partial_R S_A$,
constructed from the entanglement entropy of the interval $A$ on the line 
when the entire system is in its ground state.
Similarly to the Zamolodchikov's $C$ function \cite{Zamolodchikov:1986gt},
the quantity $C$ is UV finite and decreases monotonically along a renormalization group (RG) flow 
connecting a UV fixed point to an infrared (IR) fixed point \cite{Casini:2004bw};
hence it is often called entropic $C$ function.
The proof of this monotonic behaviour of $C$ is based both 
on the strong subadditivity property of the entanglement entropy 
and on the relativistic invariance \cite{Casini:2004bw}.
In relativistic quantum field theories, the effect of the finite density on the entanglement entropies 
has been also explored 
\cite{Wong:2013gua,Cardy-talk, Ogawa:2011bz, Liu:2012eea, Belin:2013uta, Daguerre:2020pte}.

In order to gain some new insights about the relation between the spacetime symmetry and the 
characteristic features of entanglement,
it is worth investigating the entanglement entropies in non-relativistic quantum field theories,
which have also close connections with the quantum many-body systems. 
The properties of the bipartite entanglement quantifiers depend on 
whether the quantum field theory model displays a relativistic or a non-relativistic invariance. 
For instance, 
free Fermi systems at finite density 
exhibit a well known
logarithmic violation of the area law of the entanglement entropy
due to the occurrence of a Fermi surface
\cite{Gioev:2006zz, Wolf:2006zzb}
(for numerical results in lattice models, see  e.g. \cite{Barthel-06,Li-06}).
Furthermore, 
an entropic $C$ functions for non-relativistic quantum field theories in $d=1$  is not known 
\cite{Swingle:2013zla,Daguerre:2020pte}.

In the Wilsonian approach to quantum field theory, 
a fixed point corresponds to a scale invariant model
and the scaling symmetry may not act on space and time in the same way.
Under the assumption of spatial isotropy, 
the Lifshitz scale transformation is defined by
$t \to \chi^z \,t$ and $\boldsymbol{x} \to \chi\, \boldsymbol{x}$
for any spatial position vector $\boldsymbol{x} \in \mathbb{R}^d$, 
where the parameter $z>0$ is the Lifshitz exponent or dynamical critical exponent \cite{Hertz-76}.
The Poincar\'e algebra, which characterises the relativistic field theories, has $z=1$.
The Schr\"odinger algebra has $z=2$
\cite{Niederer-72, Hagen-72, Henkel:1993sg, Nishida:2007pj, Hartong:2014pma, Hartong-15}.

We focus on $d=1$ and consider 
the fermionic free Schr\"odinger field theory at finite density $\mu$ and on the infinite line,
which is a free non-relativistic quantum field theory with $z=2$.
This model describes the dilute spinless Fermi gas in $d=1$ \cite{sachdev_book}.
The fermionic spinfull model in $d=1$ in the presence of the quartic interaction
has been studied through renormalization group methods \cite{Benfatto-book, Gallavotti_01, Gentile:2001gb}.
Free fermionic spinless models in $d=1$ with positive integer values of $z$ have been also considered \cite{Hartmann:2021vrt}.
Let us remark that, in our analysis of the entanglement entropies for this free model,
we do not approximate the dispersion relation 
with a linear dispersion relation at the Fermi points
(Tomonaga's approximation) \cite{Giamarchi_book, Glazman_12}.

In $d=2$, an interesting model with $z=2$ called quantum Lifshitz model has been introduced in \cite{Ardonne:2003wa}:
it is a free bosonic quantum field theory with the symmetries of a Lifshitz critical point with $z=2$  
and its bipartite entanglement has been studied in \cite{Fradkin:2006mb, Hsu:2008af, Fradkin:2009dus}, 
finding an area law behaviour. 
This model belongs to a class of Lifshitz theories in $d\geqslant 2$ 
having $z=d$ \cite{Keranen:2016ija}, whose bipartite entanglement entropy 
for the ground state has been investigated in \cite{Angel-Ramelli:2019nji}.
The entanglement entropy in non-relativistic Schr\"odinger models in $d\geqslant 2$ 
has been computed also through heat kernel methods \cite{Solodukhin:2009sk}.
In $d=1$, numerical studies of the entanglement entropy in bosonic lattice models with various integer $z$ 
have been reported e.g. in \cite{He:2017wla, MohammadiMozaffar:2017nri}.
In the context of the gauge/gravity correspondence, 
gravitational backgrounds dual to Lifshitz spacetimes have been introduced 
\cite{Balasubramanian-08, Kachru:2008yh, Gubser:2008px, Balasubramanian:2009rx} (see the review \cite{Taylor:2015glc})
and the holographic entanglement entropy for $z\neq 1$ has been computed in various settings
\cite{Azeyanagi:2009pr, Ogawa:2011bz, Huijse:2011ef, Tonni:2010pv, Keranen:2011xs, Fonda:2014ula, 
Alishahiha:2014cwa, Gentle:2017ywk, Cavini:2019wyb}.

In this manuscript we study the entanglement entropies of an interval $A = [-R, R]$ on the line for 
the free fermionic spinless Schr\"odinger field theory at zero temperature and  finite density $\mu$.
When $\mu = 0$, we have $S_A = 0$, as observed in \cite{Pal:2017ntk, Hason-17, Hartmann:2021vrt}.

At finite density $\mu >0$, 
we find that the entanglement entropies are finite functions of one variable 
given by the dimensionless parameter $\eta \equiv R \,k_{\textrm{\tiny F}}$,
where $p_{\textrm{\tiny F}} \equiv \hbar \,k_{\textrm{\tiny F}}$ is the Fermi momentum.
This parameter is proportional to the area of a limited rectangular region in the phase space
that can be naturally identified from the interval $A$ and the Fermi momentum.
The finiteness of the entanglement entropies 
is proved by exploiting the properties of the 
solution of the spectral problem associated to the sine kernel in the interval $A$,
first reported in a series of seminal papers by Slepian, Pollak and Landau
\cite{Slepian-part-1, Slepian-part-2, Slepian-part-3, Slepian-part-4}
in terms of the prolate spheroidal wave functions (PSWFs) of order zero
(see also the overview \cite{Slepian-83} and the recent book \cite{Rokhlin-book}).
Efficient algorithms for the numerical evaluation of these functions have been developed
(see \cite{Rokhlin-book} and references therein).
The role of PSWFs in the context of the entanglement for free fermions
has been studied in \cite{EislerPeschelProlate}, which has inspired our work.

We find analytic results for the expansions of entanglement entropies 
in the regimes of small and large values of $\eta$.
These results are obtained by employing the method proposed in \cite{Jin_2004,Keating_04} 
for the entanglement entropies in some spin chains
(further developed in \cite{Calabrese:2009us, Calabrese-Essler-10} to include the subleading terms)
and the asymptotic expansions of the sine kernel tau function 
\cite{Gamayun:2013auu, Lisovyy:2018mnj, Bonelli:2016qwg}.
The latter expansions extend previous results \cite{JMMS, Jimbo-82, McCoy:1985, TracyWidom92, forrester-book}
and have been found by adapting to the Painlev\'e V equation
the method introduced in the seminal paper \cite{Gamayun:2012ma} 
to write the general solution of the Painlev\'e VI equation.

By applying the results of Fisher-Hartwig \cite{FHc, Basor-91, Basor-94} 
and Widom \cite{Widom1982, Sobolev_14} 
to the corresponding matrix spectral problem,
the asymptotic behaviour of the entanglement entropies 
for various models and in different space dimensions have been extensively studied 
\cite{Jin_2004,Keating_04, Calabrese:2009us, Calabrese-Essler-10, Mintchev-PRL, Calabrese:2011vh}.
For the Schr\"odinger field theory on the line considered in this manuscript, 
the spectral problem in the continuum has an explicit solution in terms of the PSWFs
and this allows to analyse the entropies in the whole range of parameters and not only in the asymptotic regime.

We study also the entanglement entropies of the interval $A$ 
in the hierarchy of free Lifshitz fermions on the line considered in  \cite{Hartmann:2021vrt},
in the massless case at zero temperature and finite density.
These models are labelled by their integer Lifshitz exponent $z \geqslant 1$.
The special cases $z=1$ and $z=2$ correspond respectively to 
the relativistic chiral fermion and to the Schr\"odinger field theory introduced above.

The outline of the manuscript is as follows.
In Sec.\,\ref{sec_model} 
the free fermionic spinless Schr\"odinger field theory on the line at finite density $\mu$ 
is briefly described. 
In Sec.\,\ref{sec_spectral} 
we present the solution of the spectral problem associated to the sine kernel in the interval
and some properties of its spectrum.
The spectrum of this kernel is employed in Sec.\,\ref{sec_entropies}
to evaluate the entanglement entropies of the interval on the line. 
In Sec.\,\ref{sec_flow} 
we discuss the behaviour of the entanglement entropies and
of the quantity analogue to the relativistic entropic $C$ function
along the flow generated by the dimensionless parameter $\eta$.
In Sec.\,\ref{sec_lifshitz_exponent} 
we explore the entanglement entropies of an interval
for a class of $d=1$ free fermionic models with integer Lifshitz exponents $z \geqslant 1$.
The expansions of the entanglement entropies in the asymptotic regimes 
of small $\eta$ and large $\eta$ are investigated 
in Sec.\,\ref{sec_small_distance} and Sec.\,\ref{sec_large_distance} respectively. 
In Sec.\,\ref{sec_schatten} we explore the Schatten norms of the sine kernel.
In Sec.\,\ref{sec_conclusions} we draw some conclusions.
The derivations of some formulas and further auxiliary results
are reported in the Appendices\;\ref{app_bounds}, \ref{app_cumulants}, 
\ref{app_finiteness-C-alpha}, \ref{app_mod-ham-flow}, \ref{app_small_distance},
\ref{app_large_distance} and \ref{sec_large_distance_lattice}.

\section{Free Schr\"odinger field theory at finite density}
\label{sec_model}

\subsection{The model}

We consider the non-relativistic spinless complex fermion field $\psi(t,x)$ of mass $m$ on the line,
which evolves with the Schr\"odinger Hamiltonian 
\begin{equation}
H = \int_{-\infty}^\infty {\cal E}(t,x) \,\rd x 
\;\;\;\qquad \;\;\;
{\cal E}(t,x) = \frac{\hbar^2}{2m} \,\partial_x \psi^*(t,x) \,\partial_x \psi (t,x)
\label{hamiltonian}
\end{equation}
and satisfies the equal-time canonical anticommutation relations 
\begin{eqnarray}
&&
\big\{\psi (t,x_1)\, ,\, \psi (t,x_2) \big\} \,=\, \big\{\psi^* (t,x_1)\, ,\, \psi^*(t,x_2)\big\} \,=\, 0 
\label{e2a} 
\\ 
\rule{0pt}{.6cm}
&&
\big\{\psi (t,x_1)\, ,\, \psi^*(t,x_2) \big\} \,=\, \delta(x_1-x_2) 
\label{e2b}
\end{eqnarray}
where $^*$ stands for Hermitian conjugation. The solution of the equation of motion 
\begin{equation}
\left (\ri \hbar\,\partial_t +\frac{\hbar^2}{2m} \,\partial_x^2\right )\psi (t,x) \,=\, 0 
\label{e1}
\end{equation} 
which satisfies also (\ref{e2a}) and (\ref{e2b}) is
\cite{Landau_book_3,sachdev_book,Dick_book} 
\begin{equation} 
\psi (t,x)  = \int_{-\infty}^{\infty} 
a (k) \, \e^{-\ri \omega (k) t }\, \e^{\ri k x}\; \frac{dk}{2\pi } 
\;\;\;\qquad \;\;\;
\omega(k) = \frac{\hbar}{2m}\, k^2
\label{e3} 
\end{equation} 
where $\{a (k),\, a^*(k)\, :\, k \in {\mathbb R} \}$  
generate the canonical anticommutation relation (CAR)  algebra $\cal A$ given by
\be
\big\{a(k)\, ,\, a(p) \big\} = \big\{a^*(k)\, ,\, a^*(p) \big\} = 0 
\;\;\;\qquad\;\;\;
\big\{a(k)\, ,\, a^*(p) \big\} = 2\pi \,\delta(k-p) \,.
\label{car1}
\ee

In order to determine the state space of the system, 
a Hilbert space representation of the algebra $\cal A$ must be fixed.
Before doing that, let us recall the symmetry content of the model. 
The system has an internal $U(1)$ symmetry that implies the current conservation 
\be 
\partial_t \varrho (t,x) + \partial_x j(t,x) = 0 
\label{curr1}
\ee 
where  $\varrho (t,x)$ and $j(t,x)$ are the particle density and current respectively,
which are written through the field as follows
\be 
\varrho (t,x) = \psi^*(t,x) \,\psi (t,x)
\;\; \qquad \;\;
j(t,x) = \frac{\ri \hbar}{2m}\,
 \big[
(\partial_x \psi^*)(t,x) \,\psi(t,x) 
- \psi^*(t,x) \,(\partial_x\psi)(t,x) 
\big] \,.
\label{curr2}
\ee

The space-time symmetries of \eqref{e1} form the Schr\"odinger group \cite{Niederer-72,Hagen-72,Nishida:2007pj},
whose Lie algebra is generated by the momentum $P$, 
the Galilean boost $G$, the dilatation $D$ and the special conformal transformation $K$. 
In terms of (\ref{curr2}), these generators read 
\bea 
& &
P = \int_{-\infty }^\infty j(t,x) \,\rd x 
\hspace{2cm}
G = \int_{-\infty }^\infty x \, \varrho (t,x) \,\rd x 
\label{charge1}
\\
\rule{0pt}{.7cm}
& &
D =\int_{-\infty }^\infty x \, j(t,x) \,\rd x 
\hspace{1.7cm}
K = \int_{-\infty }^\infty \frac{x^2}{2}\, \varrho (t,x) \,\rd x \,.
\label{charge2}
\eea
Using the canonical anti-commutation relations (\ref{e2a}) and (\ref{e2b}), one finds \cite{Nishida:2007pj}
\be 
[G\, ,\, D]= \frac{\ri \hbar}{m}\,  G 
\qquad 
[K\, ,\, D]= \frac{2\ri \hbar}{m}\,  \, K 
\qquad 
[K\, ,\, P]= \frac{\ri \hbar}{m}\, G
\label{Lie1}
\ee
\be
[G\, ,\, K] = 0 
\qquad 
[D\, ,\, P]= \frac{\ri \hbar}{m}\, \, P 
\qquad 
[G\, ,\, P]= \frac{\ri \hbar}{m}\, N   
\label{Lie2}
\ee 
where $N$ in (\ref{Lie2}) is the particle number operator  
\be 
N = \int_{-\infty }^\infty \varrho (t,x) \,\rd x \,.
\label{charge2}
\ee
This operator commutes with all Schr\"odinger generators and defines a central extension of the 
Schr\"odinger algebra associated to a non-trivial cocycle 
\cite{Niederer-72,Hagen-72}.
The local form of the central term in the r.h.s. of the commutator $[G,P]$ in (\ref{Lie2}) reads
\be 
c(t,y) \equiv \frac{\hbar}{m}\, \varrho (t,y)\,.
\label{Lie3}
\ee
For any representation of the CAR algebra $\cal A$ in (\ref{car1}), which is  
generated by a time and space invariant cyclic vector $\Omega$, the expectation value 
\be 
c_{{}_\Omega} \equiv \langle c(t,x) \rangle_{{}_\Omega} 
= 
\frac{\hbar}{m} \,\langle \varrho (t,x) \rangle_{{}_\Omega}
\label{c1}
\ee 
is a $t$ and $x$-independent dimensionless parameter 
(in our convention, $t$ and $x$ are measured in the same units) 
which characterises not only the central extension 
of the Schr\"odinger algebra, but also the representation of $\cal A$.

In order to illustrate the above structure we consider two different representations of 
$\cal A$ with well known physical applications. 
The first one is the Fock representation in which $a(k)\Omega = 0$. All correlation functions 
of $\{a (k),\, a^*(k)\}$ in this representation can be expressed 
in terms of the following expectation values in the state $\Omega$ 
\begin{equation}
\langle a^*(p)\,a(k) \rangle_{{}_{\rm F}}  = 0
\;\;\;\qquad \;\;\;
\langle a(k)\,a^*(p)\rangle_{{}_{\rm F}}  = 2\pi\, \delta (k-p)   \,.
\label{f1}
\end{equation}
Accordingly, we have
\begin{equation}
\langle \psi^*(t_1,x_1)\,\psi (t_2,x_2)\rangle_{{}_{\rm F}} = 0 
\label{f2}
\end{equation}
and 
\begin{equation}
\label{f3}
\langle \psi(t_1,x_1)\,\psi^*(t_2,x_2)\rangle_{{}_{\rm F}}
= 
\int_{-\infty}^{\infty} 
\e^{-\ri \omega(k) (t_{12}-\ri \varepsilon)}  \,\e^{\ri k x_{12}} \, \frac{\rd k}{2\pi } 
\,=\, 
\frac{\sqrt { \hbar \,m}\; 
\e^{\frac{\ri \hbar \,m x_{12}^2}{2(t_{12} -\ri \varepsilon)}}}{\sqrt {2\pi \ri \, (t_{12}-\ri \varepsilon)}} 
\end{equation}
where $t_{12} \equiv t_1-t_2$, $x_{12} \equiv x_1-x_2$ and $\varepsilon \to 0^+$. 
At equal time $t_1=t_2\equiv t$, these correlators become respectively 
\begin{equation}
\langle \psi^*(t,x_1) \, \psi (t,x_2)\rangle_{{}_{\rm F}} = 0\,  
\;\;\qquad \;\;
\langle \psi(t,x_1)\,\psi^*(t,x_2)\rangle_{{}_{\rm F}} 
=
\sqrt {\frac{\hbar\,m}{2\pi \varepsilon}}\; 
\e^{-\frac{\hbar\,m x_{12}^2}{2 \varepsilon }} 
 \underset{\varepsilon\,\rightarrow \,0}{\longrightarrow}
\delta(x_{12}) \,.
\label{f4}
\end{equation}
The two-point functions (\ref{f2}) and (\ref{f3}) are invariant 
under the dilatation transformation
\begin{equation} 
U_\chi \,\psi(t,x) \,U_\chi^* = \chi^{1/2} \,\psi (\chi^2 t,\chi x) 
\;\;\qquad \;\;
\chi > 0
\label{d1}
\end{equation}
where $U_\chi$ is the one-parameter group, which is generated by the dilatation operator $D$ (\ref{charge2}) 
and leaves invariant the Fock vacuum, i.e. $U_\chi \Omega = \Omega$.  
From (\ref{d1}), one infers the scaling dimension $[\psi] = 1/2$. 
Because of (\ref{f2}), the Fock representation 
is characterised by  
\be 
\langle {\cal E} (t,x) \rangle_{{}_{\rm F}}= 0 
\;\;\qquad \;\;
\langle \varrho (t,x) \rangle_{{}_{\rm F}} = 0 
\;\;\qquad \;\;
c_{{}_{\rm F}} = \frac{\hbar}{m}\langle \varrho (t,x) \rangle_{{}_{\rm F}} = 0 \,.
\label{c2}
\ee

Another representation of the CAR algebra $\cal A$, which implements the contact of the system with a heat bath  
at inverse temperature $\beta>0$ and chemical potential $\mu \in \RR$, is the Gibbs representation \cite{Brattelli2}. 
Since the Gibbs state $\Omega_{\beta,\mu}$ is Gaussian, 
all correlation functions of $\{a (k),\, a^*(k)\}$ in the Hilbert space ${\cal H}_{\beta,\mu}$ of this representation 
can be expressed in terms of the expectation values 
\bea
\langle a^*(p) \, a(k) \rangle_{\beta,\mu}  
&=&
\frac{1}{1 + \e^{\beta [ \hbar \,\omega(k) - \mu]}} \; 2\pi \, \delta (k-p)   
\label{e6a}
\\
\rule{0pt}{.9cm}
\langle a(k) \, a^*(p)\rangle_{\beta,\mu}  
&=& 
\frac{\e^{\beta [\hbar\,\omega(k) - \mu]}}{1+\e^{\beta [\hbar\,\omega(k)  - \mu]}} \; 2\pi\, \delta (k-p)   
\label{e6b}
\eea
where one recognises the Fermi distribution. Therefore, the non-vanishing two-point functions of the field $\psi(t,x)$ at 
finite density and temperature are given by 
\bea
\langle \psi^*(t_1,x_1) \,\psi (t_2,x_2)\rangle_{\beta,\mu} 
&=& 
\int_{-\infty}^{\infty} 
\frac{1}{1+\e^{\beta[ \hbar\,\omega(k) -\mu]}}  \;
\e^{\ri \omega(k) (t_{12}-\ri \varepsilon) } \, \e^{-\ri k x_{12}}
\;\frac{\rd k}{2\pi } 
\label{g1}
\\
\rule{0pt}{.9cm}
\langle \psi(t_1,x_1)\,\psi^*(t_2,x_2)\rangle_{\beta,\mu} 
&=& 
\int_{-\infty}^{\infty} 
\frac{\e^{\beta [ \hbar \,\omega(k)- \mu]}}{1+\e^{\beta [ \hbar\,\omega(k) - \mu]}}    
\; \e^{-\ri \omega(k) (t_{12}-\ri \varepsilon) } \, \e^{\ri k x_{12}}
\;\frac{\rd k}{2\pi } \,.
\label{g2}
\eea
Using these correlators one easily checks that 
$U_\chi\, \Omega_{\beta,\mu} \not= \Omega_{\beta,\mu}$, 
implying that the dilatations are not unitarily implemented in the Gibbs representation ${\cal H}_{\beta,\mu}$. 
Indeed, $U_\chi$ intertwines the two Gibbs representations 
\begin{equation} 
U_\chi \, :\,  {\cal H}_{\beta,\mu} \longrightarrow {\cal H}_{\chi^2\beta,\, \mu/\chi^{2}}
\;\;\;\qquad\;\;\;
U^*_\chi \, :\, {\cal H}_{\chi^2\beta,\, \mu/\chi^{2}} \longrightarrow {\cal H}_{\beta,\mu} \,.
\end{equation}
Thus, (\ref{g1}) and (\ref{g2}) are invariant under dilatations, 
provided that one transforms simultaneously the temperature and chemical potential according to 
\begin{equation} 
\label{flow-beta-mu}
\beta \longmapsto \chi^2 \beta 
\;\; \qquad \;\;
\mu \longmapsto \frac{\mu}{\chi^{2} } \,.
\end{equation}

In the zero temperature limit $\beta \to \infty$, the correlators (\ref{g1}) and (\ref{g2}) give 
\bea
\langle \psi^*(t_1,x_1) \,\psi (t_2,x_2)\rangle_{\infty,\mu} 
&=& 
\int_{-\infty}^{\infty} 
\theta \big(\mu-\hbar\,\omega(k)\big)  \,
\e^{\ri \omega(k) (t_{12}-\ri \varepsilon) } \, \e^{-\ri k x_{12}}
\;\frac{\rd k}{2\pi } 
\label{g1zeroT}
\\
\rule{0pt}{.9cm}
\langle \psi(t_1,x_1)\,\psi^*(t_2,x_2)\rangle_{\infty,\mu} 
&=& 
\int_{-\infty}^{\infty} 
\big[1-\theta \left (\mu-\hbar\,\omega(k)\right )\big]
\, \e^{-\ri \omega(k) (t_{12}-\ri \varepsilon) } \, \e^{\ri k x_{12}}
\;\frac{\rd k}{2\pi } 
\label{g2zeroT}
\eea
where $\theta$ is the Heaviside step function. For $\mu \leqslant 0$ the two-point functions (\ref{g1zeroT}) and (\ref{g2zeroT}) 
reproduce precisely the correlators (\ref{f2}) and (\ref{f3}) in the Fock representation. 

When $\mu > 0$ and at equal time $t_1=t_2 \equiv t$, 
the correlators \eqref{g1zeroT} and \eqref{g2zeroT} become respectively \cite{EislerPeschelProlate}
\bea
\langle \psi^*(t,x_1) \, \psi (t,x_2)\rangle_{\infty,\,\mu>0} 
&=&
\int_{- k_{\rm F}}^{k_{\textrm{\tiny F}}} 
\e^{\ri k x_{12}} \, \frac{\rd k}{2\pi } 
\,=\,  
\frac{\sin ( k_{\textrm{\tiny F}} \,x_{12}  )}{\pi x_{12}}
\label{g3}
\\
\rule{0pt}{.9cm}
\langle \psi(t,x_1) \, \psi^*(t,x_2)\rangle_{\infty,\,\mu>0} 
&=&
 \int_{-\infty}^{-k_{\textrm{\tiny F}} } 
 \e^{\ri k x_{12}} \,  \frac{\rd k}{2\pi } 
 + 
\int_{k_{\textrm{\tiny F}} }^{\infty}  
\e^{\ri k x_{12}}\,  \frac{\rd k}{2\pi } 
\,=\, 
\delta(x_{12}) - \frac{\sin (k_{\textrm{\tiny F}}\,x_{12} )}{\pi x_{12}}   
\phantom{xxxxxx}
\label{g4}
\eea
where the Fermi wave number $k_{\textrm{\tiny F}}$ is defined 
in terms of the Fermi momentum $p_{\textrm{\tiny F}}$
as follows
\be
\label{kF-def}
k_{\textrm{\tiny F}} 
\equiv \frac{p_{\textrm{\tiny F}} }{ \hbar}
\;\;\;\;\qquad\;\;\;\;
p_{\textrm{\tiny F}} \equiv \sqrt {2m\mu} \,.
\ee
The two-point function (\ref{g3}) is a projector on the line; indeed, it satisfies
\be
\int_{-\infty}^{+\infty}
\frac{\sin (k_{\textrm{\tiny F}} \,x_{12}  )}{\pi x_{12}}
\;
\frac{\sin (k_{\textrm{\tiny F}} \,x_{23}  )}{\pi x_{23}}
\;
\rd x_2
\,=\,
\frac{\sin (k_{\textrm{\tiny F}} \,x_{13}  )}{\pi x_{13}}
\ee
which tells us that the state providing the correlators  (\ref{g3}) and (\ref{g4}) is pure \cite{powers-70}. 
The expectation value (\ref{g3}) provides the kernel of the integral operator
whose spectrum is employed in the evaluation of the entanglement entropies.

In the zero temperature limit, the Gibbs representation with $\mu>0$ is characterised by  
\be 
\langle {\cal E} (t,x) \rangle_{\infty,\mu} = \frac{\hbar^2 k_{\textrm{\tiny F}}^3}{6\pi \,m} 
\qquad
\langle \varrho (t,x) \rangle_{\infty,\mu}  = \frac{k_{\textrm{\tiny F}}}{\pi} 
\qquad
c_\mu 
=  \frac{\hbar}{m} \langle \varrho (t,x) \rangle_{\infty,\mu} 
= \frac{\hbar\, k_{\textrm{\tiny F}}}{\pi \,m}  
= \frac{1}{\pi}\, \sqrt {\frac{2\mu}{m}}\,.
\label{c3}
\ee
In particular, $c_\mu $ is proportional to the Fermi velocity $v_{\textrm{\tiny F}}= p_{\textrm{\tiny F}}/m$, 
i.e. the velocity at the Fermi surface, which is dimensionless in our conventions.

In the limit $k_{\textrm{\tiny F}} \to  0$, the correlators  (\ref{g3}) and (\ref{g4}) become respectively
\be
\label{2-point-mu-neg}
\langle \psi^*(t,x_1) \, \psi (t,x_2)\rangle_{\infty,\mu}  \,\longrightarrow\,0
\qquad
\langle \psi(t,x_1) \, \psi^* (t,x_2)\rangle_{\infty,\mu}  \,\longrightarrow\,\delta(x_{12})
\;\;\;\qquad\;\;\;
k_{\textrm{\tiny F}} \to  0 \,.
\ee
In the opposite limit $k_{\textrm{\tiny F}} \to  \infty$,
by employing the delta sequence given by $\frac{\sin(kx)}{\pi x} \to \delta(x)$ as $k \to \infty$ 
\cite{Gelfand-book1},
for the correlators  (\ref{g3}) and (\ref{g4}) one obtains respectively
\be
\label{2-point-mu-infty}
\langle \psi^*(t,x_1) \, \psi (t,x_2)\rangle_{\infty,\mu}  \,\longrightarrow\,\delta(x_{12})
\qquad
\langle \psi(t,x_1) \, \psi^* (t,x_2)\rangle_{\infty,\mu}  \,\longrightarrow\,0
\;\;\;\qquad\;\;\;
k_{\textrm{\tiny F}} \to  \infty \,.
\ee


\subsection{The parameter $\eta$}

In this manuscript we consider the bipartition of the infinite line given by an interval $A$
 and its complement. 
 The translation invariance allows to choose $A=[-R\, , R]$ without loss of generality.

The mean value of the particle number operator
$N_A \equiv \int_{A} \varrho (t,x) \,\rd x $ in $A$ is \cite{Abanov-2011}
\be 
\label{N_A-segment}
\langle N_A \rangle_{\infty , \mu} 
\, = 
\int_{-R}^R \langle \varrho(t,x) \rangle_{\infty , \mu} \, {\rm d} x 
\,=\,
\frac{k_{\textrm{\tiny F}} }{\pi}\, 2R
\,  \equiv \frac{2}{\pi} \,\eta 
\,=\,
\frac{\mathsf{a}}{2\pi}
\,=\,
\frac{\tilde{\mathsf{a}}}{2\pi  \hbar}
\ee
where we have introduced the dimensionless parameter
\be
\label{eta-def}
\eta \equiv R \,k_{\textrm{\tiny F}} 
\ee
and also its rescalings 
\be
\label{a-parameters-def}
\mathsf{a}  \equiv 4\, \eta  = 4 \,R \,k_{\textrm{\tiny F}} = \frac{\tilde{\mathsf{a}}}{\hbar}
\;\;\;\;\qquad\;\;\;\;
\tilde{\mathsf{a}}  \equiv 4 \,R \,p_{\textrm{\tiny F}} \,.
\ee
Notice that $\tilde{\mathsf{a}}$ is the area of the following region in the phase space
\be
\label{def-red-phae-space}
\Gamma_{A,\,p_{\textrm{\tiny F}}} \equiv \big\{ (x,p)\,;\, x\in A \,,\, -\,p_{\textrm{\tiny F}} \leqslant p \leqslant p_{\textrm{\tiny F}} \big\}
\ee
obtained through a {\it space-momentum limiting} process,
which is the phase space analogue of the {\it time-frequency limiting} process
considered for signals in the seminal papers 
\cite{Slepian-part-1, Slepian-part-2, Slepian-part-3, Slepian-part-4, Slepian-83}.
In this analogy, the Heisenberg principle corresponds to the 
impossibility of the simultaneous confinement of a signal and of the amplitude of its frequency spectrum.
Thus, non-zero band-limited signals
(i.e. signals whose frequencies belong to the finite band $(-\mathcal{W},\mathcal{W})$, where $\mathcal{W}$ is called bandwidth)
cannot be also time-limited (i.e. non-vanishing only in a finite interval of time). 
However, band-limited signal can be observed for a finite amount of time $\mathcal{T}$.
Within this parallelism, 
$p_{\textrm{\tiny F}}$ and $R$ are the analogues of $\mathcal{W}$ and  $\mathcal{T}$ respectively. 
Following this terminology, we call (\ref{def-red-phae-space}) limited phase space
and its area is $\tilde{\mathsf{a}} = 4\hbar \,\eta$.

The dimensionless parameter $\eta$ (or $\mathsf{a}$, equivalently)
plays a crucial role in the analysis of the entanglement entropies of $A$
described in the subsequent sections.
We always keep $\eta$ finite and non-vanishing.
The limits $\eta \to 0$ and $\eta \to \infty$ are taken only in the final results.


\section{The spectral problem for the sine kernel}
\label{sec_spectral}

In this section we describe the spectral problem corresponding to the sine kernel
in the interval $A = [-R, R]$, whose spectrum determines the entanglement 
entropy of the bipartition of the line given by $A$ and its complement,
as discussed in Sec.\,\ref{sec_entropies}.

The spectral problem associated to the correlator \eqref{g3} restricted to the interval $[-1,1]$ is \cite{Slepian-part-1, Rokhlin-book}
\be
\label{spectral-problem-v2}
\int_{-1}^{1} 
K(\eta;x-y) 
\; f_n(\eta;y) \, \rd y \,=\, \gamma_n \, f_n(\eta;x)
\ee
where $x\in [-1, 1]$, $n\in\mathbb{N}_0$
and $K(\eta;x-y) $ is the sine kernel
\be 
\label{sine-kernel def}
K(\eta;x-y) \equiv \frac{\sin[\eta(x-y)]}{\pi(x-y)} \,.
\ee 
Since the sine kernel is parametrised by the dimensionless parameter $\eta>0$,
its eigenvalues $\gamma_n$ depend only on $\eta$.
The normalization condition for the eigenfunctions in \eqref{spectral-problem-v2} can be fixed 
through the standard inner product of $L^2[-1,1]$, 
i.e. $\int_{-1}^1 \overline{f_p(\eta;x)} f_{q}(\eta;x) \,\rd x = \delta_{p,q}$,
where the bar indicates the complex conjugation.

The spectral problem (\ref{spectral-problem-v2}) has been studied and solved 
in a series of seminal papers by Slepian, Pollak and Landau
\cite{Slepian-part-1, Slepian-part-2, Slepian-part-3, Slepian-part-4}
(see also the overview \cite{Slepian-83} and the recent book \cite{Rokhlin-book}).

The spectral problem defined by the correlator \eqref{g3} restricted to the interval $A=[-R,R]$
is equivalent to (\ref{spectral-problem-v2}); indeed, it reads
\be
\label{spectral-problem-R}
\int_{-R}^{R} \frac{\sin\!\big[ k_{\textrm{\tiny F}} (x-y)\big]}{\pi\, (x-y)} \; 
\frac{f_n(\eta;y/R)}{\sqrt{R}}  \; \rd y 
\,=\, 
\gamma_n \, \frac{f_n(\eta;x/R)}{\sqrt{R}}
\ee
where $x\in [-R, R] $ and the dimensionless parameter $\eta$ has been introduced in (\ref{eta-def}).
In particular, the same eigenvalues occur in 
(\ref{spectral-problem-v2}) and (\ref{spectral-problem-R})
and the factor $1/\sqrt{R}$ in the eigenfunctions of \eqref{spectral-problem-R} 
has been introduced to satisfy the normalisation condition
imposed by the standard inner product of $L^2[-R,R]$.

In the two limits $\eta \to 0$ and $\eta \to \infty$,
the kernel (\ref{sine-kernel def}) degenerates to 
the vanishing kernel and to the Dirac delta respectively.

The sine kernel \eqref{sine-kernel def} is real, symmetric (i.e. $K(\eta; x-y) = K(\eta; y-x)$) and satisfies
\be 
\label{s3}
\int_{-1}^1 \int_{-1}^1 K(\eta;x-y)\,  \rd x \, \rd y
 \,= \,
 \frac{2}{\pi \eta}\,  \Big( \cos(2\eta) + 2\eta\, {\rm Si}(2\eta) -1 \Big)
\ee
where ${\rm Si}\, (\xi ) = \int_0^\xi  \sin(t) \, \rd t$ is the sine integral function.
This implies the following estimate 
\be 
\label{s5}
\quad 0 \leqslant \int_{-1}^1 \int_{-1}^1 K(\eta;x-y) \,  \rd x \, \rd y \,\leqslant 2 
\ee
From \eqref{s3} and \eqref{s5},
one infers that the sine kernel \eqref{sine-kernel def} 
defines a positive trace-class operator. 
The spectrum $\{\gamma_n\, |\, n \in \mathbb{N}_0 \}$  
of an operator belonging to this class is compact, countable, real positive 
and satisfies $\sum_{n=0}^\infty \gamma_n < \infty$ \cite{ReedSimon-book,CH_book}. 
The spectrum of the sine kernel \eqref{sine-kernel def} is non-degenerate,
it satisfies the following bounds (see \cite{CadaMoore04} and Eqs.~(3.49)~and~(3.55) in \cite{Rokhlin-book})
\be
\label{spectrum-01}
0 < \gamma_n < 1
\ee
and its trace reads
\be
\label{sum-gamma-n}
\sum_{n=0}^{\infty} \gamma_n 
= \frac{2}{\pi}\, \eta
= \frac{\mathsf{a} }{2\pi} \,.
\ee
Notice that the trace of the sine kernel
coincides with the mean value of the particle number in $A$ given in \eqref{N_A-segment} (see also \eqref{mean-QA}).

From \eqref{spectrum-01} and \eqref{sum-gamma-n}, it is straightforward to realise that  $\gamma_n \to 0$ when $\eta \to 0$.

We remark that,
in the relativistic model given by the massless Dirac field in one spatial dimension,
the kernel defined by the two-point function of the massless Dirac field 
restricted to an interval (see (\ref{gLL}) for $z=1$) is not a compact operator.
As a consequence, its spectrum is not discrete and spans the set $[0,1]$
(see e.g. \cite{Casini-08,Casini:2009vk,Arias:2017dda,Arias:2018tmw}).

The eigenvalues and eigenfunctions in \eqref{spectral-problem-v2} 
can be expressed in terms of the prolate spheroidal wave functions (PSWFs), 
which have been introduced as solutions of the Helmholtz wave equation in spheroidal coordinates \cite{Morse-Feshbach-book, Flammer-book, abramowitz}. 
In particular, the eigenvalues can be written through
the radial PSWFs of zero order $\mathcal{R}_{0n}$ as follows 
\cite{Slepian-part-1,Rokhlin-book,CadaMoore04}
\be
\label{eigenvalues} 
\gamma_n(\eta) 
\,=\, 
\frac{2\eta}{\pi}\; \mathcal{R}_{0n}(\eta,1)^2 
\;\;\;\qquad\;\;\;
n \in \mathbb{N}_0 
\ee
and the eigenfunctions  in terms of the
angular PSWFs of zero order $\mathcal{S}_{0n}$ as
\be
 \label{PSWF-def}
f_n(\eta;x) \,=\, \sqrt{n+\frac{1}{2}} \; \mathcal{S}_{0n}(\eta, x) \,.
\ee

The functions $\mathcal{R}_{0n}(\eta,x)$ and $\mathcal{S}_{0n}(\eta,x)$ in
\eqref{eigenvalues} and \eqref{PSWF-def} can be studied e.g. through
Wolfram Mathematica, where for the radial and angular PSWFs 
correspond to the built-in symbols
$ \mathcal{R}_{mn}(x_1,x_2) = \mathrm{SpheroidalS1}[n, m, x_1, x_2] $
and $\mathcal{S}_{mn}(x_1, x_2) = \mathrm{SpheroidalPS}[n, m, x_1,x_2]$ 
respectively \cite{spheroidalWR},
where $m$ is the order of the PSWF
and a particular normalisation is adopted.
In \eqref{eigenvalues} and \eqref{PSWF-def},
the normalisation chosen by Wolfram Mathematica
for $\mathcal{R}_{0n}(\eta,x)$ and $\mathcal{S}_{0n}(\eta,x)$
is compatible with the normalisation induced
by the standard inner product of $L^2[-1,1]$
for the eigenfunctions $f_n(\eta;x)$,
which has been imposed above.
Notice that different normalisations for the PSWFs have been introduced 
in the literature \cite{Slepian-part-1, CadaMoore04}. 

\begin{figure}[t!]
\vspace{-.2cm}
\hspace{-.8cm}
\includegraphics[width=1.05\textwidth]{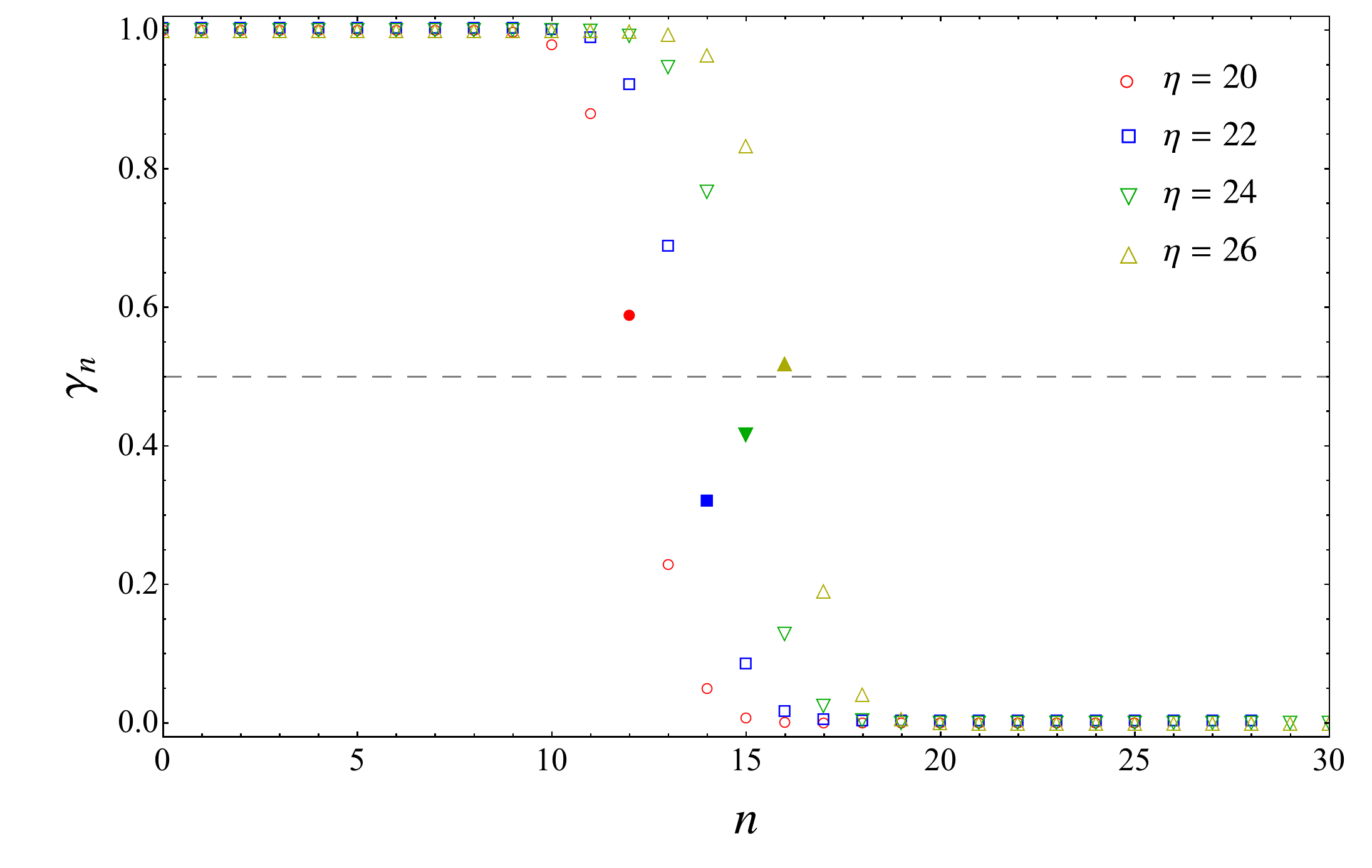}
\vspace{-.7cm}
\caption{
The first eigenvalues of the spectrum for the sine kernel
(see \eqref{spectral-problem-v2} and \eqref{spectral-problem-R})
for some values of the dimensionless parameter $\eta$.
The sets I, II and III providing a natural partition of the spectrum can be 
easily recognised. 
For any $\eta$, the eigenvalue closest to $1/2$ (full marker) provides the largest contribution 
to the entanglement entropies \eqref{entropies-def-sums} and (\ref{single-copy-ent-sum}).
}
\label{fig:eigenvalues}
\end{figure}

The eigenvalues \eqref{eigenvalues} are arranged in decreasing order.
The eigenfunction $f_n(\eta;x)$ corresponding to $\gamma_n$
has a definite parity under $x \to - x$, which is equal to the parity of $n$.
Furthermore, $f_n(\eta;x)$ has $n$ simple roots in $(-1,1)$ (see Theorem 2.3 in \cite{Rokhlin-book}).

In our numerical analysis we have employed an optimised Fortran code
provided to us by Vladimir Rokhlin\footnote{We are deeply grateful to 
Vladimir Rokhlin for having shared his optimised Fortran code with us.},
which is based on the results discussed in \cite{Rokhlin-07,Rokhlin-14,Rokhlin-book} 
regarding the numerical evaluation of the PSWFs. 
This code is faster than Mathematica and 
it also provides reliable results for the spectrum in the large $\eta$ regime ($\eta \gtrsim 350$).

\begin{figure}[t!]
\vspace{-.2cm}
\hspace{-.8cm}
\includegraphics[width=1.05\textwidth]{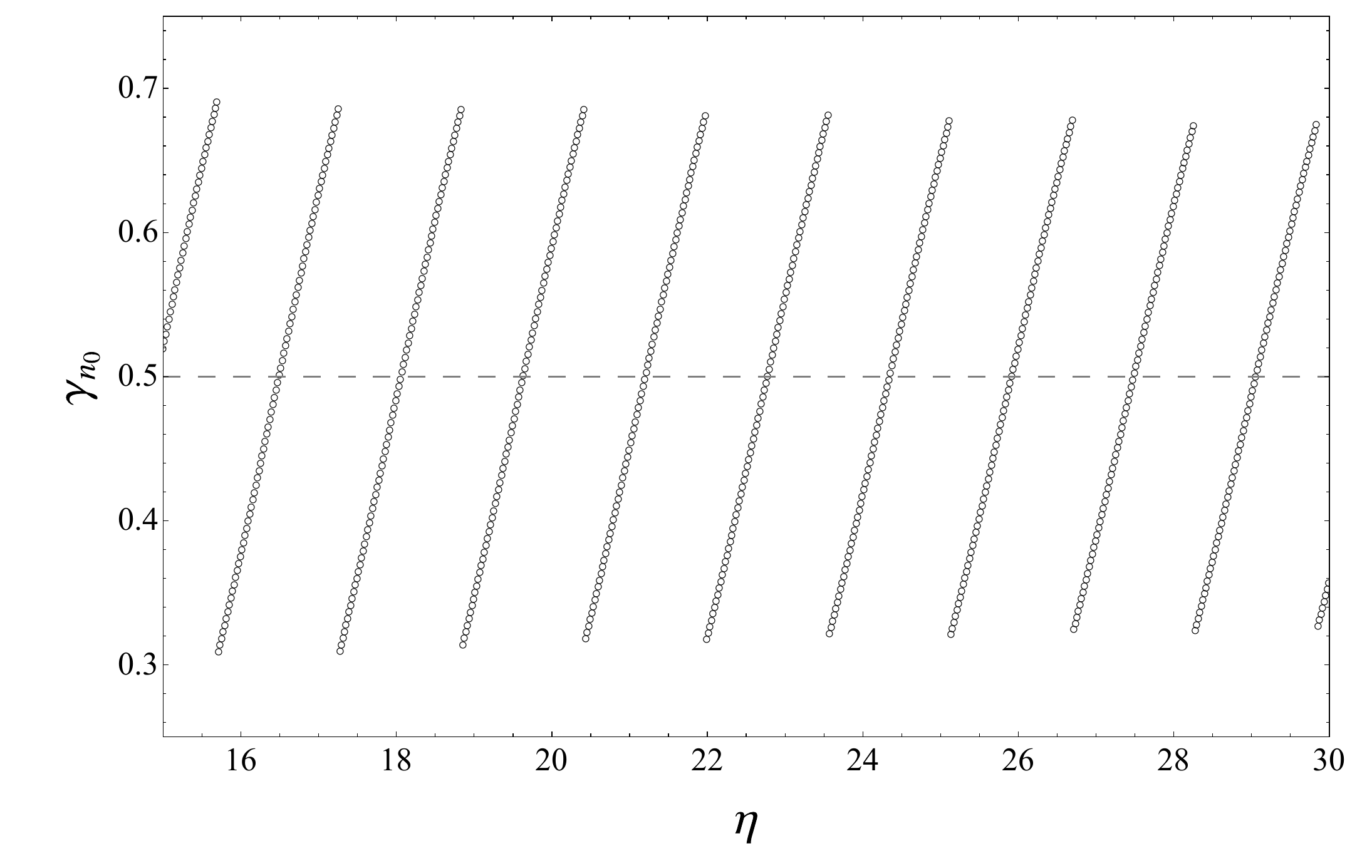}
\vspace{-.7cm}
\caption{
The eigenvalue $\gamma_{n_0}$ corresponding to the critical index (\ref{critical-n0-def}) 
in terms of $\eta$.
}
\label{fig:critic-eigenvalue-n0}
\end{figure}

The arrangement of the eigenvalues in the decreasing order
leads to identify a natural partition of the spectrum 
in the following three subsequent sets
(see e.g. \cite{Landau-65, bonami-21}):
\be
\begin{array}{ccl}
\textrm{I}
&\;&
\textrm{
a finite slow evolution region corresponding to $ 2\eta/\pi - n \gtrsim \log(\eta) $ where $\gamma_n \approx 1$;
}
\\
\rule{0pt}{.6cm}
\textrm{II}
& &
\textrm{
a finite plunge region for $ | n - 2\eta/\pi | \lesssim \log(\eta)$
where the eigenvalues rapidly change;
}
\\
\rule{0pt}{.6cm}
\textrm{III}
& &
\textrm{
an infinite fast decay region corresponding to  $ n - 2\eta/\pi  \gtrsim \log(\eta) $ 
where $\gamma_n \approx 0$
}
\\
&&
\textrm{
and $\gamma_n \to 0$ as $n \to +\infty$ 
at a super-exponential rate.
}
\end{array}
\nonumber
\ee
In Fig.\,\ref{fig:eigenvalues} we show the first $31$ eigenvalues 
of the spectra for four  values of $\eta$
such that the three parts I, II and III of the spectrum can be neatly identified.

The partition of the spectrum in its three subsets I, II and III 
naturally leads to introduce the critical index $n_0$ as follows (see Eq.\,(3.9) of \cite{Rokhlin-book})
\be
\label{critical-n0-def}
n_0 \equiv \left\lfloor \frac{2\eta}{\pi} \right\rfloor = \bigg\lfloor \frac{\mathsf{a}}{2\pi} \bigg\rfloor
\ee
where $\lfloor x \rfloor$ denotes the integer part of $x$.
We remark that $n_0$ is a function of $\eta$.
In a numerical inspection based on $50000$ values of $\eta \in [0.02, 1000]$,
we have observed that  $\gamma_{n} > 1/2$ for all $n<n_0$ and $\gamma_{n} < 1/2$ for all $n>n_0$,
 while whether $\gamma_{n_0}$ is greater or smaller than $1/2$ depends on $\eta$. 
In this analysis we have also found that  the eigenvalue closest to $1/2$ is  $\gamma_{n_0}$ for the vast majority of the values of $\eta$ explored, 
 while for few cases it is given by $\gamma_{n_0-1}$, where we observed that
 $\gamma_{n_0-1} > 1/2 > \gamma_{n_0}$ and $\gamma_{n_0-1} - 1/2 \approx 1/2 - \gamma_{n_0}$.

From these observations and the discreteness of the spectrum,
one realises that the behaviour of $\gamma_{n_0}$ in terms of $\eta$
is oscillatory. 
This is shown in Fig.\,\ref{fig:critic-eigenvalue-n0}, 
which displays the sawtoothed profile of $\gamma_{n_0}$ 
in a typical finite range of $\eta$.

\newpage
\section{Entanglement entropies}
\label{sec_entropies}

In this section we study the entanglement entropies of the interval $A=[-R, R]$ on the line
for the Schr\"odinger field theory at zero temperature and finite density described in Sec.\,\ref{sec_model}.

The Schr\"odinger field theory  is a free fermionic model;
hence its entanglement entropies (\ref{renyi-def-intro}) and (\ref{ee-def-intro})
can be evaluated from the eigenvalues of the spectral problem \eqref{spectral-problem-R} 
(or (\ref{spectral-problem-v2})) as follows
\cite{Peschel:2003rdm, EislerPeschel:2009review,Casini:2009sr}
\be
\label{entropies-def-sums}
S_A = \sum_{n = 0}^{\infty} s(\gamma_n)
\;\;\;\;\qquad\;\;\;\;
S_A^{(\alpha)} = \sum_{n = 0}^{\infty} s_\alpha(\gamma_n) 
\ee
where 
\be
\label{entropies_func}
s(x) \equiv -\,x \log(x) - (1-x) \log(1-x) 
\;\;\qquad\;\;
s_\alpha(x) \equiv\, \frac{1}{1-\alpha}\, \log\!\big[ x^\alpha +(1-x)^\alpha\big]
\ee
with finite $\alpha>0$ and $\alpha \neq 1$.
The limit $S_A^{(\alpha)} \to S_A^{(\infty)}$ as $\alpha \to +\infty$ 
defines the single-copy entanglement $S_A^{(\infty)} = - \log (\lambda_{\textrm{\tiny max}})$,
which gives the largest eigenvalue 
$\lambda_{\textrm{\tiny max}}\in (0,1)$ of the reduced density matrix
\cite{peschel-05,eisert-05,orus-06}.
For the model that we are considering, 
from \eqref{entropies-def-sums} and \eqref{entropies_func}, one obtains
\be
\label{single-copy-ent-sum}
S_A^{(\infty)}
=
 \sum_{n = 0}^{\infty} s_\infty(\gamma_n)
\;\;\qquad\;\;
s_\infty(x) \equiv 
\left\{\begin{array}{ll}
- \log(1-x) 
\hspace{.5cm}
& x\in [0,1/2]
\\
\rule{0pt}{.5cm}
- \log(x)
& x\in (1/2,1] \,.
\end{array}\right. 
\ee

Since the spectrum of the sine kernel depends only 
on the dimensionless parameter $\eta$  in \eqref{eta-def}, 
also the entanglement entropies in \eqref{entropies-def-sums} and \eqref{single-copy-ent-sum} 
depend only on $\eta >0$. 
Let us remark that,
denoting by $\{\gamma_n\, ; n \in \mathbb{N}_0\}$ the spectrum
corresponding to the correlator (\ref{g3}) restricted to $A$ 
(see (\ref{spectral-problem-R})),
the spectrum obtained from the correlator (\ref{g4}) restricted to $A$
is $\{1-\gamma_n\, ; n \in \mathbb{N}_0\}$.
Both these spectra provide the same entanglement entropies
in (\ref{entropies-def-sums}) and (\ref{single-copy-ent-sum})
because the functions of $x$ defined in \eqref{entropies_func} and (\ref{single-copy-ent-sum})
are invariant under $x \leftrightarrow 1-x$.

Following \cite{Peschel:2003rdm}, 
the single-particle entanglement energies $\varepsilon_n $ can be introduced as 
\be
\label{ent-eps-def}
\gamma_n = \frac{1}{\e^{\varepsilon_n} + 1}
\ee
i.e. $\varepsilon_n \equiv  \log(1/\gamma_n - 1)$, which are negative for a finite number of $n$'s.
This allows to 
write the entanglement entropy in (\ref{entropies-def-sums}) 
in the following suggestive form 
 \cite{Peschel_2004}
\be
\label{ee-thermo-form}
S_A 
=
E_A - \Omega_A
\ee
where
\be
\label{EA-OmegaA-def}
E_A  \,\equiv\,
\sum_{n = 0}^{\infty} 
\frac{\varepsilon_n}{\e^{\varepsilon_n} + 1}
\;\;\;\;\qquad\;\;\;\;
\Omega_A  \,\equiv\,
- \sum_{n = 0}^{\infty} 
\log\big(1+ \e^{-\varepsilon_n}\big) \,.
\ee
These expressions have the same form 
of  the internal energy and of the thermodynamic potential
of a quantum ideal Fermi gas respectively.
Notice that $\Omega_A < 0$; hence $S_A > E_A$.

In the Schr\"odinger field theory at finite density, 
the entanglement entropies in \eqref{entropies-def-sums} and \eqref{single-copy-ent-sum} 
are finite. 
This crucial feature is due to the fast decay of $\gamma_n \to 0^+$ as $n \to +\infty$.
Indeed, for the eigenvalues $\gamma_n$ of the sine kernel,
the following upper bound has been obtained (see Theorem 3.20 in \cite{Rokhlin-book})
\be
\label{gamma-upper-bound}
0 \leqslant \gamma_n \leqslant  \tilde{\gamma}_n 
\ee
where
\be
\label{gamma-tilde-def}
\tilde{\gamma}_n 
\equiv\,
\tilde{g}_n \,\eta^{2n+1} 
\;\;\;\qquad\;\;\;
\tilde{g}_n 
\equiv \frac{1}{2}\left(  \frac{(n!)^2}{(2n)! \; \Gamma(n+3/2)} \right)^2
=
\frac{2}{\pi}\left(  \frac{2^{2n} \,(n!)^3}{(2n)! \, (2n+1)!} \right)^2  .
\ee
By employing the Stirling's approximation formula in (\ref{gamma-tilde-def}),
for the asymptotic behaviour of $\tilde{\gamma}_n $ as $n \to +\infty$
one finds (see \cite{widom-94,bonami-21} and Eq.\,(3.76) in \cite{Rokhlin-book})
\be
\label{gamma-tilde-infty}
\tilde{\gamma}_n  =  
\tilde{\gamma}_{\infty ,n} 
\big[ 1 + O(1/n)\big]
\;\;\;\qquad\;\;\;
\tilde{\gamma}_{\infty ,n}  \equiv \frac{\textrm{e}^{2n} \, \eta^{2n+1}}{(4n)^{2n+1}} 
\ee
which tells us that $\gamma_n $ vanishes as $n \to +\infty$ 
with a super-exponential decay rate.
In the following we show that
this important feature of the spectrum leads to finite entanglement entropies 
(\ref{entropies-def-sums}) and (\ref{single-copy-ent-sum}).

From (\ref{entropies_func}) and (\ref{single-copy-ent-sum}), 
one first realises that
$0\leqslant s_\infty(x) \leqslant s_{\alpha_2}(x) \leqslant s(x) \leqslant s_{\alpha_1}(x)$
for any $x\in[0,1]$ and $0<\alpha_1<1<\alpha_2$;
hence
\be
\label{ent_func_ine}
0\leqslant S_A^{(\infty)} \leqslant S_A^{(\alpha_2)} \leqslant S_A \ \leqslant S_A^{(\alpha_1)}
\;\;\; \qquad \;\;\;
 0<\alpha_1<1<\alpha_2  \,.
\ee
This sequence of inequalities tells us that 
the finiteness of $S_A^{(\alpha)}$ with $\alpha\in(0,1)$
implies the finiteness of the remaining entanglement entropies,
i.e. the entanglement entropy, 
the R\'enyi entropies $S_A^{(\alpha)}$ with $\alpha > 1$ 
and the single-copy entanglement entropy $S_A^{(\infty)}$.
Considering a R\'enyi entropy with index $\alpha\in(0,1)$, 
the corresponding  function $s_\alpha(x)$ is increasing for $x \in [0,1/2]$;
therefore we have
\be
\label{ineq-SA-ren}
0 \, \leqslant \, S_A^{(\alpha)} \, \leqslant \, 
\sum_{n =0}^{n_0} s_\alpha(\gamma_n) +  \sum_{n > n_0} s_\alpha(\tilde{\gamma}_n) 
\ee
where $n_0$ is the critical index \eqref{critical-n0-def} 
and $\tilde{\gamma}_n$ is the upper bound introduced in \eqref{gamma-tilde-infty}. 
In order to employ the ratio test for the convergence of a series 
for the one occurring in the last expression of (\ref{ineq-SA-ren}),
we consider the following limit
\be
\label{ratio-test}
\lim_{n \to \infty}
\frac{s_\alpha(\tilde{\gamma}_{n+1}) }{s_\alpha(\tilde{\gamma}_{n}) } 
\,=\,
\lim_{n \to \infty}
\frac{s_\alpha(\tilde{\gamma}_{\infty ,n+1}) }{s_\alpha(\tilde{\gamma}_{\infty ,n}) } 
\,=\,
\lim_{n \to \infty}
\left(\frac{\tilde{\gamma}_{\infty ,n+1} }{\tilde{\gamma}_{\infty ,n} } \right)^\alpha
\,=\,
0 
\ee
where the first equality comes from \eqref{gamma-tilde-infty},
while the second equality has been obtained by using that
$\tilde{\gamma}_{\infty,n} \to 0$ as $n \to \infty$ 
and that $s_\alpha(x) = \tfrac{x^\alpha}{1 - \alpha }  \big(1+o(1)\big)$ as $x \to 0^+$. 
According to the ratio test,
the vanishing of the limit in (\ref{ratio-test}) implies the finiteness of
the series occurring in the last expression of (\ref{ineq-SA-ren});
hence (\ref{ineq-SA-ren}) tells us that also $S_A^{(\alpha)}$ is finite.

The finiteness of the R\'enyi entropies $S_A^{(\alpha)} $ with index $\alpha > 1$ 
can be proved also through trace class condition of the sine kernel \eqref{sine-kernel def}.
Indeed, given a trace-class operator, 
its spectrum $\{ \xi_n\, ;  n\in\mathbb{N}_0\}$ is countable
and $\sum_{n=0}^\infty \xi_n $ is finite.
From (\ref{entropies_func}), we have
$s_\alpha(x) = \tfrac{\alpha}{\alpha - 1} \,x \,(1+o(1))$ as $x\rightarrow 0^+$ when $\alpha>1$;
hence  $\tfrac{s_\alpha(\xi_n)}{\xi_n} \to  \tfrac{\alpha}{\alpha-1}$ as $n \to \infty$.
Combining this observation with the finiteness of $\sum_{n=0}^\infty \xi_n $,
one realises that $\sum_{n=0}^\infty s_\alpha({\xi_n})$ is finite when $\alpha > 1$.
This result and \eqref{ent_func_ine} imply 
the finiteness of the single copy entanglement $S^{(\infty)}_A$.
Notice that, instead, this argument fails for the entanglement entropy and for the 
R\'enyi entropies with index $0<\alpha<1$.

In Sec.\,\ref{sec_spectral}, we mentioned the natural partition of the spectrum in the regions I, II and III.
From a numerical inspection,
we have observed that the contribution to the entanglement entropy 
coming from each part of this partition grows logarithmically for large values of $\eta$.
In the Appendix\;\ref{app_bounds} we discuss bounds for these three terms.

The entanglement entropy $S_A$ is a concave function of $\eta$.
This important feature is a consequence of 
the strong subadditivity property of the entanglement entropy \cite{Lieb-73} 
\be
\label{ssa-def}
S_{A_1} + S_{A_2} \geqslant S_{A_1 \cup A_2} + S_{A_1 \cap A_2} 
\ee
which holds for any choice of two spatial regions $A_1$ and $A_2$.
Since the model that we are considering is defined in one spatial dimension 
and invariant under spatial translations,
the entanglement entropy of an interval is a function of the length of the interval,
for any given value of $k_{\textrm{\tiny F}}$.
Choosing $A_1 = (-R, R)$ and $A_2 = (-R+\epsilon, R+\epsilon)$ for finite $\epsilon >0$,
we have $S_{A_1} = S_{A_2} $ 
because of the translation invariance; 
hence (\ref{ssa-def}) becomes
\cite{Wehrl-78,Casini:2003ix}
\be
2 \,S_A(R)
\,\geqslant\, 
S_A\!\left( \frac{R+\epsilon-(-R)}{2}\right)
+
S_A\!\left( \frac{R-(-R+\epsilon)}{2}\right)\,.
\ee
Assuming that the limit $\epsilon \to 0^+$ of this inequality exists, one finds
\be
\label{EE-prime-decreasing}
\partial_R^2 S_A \,\leqslant \,0 
\ee
which implies that $S_A$ is a concave function of $\eta$.

\begin{figure}[t!]
\vspace{-.2cm}
\hspace{-.8cm}
\includegraphics[width=1.05\textwidth]{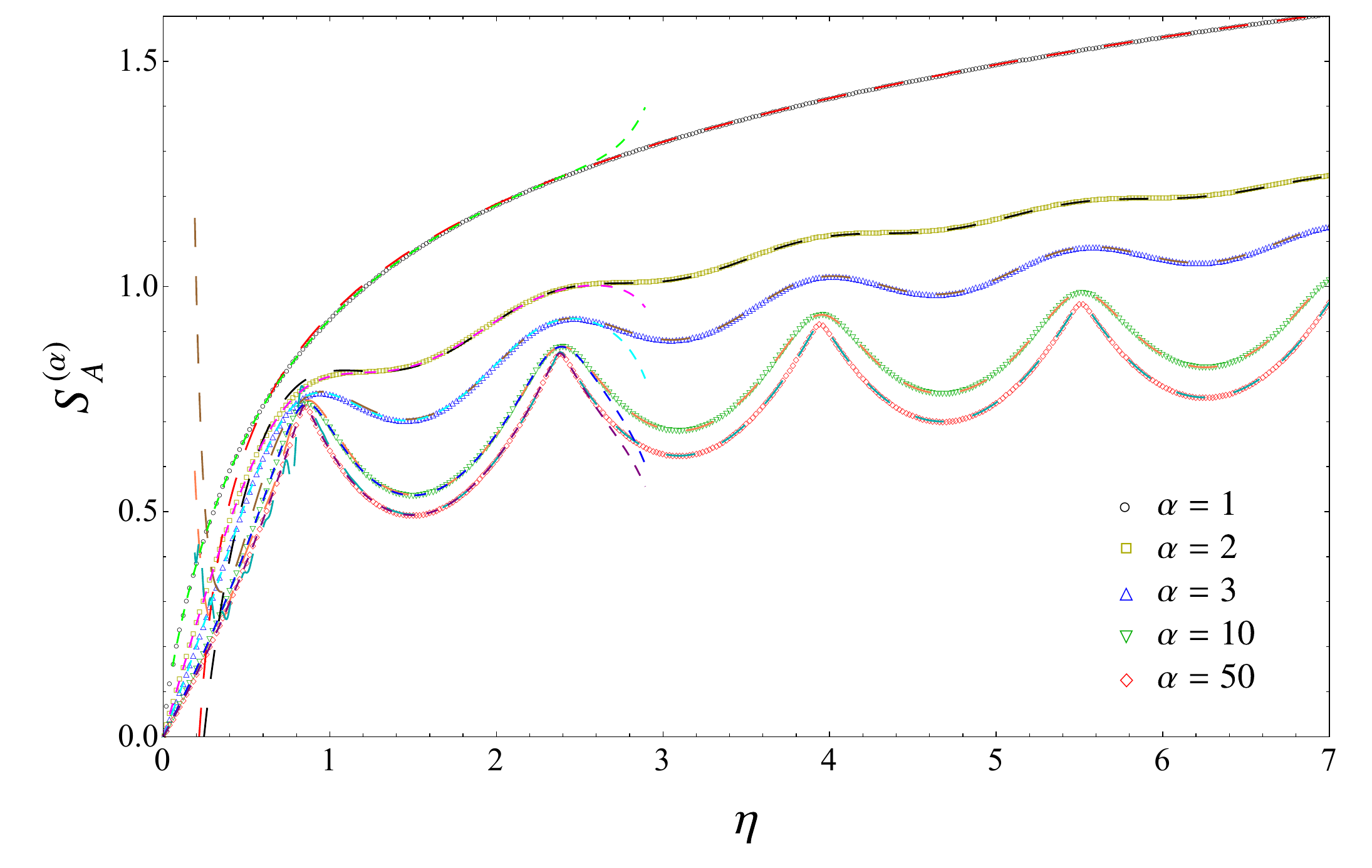}
\vspace{-.7cm}
\caption{
Entanglement entropies:
The data points have been obtained from (\ref{entropies-def-sums}),
while the dashed lines correspond to the small interval expansion
(see \eqref{approx-entropies-small-eta})
and to the large interval expansion 
(see \eqref{entropies-tau-dec-large-eta}, 
\eqref{S_alpha_main_large_eta_0} and \eqref{exp-S-infty-123}).
}
\label{fig:ee-functions-gen}
\end{figure}

The data points displayed in Fig.\,\ref{fig:ee-functions-gen} are numerical results 
for the entanglement entropies of the interval $A = (-R,R)$ for some values of $\alpha$
and the data for the entanglement entropy correspond to $\alpha=1$.
They have been obtained 
by using \eqref{entropies-def-sums} and \eqref{single-copy-ent-sum},
with $\gamma_n$ computed through  the Fortran code developed by Rokhlin
and by evaluating the finite sums made by the contributions corresponding to the first
$n_{\textrm{\tiny max}}=5 (n_0+2)$ eigenvalues,
where $n_0$ is the critical value  \eqref{critical-n0-def}.
 We checked numerically that the entanglement entropies 
 do not change significantly 
by increasing  $n_{\textrm{\tiny max}}$.
This is due to the fact that $n_{\textrm{\tiny max}}$ is linear in $n_0$,
while the width of the region II depends logarithmically on $n_0$ (see \eqref{size_plunge}).
 %
 This criterion for the truncation of the infinite sums have been adopted also for 
 in the numerical determination of other quantities considered in this manuscript.

An analytic formula for the entanglement entropies as functions of $\eta$
is not known in the entire domain $\eta > 0$;
hence it is worth studying analytic expressions for their expansions 
in the regimes of small and large $\eta$.
These analyses are discussed 
in Sec.\,\ref{sec_small_distance}  and Sec.\,\ref{sec_large_distance}
respectively.
We study these expansions through two different approaches:
one based on the expansions of the eigenvalues $\gamma_n$ 
(employed only in the regime of small $\eta$) \cite{Rokhlin-book,Slepian-expansions}
and another one based on the tau function of the sine kernel 
\cite{TracyWidom92,Budylin1996,Gamayun:2013auu,Lisovyy:2018mnj},
which allows to obtain results both for $\eta \to 0$ and for $\eta \to \infty$.

The approach involving the sine kernel tau function 
is based on the method first employed in \cite{Jin_2004, Keating_04}
for the entanglement entropies in some spin chains,
where the spectrum of a Toeplitz matrix is involved. 
This method exploits the possibility to compute the entanglement entropies
in \eqref{entropies-def-sums} and \eqref{single-copy-ent-sum} as specific contour integrals in the complex plane. 
Since $0<\gamma_n <1$ in our case, the closed path to consider 
must encircle the interval $[0,1]$ on the real axis.
A natural choice is
$\mathfrak{C} = \mathfrak{C}_0 \cup \mathfrak{C}_- \cup \mathfrak{C}_1 \cup \mathfrak{C}_+ $,
where $\mathfrak{C}_0$ and $\mathfrak{C}_1$ are two arcs of radius $\epsilon/2$ 
centered in $0$ and $1$ respectively,
while $\mathfrak{C}_\pm$ are the segments 
belonging to the horizontal lines $ x \pm \textrm{i} \delta$ with $x\in \mathbb{R}$ 
and intersecting $\mathfrak{C}_0$ and $\mathfrak{C}_1$
(see e.g. in Fig.\,1 of \cite{Jin_2004}, where a similar path is shown).
Thus, the closed path $\mathfrak{C}$ is parametrised by the infinitesimal parameters $\epsilon$ and $\delta$.
The entanglement entropies in \eqref{entropies-def-sums} and \eqref{single-copy-ent-sum} (we assume that $\alpha =1$ corresponds to the entanglement entropy)
can be written as 
\bea
\label{renyi-det-tau}
S_A^{(\alpha)} 
&=&
 \lim_{\epsilon, \delta \to 0}\,
  \sum_{n=0}^{+\infty} 
   \,\frac{1}{2\pi \textrm{i}} 
 \oint_{\mathfrak{C}}
 \frac{ s_\alpha(z) }{z - \gamma_n}\, \textrm{d}z
  \\
 \rule{0pt}{.7cm}
 &=&
    \lim_{\epsilon, \delta \to 0}\,
    \sum_{n=0}^{+\infty} 
    \, \frac{1}{2\pi \textrm{i}} 
   \oint_{\mathfrak{C}}
 s_\alpha(z) 
 \left\{ \,
\frac{1}{z} + \partial_z  \log(1 - \gamma_n / z) 
 \right\}
 \textrm{d}z
 \\
 \rule{0pt}{.7cm}
 \label{complex-integral-sum}
 &=&
     \lim_{\epsilon, \delta \to 0}\,
    \sum_{n=0}^{+\infty} 
     \,\frac{1}{2\pi \textrm{i}} 
   \oint_{\mathfrak{C}}
 s_\alpha(z) \; \partial_z \log(1 - \gamma_n / z) \, \textrm{d}z
\eea
where the last expression has been obtained by using that $s_\alpha(0) = 0$
for the functions $s_\alpha(x)$ defined in \eqref{entropies_func} and  \eqref{single-copy-ent-sum}.
By exchanging the summation with the integration along $\mathfrak{C}$ in \eqref{complex-integral-sum},
one finds that the entanglement entropies can be computed as follows
\be
\label{entropies-tau}
S_A^{(\alpha)} 
= 
  \lim_{\epsilon, \delta \to 0}\,
   \frac{1}{2\pi \textrm{i}} 
  \oint_{\mathfrak{C}}
 s_\alpha(z) \;
 \partial_z  \log( \tau )\, \textrm{d}z
\ee
where $s_\alpha(x)$ are the functions in \eqref{entropies_func} and  \eqref{single-copy-ent-sum}
and $\tau$ is the sine kernel tau function
\be
\label{tau-function-sine}
\tau\,\equiv\,  \textrm{det}\big( I - z^{-1} K \big) 
\ee
($I$ denotes the identity operator)
which corresponds to the 
Fredholm determinant of the sine kernel $K$ in \eqref{sine-kernel def}.

The expression \eqref{single-copy-ent-sum} is  obtained for holomorphic functions $s_\alpha(z)$.
This is not the case for the function $s_\infty(z)$ in \eqref{single-copy-ent-sum} 
providing the single copy entanglement. 
However, we apply \eqref{single-copy-ent-sum} also for $S_A^{(\infty)}$ 
and then we check the outcomes with the corresponding numerical results.

The sine kernel tau function (\ref{tau-function-sine}) has a long history
and its expansions have been widely studied \cite{JMMS, Jimbo-82, McCoy:1985, TracyWidom92, forrester-book}.
Its auxiliary function known as $\sigma$-form is the solution of a particular Painlev\'e V differential equation 
\cite{JMMS}, as discussed in Sec.\,\ref{subsec-CH-c-function}.
Recent important advances in the analysis of the solutions 
of the Painlev\'e equations started with \cite{Gamayun:2012ma}
have lead to find expansions of the sine kernel tau function to all orders 
\cite{Gamayun:2013auu, Lisovyy:2018mnj, Bonelli:2016qwg}.
In Sec.\,\ref{sec_small_distance} and Sec.\,\ref{sec_large_distance},
we employ these results to obtain
analytic expressions for the expansions in the small and large $\eta$ regimes
which correspond to the dashed curves in Fig.\,\ref{fig:ee-functions-gen}
(see \eqref{approx-entropies-small-eta} for small $\eta$
and  the combination of \eqref{entropies-tau-dec-large-eta}, 
\eqref{S_alpha_main_large_eta_0} and \eqref{exp-S-infty-123} for large $\eta$).

The curves in Fig.\,\ref{fig:ee-functions-gen}
indicate that $S_A^{(\alpha)}$ vanish when $\eta \to 0$.
This is expected from \eqref{2-point-mu-neg}, as emphasised in \cite{Pal:2017ntk, Hason-17, Hartmann:2021vrt};
hence the limit $k_{\textrm{\tiny F}} \to  0$ and the evaluation of $S_A^{(\alpha)}$ commute.

We find it worth anticipating  that the leading term of the large $\eta$ expansion 
for the entanglement entropy  is
$S_A^{(\alpha)} = \tfrac{1}{6}\big(1+\tfrac{1}{\alpha}\big) \log(\eta) + \dots$ as $\eta \to \infty$
(see (\ref{entropies-tau-dec-large-eta}) and (\ref{S_alpha_main_large_eta_0})),
in agreement with the one dimensional case of
the general result found in \cite{Gioev:2006zz},
obtained for fixed $k_{\textrm{\tiny F}}$ and $R \to \infty$. 

In Fig.\,\ref{fig:ee-functions-gen} one observes that $S_A^{(\alpha)}$ with $\alpha \neq 1$ display oscillations, while the entanglement entropy does not oscillate.
The origin of the oscillatory behaviour can be identified with the sawtoothed behaviour of $\gamma_{n_0}$ as function of $\eta$ (see Fig.\,\ref{fig:critic-eigenvalue-n0}), which provides the largest contribution to the entanglement entropies. 
This argument leads to expect oscillations in the entanglement entropy as well,
which, instead, are not observed. 
By numerical inspection, we noticed that this lack of oscillations in the entanglement entropy occurs only when all the infinite sum in \eqref{entropies-def-sums} is taken into account. Indeed, by considering $s(\gamma_{n_0})$ or finite sums $\sum_{k=-p}^{p}  s(\gamma_{n_0+k} )$  for some finite integer $p \geqslant 1 $ we observe oscillating curves and only for large values of $p$ the oscillations disappear. It would be insightful to explore this feature further through the properties of the PSWFs.

Oscillations in the R\'enyi entropies have been investigated 
earlier also for a relativistic  free massive fermion at finite density 
\cite{Swingle-13,Daguerre:2020pte}. They could arise from localised terms on the defect that defines the R\'enyi entropy as a partition function on a sheeted Riemann manifold \cite{Swingle-13}.
A further interpretation of this phenomenon has been provided in \cite{Daguerre:2020pte}
through the defect operator product expansion  in a relativistic setting.  
It would be interesting to apply the same method also in our non-relativistic scenario. 
However, we expect that
the oscillations in the large $\eta$ regime discussed in this manuscript 
(see Sec.\,\ref{sec_large_distance}) are recovered
in the non-relativistic limit considered in \cite{Daguerre:2020pte}.

In the Appendix\;\ref{app_cumulants} we discuss also the cumulants
of the entanglement spectrum, which constitute an alternative to the 
moments of the reduced density matrix. 
We find that these quantities are finite and 
display an oscillatory behaviour (see Fig.\,\ref{fig:app-cumulants}),
similarly to the R\'enyi entropies. 

It is worth investigating  the limits $k_{\textrm{\tiny F}} \to 0$ and $k_{\textrm{\tiny F}} \to \infty$.
In these limits, the correlators in (\ref{g3}) and (\ref{g4})
become (\ref{2-point-mu-neg}) and (\ref{2-point-mu-infty}) respectively,
and both these results give $S_A^{(\alpha)} \to 0$.
Instead, fixing a finite value of $R>0$
and taking $k_{\textrm{\tiny F}} \to 0$ and  $k_{\textrm{\tiny F}} \to \infty$ of $S_A^{(\alpha)}$
in (\ref{entropies-def-sums}) and (\ref{single-copy-ent-sum}),
one obtains $S_A^{(\alpha)}\to 0$ and $S_A^{(\alpha)}\to +\infty$ respectively.
The discrepancy of the limits in the large $k_{\textrm{\tiny F}} $ regime
tells us that the limiting procedure $k_{\textrm{\tiny F}} \to \infty$ and the evaluation of the entanglement entropies do not commute. 
In other words, by writing
$ S_A^{(\alpha)} = \lim_{N \to \infty} \mathcal{S}_{A,N}$,
where $\mathcal{S}_{A,N}^{(\alpha)}  \equiv \sum_{n = 0}^{N} s_\alpha(\gamma_n)$,
we have that the two limits $N \to \infty$ and $\eta \to \infty$ do not commute.
Indeed, by taking $\eta \to \infty$ first and then $N \to \infty$,
one finds  the expected result for $k_{\textrm{\tiny F}} \to \infty$ mentioned at the beginning of this paragraph
because $\mathcal{S}_{A,N} \to 0$ as $\eta \to \infty$ for any finite $N$.
This tells us that entanglement entropies 
in (\ref{entropies-def-sums}) and (\ref{single-copy-ent-sum}) are not uniformly convergent.

It is worth considering the limit $\hbar \to 0$ of our results
(see also the recent analysis in \cite{Mussardo:2021gws}),
which is determined by the behaviour of the dimensionless parameter $\eta$ in this limit. 
From (\ref{eta-def}) and (\ref{kF-def}), one realises that
different results can be obtained
for entanglement entropies when $\hbar \to 0$, 
depending on the quantities that are kept constant in this limit. 
For instance, for fixed $m$ and $R$, 
the limit $\hbar \to 0$ depends on the behaviour of $\mu/\hbar^2$.
In particular, if $\mu = O(\hbar^2)$ then $\eta$ remains finite and non-vanishing
\cite{fujita-90}.

\section{Entanglement along the flow generated by $\eta$}
\label{sec_flow}

Quantifying the loss/gain of information along a flow in the space of parameters
(masses, coupling constants, etc.) characterising a given model 
is a challenging task.
For instance, along a RG flow a loss of information is expected, in some heuristic sense \cite{Preskill-00}. 
In this section we discuss this issue in the specific non-relativistic free field theory that we are exploring
and for the flow parameterised by $\eta$, which is not a RG flow.
The information quantifiers that we consider are
the entanglement entropy (Sec.\,\ref{subsec_ent_loss})
and the quantity analogue to the entropic $C$ function introduced in \cite{Casini:2004bw}
for the relativistic field theories in $d=1$ (Sec.\,\ref{subsec-CH-c-function}).

\subsection{Entanglement entropy loss for decreasing $\eta$}
\label{subsec_ent_loss}

In the free fermionic Schr\"odinger field theory on the line, at zero temperature and finite density, 
we found that the entanglement entropy $S_A$ of an interval $A$ is finite (see Sec.\,\ref{sec_entropies}).
This crucial property makes $S_A$ a natural candidate
to quantify the amount of information shared by the two parts of this bipartition. 
Let us recall that, for any quantum system in a pure state and for any bipartition $A \cup B$ of the space, 
$S_A = S_B$ and it measures the bipartite entanglement associated to the state and to the bipartition.

Consider two finite and non-vanishing values of $\eta$, 
assuming $0< \eta_1 < \eta_2$ without loss of generality.
The area of the limited phase space (\ref{def-red-phae-space}),
given in (\ref{a-parameters-def}),
is proportional to the dimensionless parameter $\eta$
and, in particular, it decreases as $\eta$ decreases.

\begin{figure}[t!]
\vspace{-.2cm}
\hspace{-.8cm}
\includegraphics[width=1.05\textwidth]{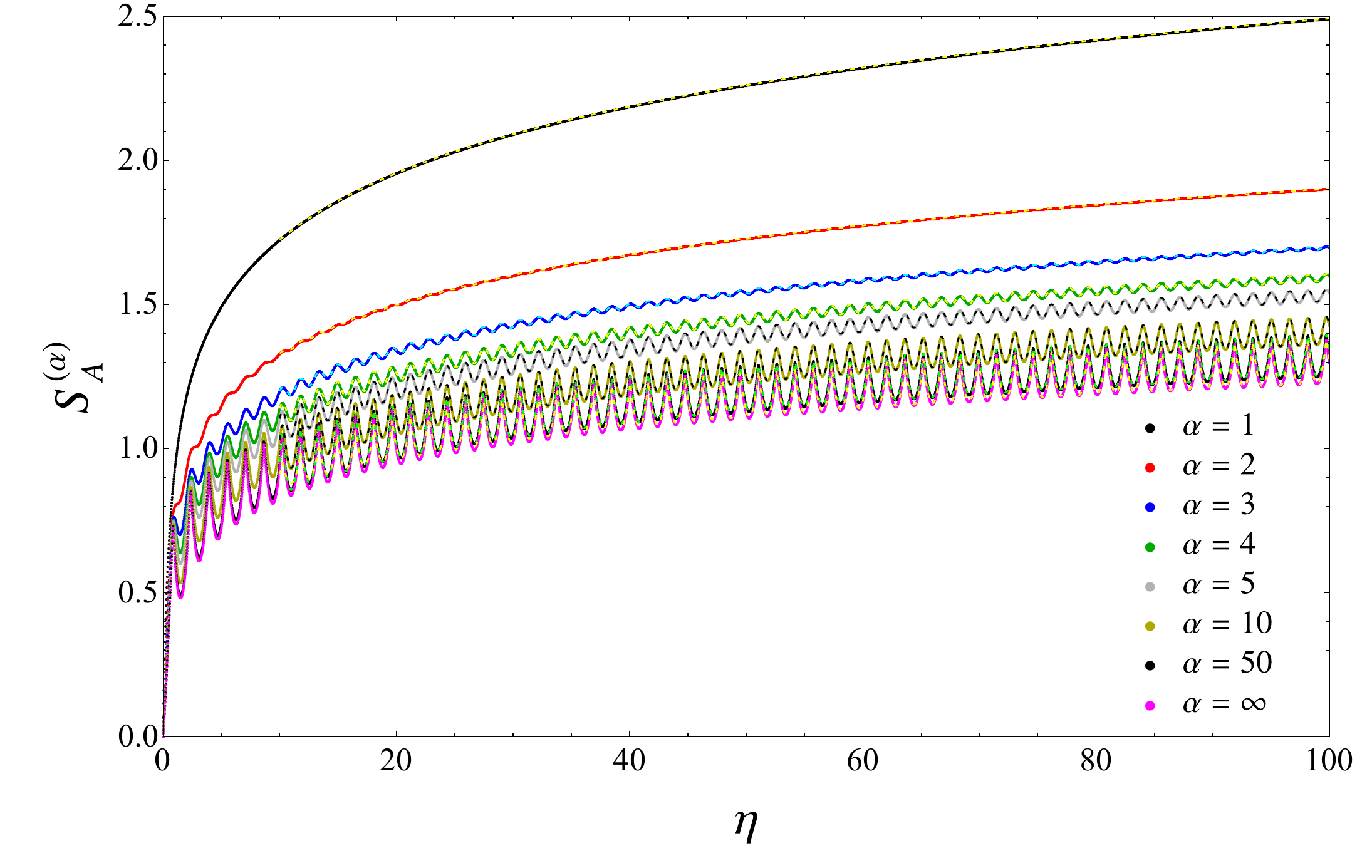}
\vspace{-.7cm}
\caption{
Entanglement entropies $S_A^{(\alpha)}$ for different values of $\alpha$.
The dashed lines (shown only for $\eta > 10$)
correspond to the large $\eta$ expansion given by 
\eqref{entropies-tau-dec-large-eta}, 
\eqref{S_alpha_main_large_eta_0} and \eqref{exp-S-infty-123}.
}
\label{fig:ee-large-100}
\end{figure}

As for the entanglement entropy,
the curves corresponding to $\alpha =1$ in Fig.\,\ref{fig:ee-functions-gen} and Fig.\,\ref{fig:ee-large-100}
show that $S_A$ increases monotonically as $\eta$ increases.
This important feature is not observed for entanglement entropies with $\alpha \neq 1$.

For a given interval, changing $\mu$ modifies the area of the limited phase space.
In this case, $S_A$ increases as $\mu$ increases. 
For the states at zero temperature and different finite density that we are exploring, 
this behaviour is observed also for the mean energy 
$\langle {\cal E}(t,x)\rangle_{\infty, \mu}$,
the mean density $\langle \varrho (t,x) \rangle_{\infty,\mu} $ 
and the central term $c_\mu$ of the Sch\"odinger algebra,
given in (\ref{c3}).

For the translational invariant system we are considering, one can prove that $S_A$ is a 
strictly increasing function of $\eta$. 
The argument \cite{Spitzer-private-comm} is based on the concavity of $S_A$ 
(see (\ref{EE-prime-decreasing}), which is a consequence of the strong subadditivity (\ref{ssa-def}) and of the translation invariance) 
and on the logarithmic growth of $S_A$ for $\eta \to \infty$
(see e.g. Fig.\,\ref{fig:ee-large-100} and Sec.\,\ref{sec_large_distance}).
The proof is by contradiction. 
Assuming that $S_A$ is not strictly increasing,
there exist two points $0\leqslant \eta_1<\eta_0$ such that $S_A(\eta_1) \geqslant S_A(\eta_0)$. 
Because of the logarithmic divergence at infinity, there exists also $\eta_2 > \eta_0$ such that $S_A(\eta_2)>S_A(\eta_0)$. 
Let us fix the parameter $t$ by requiring that $\eta_0 = t \,\eta_1 +(1-t) \eta_2$. 
Then one has $0<t<1$ and 
concavity implies therefore  that
\be 
S_A(\eta_0) \geqslant \,
t \,S_A(\eta_1) +(1-t) S_A(\eta_2) >
t \,S_A(\eta_0) +(1-t)S_A(\eta_0) = 
S_A(\eta_0) \,.
\ee 
This contradiction concludes the argument and proves that $S_A$ 
is strictly increasing with $\eta$.

This feature of $S_A$, combined with its differentiability (that we assume here),
provides some inequalities. 
Taking the derivative of $S_A$ in (\ref{entropies-def-sums}) w.r.t. $\eta$, we obtain
\be
\label{S_A-prime-forms}
S_A'\,
= \sum_{n = 0}^{\infty}  \varepsilon_n\, \gamma_n'
\,=\,
\frac{2}{\eta} \sum_{n = 0}^{\infty}  \, \gamma_n \log\! \big( 1/\gamma_n - 1 \big)  \, f_n(\eta;1)^2
\,=\,
\frac{2}{\eta}
\sum_{n = 0}^{\infty}  
\frac{\varepsilon_n}{\e^{\varepsilon_n} + 1}\, f_n(\eta ; 1)^2 \geqslant 0
\ee
where $\gamma'_n \equiv \partial_{\eta} \gamma_n$
and the single-particle entanglement energies (\ref{ent-eps-def}) have been used. 
The last two expressions in (\ref{S_A-prime-forms}) have been found by employing 
the following remarkable relation occurring 
between any eigenvalue $\gamma_n$ and the corresponding eigenfunction $f_n(\eta; x)$
in the sine kernel spectral problem \eqref{spectral-problem-v2} (see Eq.\,(3.51) in \cite{Rokhlin-book})
\be
\label{gamma-n-prime}
\gamma'_n = \frac{2}{\eta}\, \gamma_n \,f_n(\eta;1)^2 \,.
\ee
The negative terms in the series of $S_A'$ in (\ref{S_A-prime-forms})
correspond to the eigenvalues $\gamma_n >1/2$, which have $\varepsilon_n < 0$ and are a finite number.

We find it suggestive to write (\ref{S_A-prime-forms}) also in the following form
\be
\label{SA-prime-eps-pn}
S_A ' 
=
\frac{2}{\pi} \sum_{n = 0}^{\infty} 
\varepsilon_n\, p_n
\geqslant 0
\ee
where $p_n$ is defined by taking the derivative  of (\ref{sum-gamma-n}) w.r.t. $\eta$,
that gives $ \sum_{n = 0}^{\infty} p_n = 1 $ with 
\be
\label{S_A-prime-pn-def}
p_n 
\,\equiv \,
\frac{\pi}{2}\, \gamma'_n
\,=\,
\frac{\pi}{\eta}\, \gamma_n\, f_n(\eta ; 1)^2 
\,=\,
(2n+1) \big[ \mathcal{R}_{0n}(\eta,1)\,\mathcal{S}_{0n}(\eta,1)\big]^2 
\geqslant 0
\ee
(obtained by exploiting first (\ref{gamma-n-prime}) and then (\ref{eigenvalues}) and (\ref{PSWF-def}));
hence, $\big\{ p_n ; n\in \mathbb{N}_0 \big\}$ can be interpreted as a probability distribution. 
Since the $\gamma_n$ closest to $1/2$ has $n$ next to $n_0$,
(\ref{SA-prime-eps-pn}) implies that, for any given $\eta$, 
a finite integer $\tilde{n}_0 > n_0$ exists such that
$ \sum_{n = 0}^{\tilde{n}_0}  \varepsilon_n\, p_n  \geqslant 0 $.

From (\ref{ee-thermo-form}),
the property $\Delta S_A \equiv S_A(\eta_2) - S_A(\eta_1) > 0$ when  $\eta_2 >  \eta_1$
implies also that $\Delta E_A  > \Delta \Omega_A$,
where $\Delta E_A$ and $\Delta \Omega_A$ are the variations of the corresponding quantities introduced in (\ref{EA-OmegaA-def}),
which do not have a definite sign. 

An interesting inequality involves the relative entropy of the two reduced density matrices 
$\rho_{A,2} \equiv \e^{-K_{A,2}}$ and $\rho_{A,1} \equiv \e^{-K_{A,1}}$ 
(normalised to $\textrm{Tr} (\rho_{A,2}) = \textrm{Tr} (\rho_{A,1}) = 1$)
associated to the same interval $A$ and to two different values $\mu_1 < \mu_2$,
where the operator $K_{A} \equiv - \log (\rho_A)$ is the entanglement (or modular) Hamiltonian
\cite{Haag:1992hx, Hislop:1981uh, EislerPeschel:2009review, Peschel:2003rdm, Casini:2009vk, Casini:2011kv, Cardy:2016fqc}.
The relative entropy is \cite{Vedral:2002zz, Blanco:2013joa}
\be
\label{rel-ent-def}
S(\rho_{A,2} | \rho_{A,1})
\equiv
\textrm{Tr}\big( \rho_{A,2} \log \rho_{A,2} \big) - \textrm{Tr}\big( \rho_{A,2} \log \rho_{A,1} \big) 
=
\Delta \langle K_{A,1} \rangle_{_A} - \Delta S_A
\ee
where
\be
\Delta \langle K_{A,1} \rangle_{_A}
\equiv
\textrm{Tr}\big( \rho_{A,2} \,K_{A,1} \big) - \textrm{Tr}\big( \rho_{A,1} \,K_{A,1} \big) \,.
\ee
We remind that $S(\rho_{A,2} | \rho_{A,1})  \geqslant 0$,
where the equality holds if and only if $\rho_{A,1} = \rho_{A,2}\,$.
This inequality and the assumption $\Delta S_A >0$ for $\eta_2 >  \eta_1$
imply respectively that
\be
\label{ineq-bek}
\left\{\,
\begin{array}{l}
\Delta \langle K_{A,1} \rangle_{_A} > \Delta S_A
\\
\rule{0pt}{.55cm}
\Delta \langle K_{A,1} \rangle_{_A} > S(\rho_{A,2} | \rho_{A,1}) 
\end{array}
\right.
\;\;\;\;\qquad\;\;\;\;
\eta_2 >  \eta_1
\ee
where the first inequality is the Bekenstein bound in the form discussed  in \cite{Casini:2008cr}.
In (\ref{ineq-bek}) both $\Delta S_A$ with $S(\rho_{A,2} | \rho_{A,1})$ are strictly positive
and it would be interesting to compare them.

The loss of information along a RG flow
has been explored also through
the majorization condition
\cite{vidal-03,Latorre-04,Latorre-05,Latorre-06,Orus-05}.

Consider two reduced density matrices $\rho_1$ and $\rho_2$ normalised by
$\textrm{Tr}(\rho_1) = \textrm{Tr}(\rho_2) = 1$ and  their corresponding spectra
$\boldsymbol{\lambda}_1$ and $\boldsymbol{\lambda}_2 $
(which can be interpreted as probability distributions)
with the elements sorted in decreasing order
$\lambda_{\textrm{\tiny max}} \geqslant \lambda_1 \geqslant \lambda_2 \geqslant \dots$, 
where $\lambda _0 \equiv \lambda_{\textrm{\tiny max}} $ is the largest eigenvalue.
By definition, $\boldsymbol{\lambda}_1$ is majorised by $\boldsymbol{\lambda}_2$
and denoted by $\boldsymbol{\lambda}_2\succ \boldsymbol{\lambda}_1$,
when \cite{bengtsson-book,bhatia-book} 
\be
\label{majorisation-condition}
\sum_{j=0}^p \lambda_{2,j} \,\geqslant\,  \sum_{j=0}^p \lambda_{1,j} 
\;\;\;\;\qquad\;\;\;\;
\forall\, p \in \mathbb{N}_0 \,.
\ee
Accordingly, for two density matrices $\rho_1$ and $\rho_2$, 
one defines $\rho_2 \succ \rho_1$ when the same relation holds between the corresponding spectra. 
It is not likely that all the infinitely many inequalities (\ref{majorisation-condition}) are simultaneously satisfied.
However, when (\ref{majorisation-condition}) holds, insightful inequalities can be written. 
For instance, denoting by $S(\rho) \equiv - \,\textrm{Tr} (\rho \log \rho)$
the von Neumann entropy of a density matrix,
$\rho_2 \succ \rho_1$ implies $S(\rho_2) < S(\rho_1)$
because $S(\rho)$ is a Schur concave function of the entanglement spectrum.
This property holds also for the R\'enyi entropies. 

In the model we are exploring, it is natural to explore the majorization condition between
two reduced density matrices $\rho_{A,2}$ and $\rho_{A,1}$
whose spectra are associated to two values $\eta_1$ and $\eta_2$, 
which can be chosen $\eta_2 > \eta_1$ without loss of generality.
This corresponds to either two different intervals with $R_2 > R_1$ at fixed $k_{\textrm{\tiny F}}$ 
or to the reduced density matrices of the same interval for two different states of the entire system 
characterised by Fermi momenta $k_{\textrm{\tiny F}}\big|_{\eta_2} > k_{\textrm{\tiny F}} \big|_{\eta_1}$.
The single copy entanglement $S_A^{(\infty)}= - \log (\lambda_{\textrm{\tiny max}})$ shown in 
Fig.\,\ref{fig:ee-large-100}
(see also Fig.\,\ref{fig:small-eta-panels},  Fig.\,\ref{fig:ee-large-900} and Fig.\,\ref{fig:E-infty-large-eta})
allows to explore the possible occurrence of a majorization relation
between these two reduced density matrices.
In particular, the oscillatory behaviour of $S_A^{(\infty)}$ rules out a majorization relation
between $\rho_{A,2}$ and $\rho_{A,1}$ for two generic  $\eta_1$ and $\eta_2$
because the validity of the inequality (\ref{majorisation-condition}) for $p=0$ 
depends on the specific choice of $\eta_1$ and $\eta_2$.

\subsection{The analogue of the relativistic entropic $C$ function}
\label{subsec-CH-c-function}

In the class of the $d=1$ relativistic quantum field theories,
Zamolodchikov \cite{Zamolodchikov:1986gt}
constructed a finite $C$ function that 
monotonically decreases along the RG flow
and takes finite values at its fixed points 
equal to the central charges of the corresponding conformal field theories.

Considering the entanglement entropy associated to 
an interval of length $\ell$ in the infinite line when the system is in its ground state,
Casini and Huerta \cite{Casini:2004bw} 
introduced a different function $C \equiv \ell \, \partial_\ell S_A$,
proving that it is UV finite, 
it takes finite values proportional to the central charge at the fixed points
and it monotonically decreases along the RG flow;
hence this function is usually called entropic $C$ function. 
The proof of the monotonicity of this entropic $C$ function is based on 
the relativistic invariance and on the  strong subadditivity property of the entanglement entropy.
We refer the interested reader to the review \cite{Nishioka:2018khk} 
for further discussions and references about this entropic $C$ function 
and its generalisations to higher dimensions \cite{Casini:2012ei, Liu:2012eea, Jafferis:2011zi, Casini:2017vbe}.
Explicit examples of entropic $C$ functions have been studied 
for relativistic free massive boson and Dirac fermion
where $M R$ is the dimensionless parameter generating the flow,
with $M$ being the relativistic mass of the field
\cite{Casini:2005rm, Casini:2005zv, Casini:2009sr}.
Both the Zamolodchikov's $C$ function and the entropic $C$ function are
constructed through the ground state of the model along the RG flow.

In the non-relativistic model that we are exploring,
it is worth considering the analogue of the entropic $C$ function
introduced for the relativistic models.
This has been done in other settings e.g. in \cite{Daguerre:2020pte, Boudreault:2021pgj}.

The entanglement entropies discussed in Sec.\,\ref{sec_entropies} allow us to introduce
\be
\label{c-function-alpha}
C
\,\equiv\, 
 \eta\, \partial_{\eta} S_A
\;\; \;\;\;\qquad\;\;\;\;\;
C_{\alpha} 
\,\equiv\, 
 \eta\, \partial_{\eta} S_A^{(\alpha)}
\ee
where $C$ is the analogue of the relativistic entropic $C$ function in our model. 
From \eqref{entropies-def-sums} and the fact that $\gamma_n$ depend only on $\eta$,
it is straightforward to find that the functions \eqref{c-function-alpha} 
can be written respectively as 
\be
\label{C-functions-gamma-prime}
C\,= \sum_{n = 0}^{\infty}  \big[ \log\! \big(1/\gamma_n - 1\big) \big] \, \eta \,\gamma'_n
\;\;\qquad\;\;
C_\alpha\,=\, \frac{\alpha}{\alpha - 1}\,
\,\sum_{n = 0}^{\infty}   
\frac{(1-\gamma_n)^{\alpha-1} - \gamma_n^{\alpha-1}}{(1-\gamma_n)^{\alpha} + \gamma_n^{\alpha}}
\;\eta \, \gamma'_n
\ee
By employing (\ref{gamma-n-prime}), these expressions become respectively
\be
\label{C-function-def}
C = 
2 \sum_{n = 0}^{\infty}  \, \gamma_n \log\! \big( 1/\gamma_n - 1 \big)  \, f_n(\eta;1)^2
\;\qquad\;
C_\alpha = 
\frac{2\,\alpha}{\alpha - 1}
\sum_{n = 0}^{\infty} \gamma_n\,
\frac{(1-\gamma_n)^{\alpha-1} - \gamma_n^{\alpha-1}}{(1-\gamma_n)^{\alpha} + \gamma_n^{\alpha}}  
\; f_n(\eta;1)^2\,.
\ee
In the Appendix\;\ref{app_finiteness-C-alpha} we show that
these series are well defined functions of the dimensionless parameter $\eta$.
In our numerical analyses of $C$ and $C_\alpha$,
we have employed (\ref{C-function-def}) 
by truncating the infinite sums as discussed in Sec.\,\ref{sec_entropies}.

\begin{figure}[t!]
\vspace{-.2cm}
\hspace{-.8cm}
\includegraphics[width=1.05\textwidth]{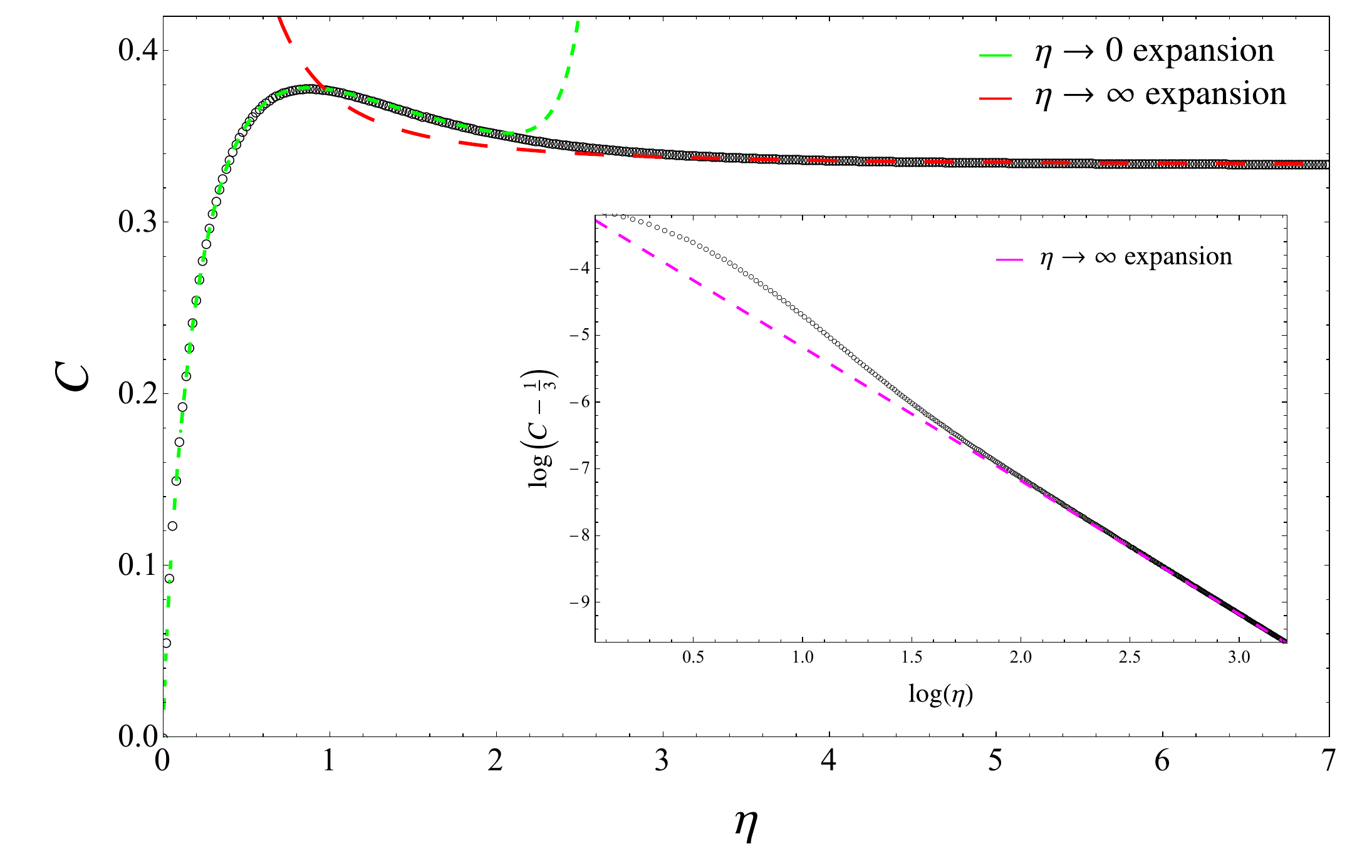}
\vspace{-.7cm}
\caption{
The quantity $C$ defined in (\ref{c-function-alpha}):
The data points have been obtained from \eqref{C-function-def},
while the dashed lines correspond to the small and large interval expansions,
found from (\ref{approx-ch-function-small-eta}) and (\ref{large_eta_c_fun}) respectively.
The inset focuses on the large interval regime in logarithmic scale
and the straight dashed magenta line corresponds to (\ref{large_eta_c_fun}).
}
\label{fig:c-functions-gen}
\end{figure}

The expression \eqref{entropies-tau} allows us to write
the functions $C$ and $C_\alpha$  in \eqref{c-function-alpha} in the form
\be
\label{C-from-sigma}
C_\alpha 
\,=\,
\frac{1}{2\pi \textrm{i}} \oint_{\mathfrak{C}}
s_\alpha(z)\, 
\partial_z \sigma\, \textrm{d}z
\ee
in terms of the auxiliary function $\sigma$ associated to a tau function
\be
\label{sigma-function-def}
\sigma \,\equiv\, \eta\,\partial_{\eta}  \log(\tau)
\ee
and in our case $\tau$ is the sine kernel tau function (\ref{tau-function-sine}).

It is well known that the auxiliary function $\sigma$ 
associated to the sine kernel tau function 
is the solution of the following Painlev\'e\;V differential equation (written in the $\sigma$-form)
\cite{JMMS,Jimbo-82,McCoy:1985, TracyWidom92}
\be
\big( \eta \,\partial^2_{\eta} \sigma \big)^2
+
16 \,
\big(\,  \eta \,\partial_{ \eta} \sigma  -\sigma \,\big) \,
\bigg[\,  \eta \,\partial_{ \eta} \sigma  -\sigma + \frac{1}{4}\,( \partial_{ \eta} \sigma )^2 \, \bigg]
=0
\ee
equipped with the following boundary condition
\be
\label{sigma-P5-bc}
\sigma \,=\, 
- \, \frac{2\,\eta}{\pi \, z} - \left( \frac{2\,\eta}{\pi \, z} \right)^2 + O\big(\eta^3\big)
\;\;\;\qquad\;\;\;
\eta \,\to \, 0 \,.
\ee

\begin{figure}[t!]
\vspace{-.2cm}
\hspace{-.8cm}
\includegraphics[width=1.05\textwidth]{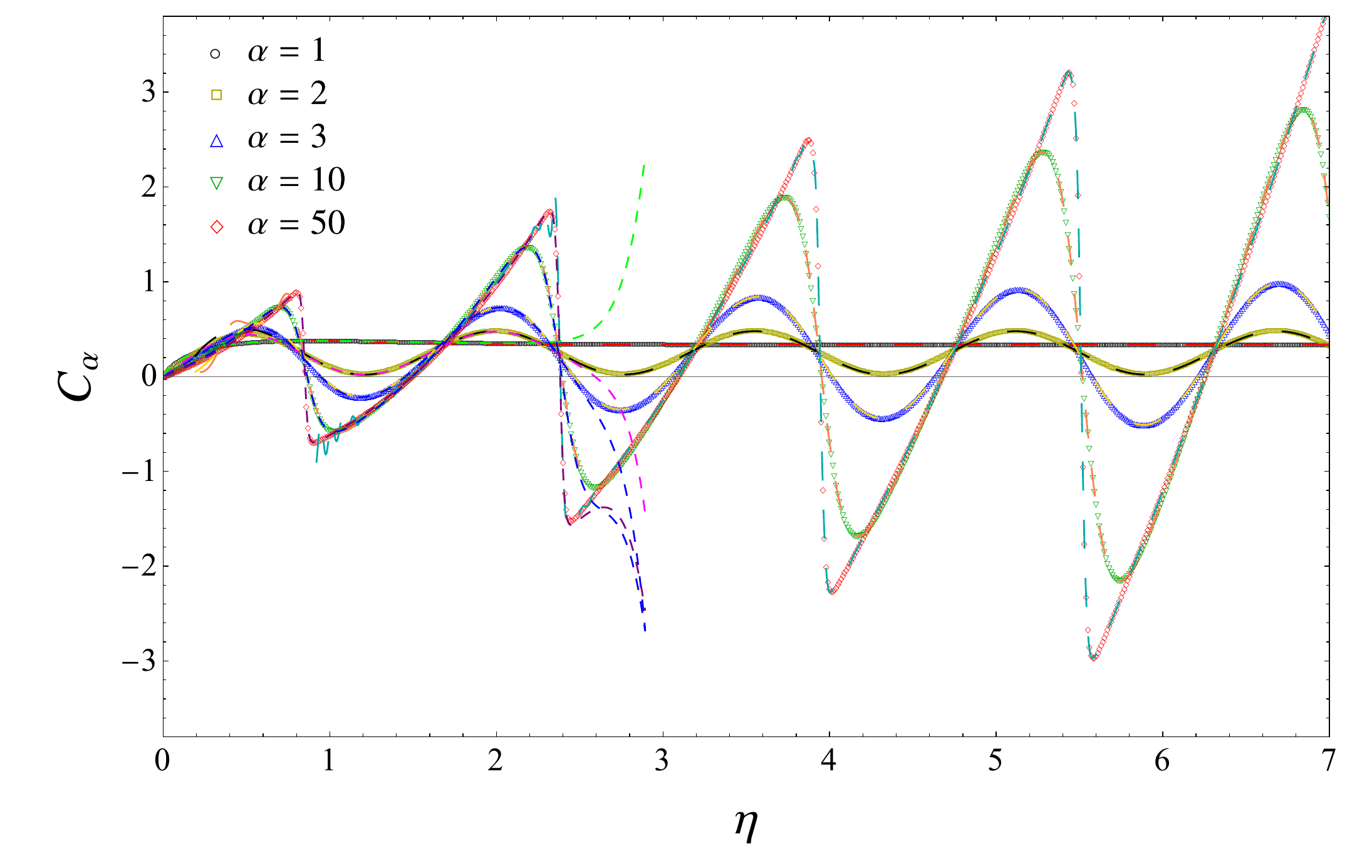}
\vspace{-.7cm}
\caption{
The quantity $C_\alpha$ defined in (\ref{c-function-alpha}):
The data points correspond to \eqref{C-function-def}
and the dashed lines are obtained from the small and large interval expansions
derived in Sec.\,\ref{sec_small_distance} (see \eqref{approx-ch-function-small-eta}) 
and Sec.\,\ref{sec_large_distance} 
(see \eqref{entropies-tau-dec-large-eta}, \eqref{S_alpha_main_large_eta_0} and \eqref{exp-S-infty-123}) 
respectively.
}
\label{fig:c-functions-alpha}
\end{figure}

In Fig.\,\ref{fig:c-functions-gen} 
we show the quantity $C$ defined in \eqref{c-function-alpha}
in terms of $\eta$.
The empty circles denote the data points obtained from \eqref{C-function-def},
while the coloured dashed lines correspond to the 
asymptotic behaviours of $C$ 
when $\eta \to 0$ (green line)
and when $\eta \to \infty$ (red and magenta lines),
which are analytic expressions derived respectively 
from (\ref{approx-ch-function-small-eta}) (see also (\ref{small_c_o5})) 
and from \eqref{entropies-tau-dec-large-eta}, 
\eqref{S_alpha_main_large_eta_0} and \eqref{exp-S-infty-123}.
In particular,  we find that 
 $C =  -\frac{2}{\pi}\, \eta\log(\eta) +\dots$ as $\eta \to 0$ 
 and that
\be
\label{large_eta_c_fun}
 C \,=\, \frac{1}{3}+\frac{1}{24 \,\eta^2}+O(1/\eta^{4})
 \;\;\;\qquad\;\;\;
 \eta \rightarrow \infty \,.
\ee
These two asymptotic behaviours and the assumption that $C$ is a continuous function
imply that $C$ must possess at least one local maximum;
hence it cannot be monotonous. 
This analysis cannot determine the number of local maxima. 
From the numerical data points in Fig.\,\ref{fig:c-functions-gen}
we observe that $C$ has only one local maximum.
It would be insightful to find a proof for this numerical result. 
%


In Fig.\,\ref{fig:c-functions-alpha} we show the quantities $C_\alpha$ introduced in \eqref{c-function-alpha}
(see also Fig.\,\ref{fig:c-20-large-100}).
The numerical data points (empty circles) are obtained through \eqref{C-function-def}
and the coloured dashed lines correspond to the asymptotic results derived in
Sec\,\ref{sec_small_distance} (see (\ref{approx-ch-function-small-eta}))
and Sec\,\ref{sec_large_distance} (see \eqref{entropies-tau-dec-large-eta}, 
\eqref{S_alpha_main_large_eta_0} and \eqref{exp-S-infty-123}).
We remark that, while $C$ does not oscillate, 
$C_\alpha$ with $\alpha > 1$ display an oscillatory behaviour, which 
tends to a sawtoothed curve as $\alpha \to +\infty$,
as discussed also at the end of Sec.\,\ref{sec_large_distance}.

\section{Integer Lifshitz exponents}
\label{sec_lifshitz_exponent}

In this section we consider  the family of Lifshitz fermion fields $\psi(t,x)$ 
whose time evolution is given by 
\begin{equation}
\left[ \,\ri \hbar\,\partial_t -\frac{1}{(2m)^{z-1}} \,(-\ri \hbar\, \partial_x)^z \, \right]\psi (t,x) = 0 
\;\;\;\qquad \;\;\;
z \in {\mathbb N} \,.
\label{e1L}
\end{equation} 
In addition to (\ref{e1L}), 
the equal time canonical anticommutation relations (\ref{e2a}) and (\ref{e2b})
are imposed on $\psi(t,x)$. 
For $z=1$, Eq.\,(\ref{e1L}) gives the familiar relativistic equation of motion of a chiral fermion $\psi(x-t)$.
Instead, the case $z=2$ corresponds to the Schr\"odinger equation (\ref{e1}). 
We find it worth considering the hierarchy of models corresponding to $z \in \mathbb{N}$ 
in a unified way by introducing the dispersion relation 
\begin{equation}
\omega_z(k) \equiv \frac{ \hbar^{z-1} \, k^z}{(2m)^{z-1}}  \,.
\label{Lif1}
\end{equation}

The solution of (\ref{e1L}) 
satisfying the anticommutation relations (\ref{e2a}) and (\ref{e2b}) is still given by (\ref{e3}) with the 
substitution $\omega(k) \longmapsto \omega_z(k)$. 
We observe in this respect that for even $z$ one usually employs the alternative basis 
$\big\{b(k) \equiv a(k)\,,\, c(k) \equiv a^*(-k) \, :\, k\geqslant 0 \big\}$, 
where $b(k)$ and $c(k)$ are interpreted as annihilation operators of particles and antiparticles respectively. 
Performing the substitution $\omega(k) \longmapsto \omega_z(k)$
 in (\ref{g1}) and (\ref{g2}), 
one obtains the two point Lifshitz correlation functions in the Gibbs representation at temperature $\beta$ 
and chemical potential $\mu$.

Hereafter, it is convenient to distinguish between even and odd values of $z$. 
In the zero temperature limit $\beta \to \infty$ one gets
\be
\label{gLL}
\langle \psi^*(t,x_1) \, \psi (t,x_2)\rangle_{z,\infty,\mu}
\,=\,
\left\{\begin{array}{lll}
\displaystyle
\; \frac{\sin (k_{\textrm{\tiny F}, z} \,x_{12}  )}{\pi \, x_{12}} 
\hspace{1.5cm}&
z=2n
\hspace{1cm}&
\mu > 0
\\
\rule{0pt}{1.cm}
\displaystyle
\;\frac{\e^{-\ri k_{\textrm{\tiny F}, z} \,x_{12}} }{2\pi \ri \,(x_{12} -\ri \varepsilon )} 
&
z=2n+1
\hspace{1cm}&
\mu \in \RR
\end{array}
\right.
\ee
where $n \in \mathbb{N}$ and the Fermi momentum is 
\be
\label{kF-def-L}
p_{\textrm{\tiny F},z} \equiv (2m)^{1-1/z}   \mu^{1/z}
\;\;\;\;\qquad\;\;\;\;
k_{\textrm{\tiny F},z} \equiv \frac{p_{\textrm{\tiny F},z} }{\hbar} \,.
\ee
Thus, $p_{\textrm{\tiny F},z=1} = \mu$, while  $p_{\textrm{\tiny F},z} \to 2m$ as $z\to +\infty$.
The special case $\mu = 0$ has been already discussed in \cite{Hartmann:2021vrt}
and in the following we consider the case of non-vanishing $\mu$.

The entanglement entropies of the interval $A=[-R , R]$
for the Lifshitz fermions (\ref{e1L}) can be obtained from (\ref{entropies-def-sums}), (\ref{single-copy-ent-sum})
and the spectrum of the kernel (\ref{gLL}) restricted to $A$.

When $z=2n$, it is straightforward to observe that the solution of this spectral problem is 
simply obtained by replacing $\eta \longmapsto R \,k_{\textrm{\tiny F},2n}$
in the solution of (\ref{spectral-problem-R}).
Hence all the results obtained in this manuscript 
for the entanglement entropies in the model with $z=2$ 
can be easily extended to the Lifshitz models with $z=2n$ 
through this simple replacement.

When $z=2n+1$, from (\ref{gLL}) one must analyse the spectral problem 
\be 
\int_{-R}^R \frac{\e^{-\ri k_{\textrm{\tiny F}, z} (x-y) }}{2\pi \ri\,(x-y -\ri \varepsilon )} \;\phi_s (y)\,\rd y 
\,=\, 
\gamma_s \, \phi_s(x) 
\;\;\;\qquad \;\;\;
s \in {\mathbb R} \,.
\label{Lif2}
\ee
Setting
\be 
\tilde {\phi}_s(x)\, \equiv\, \e^{\ri k_{\textrm{\tiny F},z} x}  \, \phi_s (x) 
\label{Lif3}
\ee 
the spectral problem (\ref{Lif2}) simplifies to  
\be 
\int_{-R}^R \frac{1}{2\pi \ri\,(x-y -\ri \varepsilon )} \,\tilde {\phi}_s (y)\,\rd y 
\,=\, 
\gamma_s \,\tilde {\phi}_s(x) 
\;\;\;\;\qquad \;\;\;\;
s \in {\mathbb R}
\label{Lif4}
\ee 
which has the same eigenvalues as (\ref{Lif3}).
The solution of the spectral problem (\ref{Lif4}) is well known \cite{Musk-book, Casini:2009vk, Arias:2016nip}
and, in particular, its eigenvalues read
\be 
\gamma_s = \frac{1-\tanh(\pi s)}{2} 
\;\;\;\qquad \;\;\;
s \in {\mathbb R} \,.
\label{Lif5}
\ee

Contrary to case of even $z$, for odd $z$ the spectrum (\ref{Lif5}) is continuous. 
This feature reflects the behavior of the dispersion relation (\ref{Lif1}), 
which is bounded from below for even $z$ and unbounded for odd $z$. 
Another substantial difference between 
the eigenvalues (\ref{eigenvalues}) and (\ref{Lif5}) is that the latter ones do not depend on the 
Fermi momentum (\ref{kF-def-L}) and therefore they are independent of the Lifshitz exponent. 
As a consequence, all Lifshitz fermions with odd $z$ have the same entanglement entropies, 
which coincide with the ones of the relativistic massless chiral fermion, i.e.
\cite{Holzhey:1994we, Calabrese:2004eu}
\be 
S_A^{(\alpha)} 
\,=\, 
\frac{1}{12}  \left(1+ \frac{1}{\alpha}\right)
\log (2R/\epsilon) + O(1) \,.
\label{Lif6}
\ee
We remark that the independence of the spectrum on the Fermi momentum leads 
to a well known logarithmic ultraviolet divergency, 
which induces the presence of the UV cut off $\epsilon$ in (\ref{Lif6}).

Summarising, the entanglement entropies for the Lifshitz hierarchy of models given by (\ref{e1L}) are 
ultraviolet divergent and $\mu$-independent for odd $z$,
while they are finite and $\mu$-dependent for even $z$. 
This tells us that the entanglement entropies are heavily influenced
by the global form of the dispersion relation and not only 
by its behaviour close to the Fermi momentum.

In the models characterised by odd values of $z$,
we can find entanglement quantifiers that explicitly depend on the Lifshitz exponent.
An important example is the entanglement Hamiltonian $K_A$ 
(also known as modular Hamiltonian) \cite{Haag:1992hx},
which provides the reduced density matrix $\rho_A \propto \e^{-K_A}$.
For fermionic free models, this operator can be studied through the Peschel's formula
\cite{Peschel:2003rdm, EislerPeschel:2009review}, 
which has been largely explored \cite{Casini:2009sr, Arias:2016nip, Arias:2017dda, Hollands:2019hje,
Blanco:2019xwi, Fries:2019ozf, Mintchev:2020uom, Mintchev:2020jhc},
also in its bosonic version \cite{Casini:2009sr, Banchi:2015aaa, Arias:2017dda, Arias:2018tmw, DiGiulio:2019cxv, Eisler:2020lyn}.
For Lifshitz fermion (\ref{e1L}) with odd $z$
and in the Gibbs state at zero temperature and finite density,
we find that $K_A$ for the interval $A\subset \RR$ is
(the derivation is reported in the Appendix\;\ref{app_mod-ham-flow})
\be
\label{K_A-local-def}
K_A
\,=\,
-\,2\pi 
\int_A
\beta_{\textrm{\tiny loc}}(x)  \big[ \, T_{tt}(0,x) - k_{\textrm{\tiny F},z}\, \varrho (0,x) \,\big]\, \rd x 
\;\;\;\qquad\;\;\;
\beta_{\textrm{\tiny loc}}(x) \equiv \frac{R^2 - x^2}{2R}
\ee
where $k_{\textrm{\tiny F},z}$ is defined in (\ref{kF-def-L}),
$\varrho (t,x)$ is the particle density introduced in (\ref{curr2}) and 
\be
\label{T00-lambda-def}
T_{tt}(0,x) 
\,\equiv\,
\,-\frac{\textrm{i}}{2}
\big( (\partial_x \psi^\ast)\, \psi - 
\psi^\ast\, (\partial_x \psi) \big)(x) \,.
\ee
Notice that the two operators $T_{tt}(0,x) $ and $\varrho (0,x)$ in (\ref{K_A-local-def})
are normal ordered in the basis of oscillators given by
$\big\{a(k)\,, a^*(k) \, :\, k \in \RR \big\}$.
In the special case of $z=1$, the expression (\ref{K_A-local-def}) becomes 
the entanglement Hamiltonian of the relativistic massless chiral fermion $\psi(x-t)$ 
at zero temperature and finite density \cite{Wong:2013gua}.

The reduced density matrix generates the one-parameter family of unitary operators $\{ \rho_A^{\textrm{i}\tau} :  \tau \in \RR\}$,
which defines an automorphism on the operator algebra known as modular flow \cite{Haag:1992hx}.
The modular flow of the field is defined as 
\be
\label{mod-evolution-lambda}
 \psi(\tau,x)
 \,\equiv\,
\rho_A^{\textrm{i} \tau} \, \psi(x)\, \rho_A^{-\textrm{i} \tau} 
 \,=\,
\e^{-\textrm{i} \tau K_A} \, \psi(x)\, \e^{\textrm{i} \tau K_A} 
\;\;\;\; \qquad \;\;\;\;
x\in A
\ee
where $\psi(x)$ is the initial configuration at  $t=0$. 
In Appendix\;\ref{app_mod-ham-flow},
by adapting the analysis described in \cite{Casini:2009vk, Mintchev:2020uom}, 
we find 
\be
\label{mod-flow-mu-main}
\psi(\tau,x)  
\,=\,
\e^{-\ri k_{\textrm{\tiny F},z}[ \xi(\tau,x) - x]}\;
\sqrt{ \frac{\beta_{\textrm{\tiny loc}}\big( \xi(\tau,x) \big)}{\beta_{\textrm{\tiny loc}}(x)} } \; \psi \big( \xi(\tau,x) \big)
\ee
where $\beta_{\textrm{\tiny loc}}(x)$ is defined in (\ref{K_A-local-def}) and 
\be
\label{xi-interval-line-gs}
 \xi(\tau,x) 
\equiv
R\, \frac{(x+R)\, \e^{2\pi \tau} -(R-x) }{(x+R)\, \e^{2\pi \tau} + R- x} \,.
\ee
Notice that the solution (\ref{mod-flow-mu-main})
satisfies the initial condition $\psi(0,x)  = \psi(x)$, as expected. 

Thus, both $K_A$ and the modular flow (\ref{mod-flow-mu-main}) explicitly depend on the Lifshitz exponent.

The expression for the field along the modular flow in (\ref{mod-flow-mu-main}) 
allows us to construct the corresponding correlation functions. 
For instance, we have
\bea
\label{mod-corr-1}
\langle \psi^\ast(\tau_1,x_1)  \, \psi(\tau_2,x_2)  \rangle_{\infty, \mu} 
&=&
\e^{\ri k_{\textrm{\tiny F},z}[ \xi_{12} - x_{12}]}\;
\sqrt{ \frac{\beta_{\textrm{\tiny loc}}(\xi_1)\, \beta_{\textrm{\tiny loc}}(\xi_2)}{
\beta_{\textrm{\tiny loc}}(x_1)\, \beta_{\textrm{\tiny loc}}(x_2)} } 
\;\langle \psi^\ast(\xi_1) \, \psi(\xi_2) \rangle_{\infty, \mu}
\\
\rule{0pt}{.9cm}
\label{mod-corr-2}
&=&
\e^{- \ri k_{\textrm{\tiny F},z} x_{12} }\;
\sqrt{ \frac{\beta_{\textrm{\tiny loc}}(\xi_1)\, \beta_{\textrm{\tiny loc}}(\xi_2)}{
\beta_{\textrm{\tiny loc}}(x_1)\, \beta_{\textrm{\tiny loc}}(x_2)} } \;\,
\frac{1 }{2\pi \ri \,(\xi_{12} -\ri \varepsilon )} 
\\
\rule{0pt}{.8cm}
\label{mod-corr-3}
&=&
\frac{\e^{- \ri k_{\textrm{\tiny F},z} x_{12} }}{2\pi \ri \,(x_{12} -\ri \varepsilon )}\;\,
\frac{\e^{w(x_1)} -\e^{w(x_2)} }{\e^{w(x_1)+\pi \tau_{12}} -\e^{w(x_2) -\pi \tau_{12}} }
\eea
where $\xi_j \equiv \xi(\tau_j, x_j)$ with $j\in \{1,2\}$,
$\xi_{12} \equiv \xi_1 - \xi_2$ and $\tau_{12} \equiv \tau_1 - \tau_2$.
The correlator in the r.h.s. of (\ref{mod-corr-1})
must be evaluated by employing  (\ref{gLL}) for odd values of $z$.
Notice that (\ref{mod-corr-3}) depends on the difference $\tau_{12}$ of the modular parameters, 
as expected.

We remark that 
(\ref{mod-corr-3}) has the structure identified in  \cite{Hollands:2019hje, Mintchev:2020uom} 
for the modular correlators in  other translation invariant cases.
As a consequence, the correlator (\ref{mod-corr-3}) 
satisfies the Kubo-Martin-Schwinger (KMS) condition \cite{Haag:1992hx}
\be
\langle \psi^\ast(\tau_1,x_1)  \, \psi(\tau_2+\tau + \textrm{i} ,x_2)  \rangle_{\infty, \mu} 
\,=\,
\langle \psi(\tau_2+\tau ,x_2) \, \psi^\ast(\tau_1,x_1)   \rangle_{\infty, \mu} 
\ee
which is also a non-trivial consistency check of the entanglement Hamiltonian (\ref{K_A-local-def}).
In the Appendix\;\ref{app_mod-ham-flow} 
the partial differential equation satisfied by the correlator (\ref{mod-corr-1}) is also reported.

\section{Small $\eta$ expansion}
\label{sec_small_distance}

In this section we discuss the expansion of the entanglement entropies for small values of $\eta$.
In Sec.\,\ref{sec_small_distance_P5} we employ the approach based on 
the tau function of the sine kernel (see (\ref{entropies-tau}) and (\ref{tau-function-sine})),
while in Sec.\,\ref{sec_small_distance_PSWF} the expansion is obtained 
by exploiting the properties of the eigenvalues of the sine kernel.

\subsection{Tau function approach}
\label{sec_small_distance_P5}

The expression \eqref{entropies-tau},
which is employed here also for the single copy entanglement, 
tells us that the expansion of $S_A^{(\alpha)}$ as $\eta \to 0$
can be studied through the expansion of the sine kernel tau function
(\ref{tau-function-sine}) in this regime.

The small distance expansion of the tau function of a Painlev\'e V 
has been found in \cite{Gamayun:2013auu,Lisovyy:2018mnj}.
In the Appendix\;\ref{app_tau_small_eta_GIL}
we specialise this result to the case of the 
sine kernel tau function (\ref{sine-kernel def}), finding  \cite{Gamayun:2013auu}
\be
\label{tau-small-distance-expansion}
\tau\,=\,
\sum_{n=0}^{\infty}
(-1)^n \frac{G(1+n)^6}{G(1+2n)^2}\;\frac{(4\,\eta)^{n^2}}{(2\pi \, z)^n}\,  \mathcal{B}_n(\eta)
\ee
where $G(x)$ is the Barnes $G$ function.
This expansion corresponds to Eq.\,(5.13) of \cite{Gamayun:2013auu} 
written in the notation given in \eqref{GIL-to-our-notation}.
The Taylor expansions of the functions $\mathcal{B}_n(\eta)$ as $\eta \to 0$ 
have been reported in the Appendix\;B of \cite{Gamayun:2013auu} for some values of $n$
(they extend the earlier result reported in Eq.\,(8.114) of \cite{forrester-book}, 
which improves the previous expansions given in \cite{Jimbo-82,McCoy:1985}).
In \eqref{B0-B1-expansion}-\eqref{B4-expansion}
we have reported only the terms of these expansions employed in our analyses.
We find it worth highlighting that the expansion (\ref{tau-small-distance-expansion}) can be written also in terms of 
the area $\mathsf{a}$ of the limited phase space (see (\ref{a-parameters-def}) and (\ref{def-red-phae-space})).

Since $G(1) =1$ and $\mathcal{B}_0(\eta)=1$ identically 
(see \eqref{B0-B1-expansion}) \cite{Gamayun:2013auu},
the summand corresponding to $n=0$ in \eqref{tau-small-distance-expansion} is equal to $1$ identically.

Approximate analytic expressions for the entanglement entropies \eqref{entropies-tau} 
are obtained by truncating the series \eqref{tau-small-distance-expansion} to a finite sum. 
In \eqref{tau-small-distance-expansion} a double series occurs 
because each $\mathcal{B}_n(\eta)$ can be written 
through its Taylor expansion as $\eta \to 0$.
Given a positive integer $\mathcal{N} \geqslant 1$,
the truncation condition of keeping all the term up to $O(\eta^{\mathcal{N}})$ included
leads to truncate also the series in $n$ to a sum of $N$ terms,
where $N$ satisfies $\mathcal{N} \leqslant (N+1)^2 - 1$.
Let us denote by $\tilde{\tau}_{\mathcal{N},N}$  the resulting finite sum,
where $N = N(\mathcal{N})$.
Since $o(1/z^{N})$ terms in (\ref{tau-small-distance-expansion}) have been neglected,
we have that $\tilde{\tau}_{\mathcal{N},N}= P_{N,\mathcal{N}}(z) /z^N$,
where  $P_{N,\mathcal{N}}(z)$ is a polynomial of order $N$ 
whose coefficients are polynomials in $\eta$ of different orders, up to order $\mathcal{N}$ included. 
Furthermore, 
since $\mathcal{B}_n(0) = 1$ for all the values of $n$ that we consider
(see \cite{Gamayun:2013auu} and the Appendix \ref{app_small_eta_entropies}),
the polynomial $P_{N,\mathcal{N}}(z)$ is monic.
These observations lead to
\be
\label{der-log-tau-approx}
\partial_z \log( \tilde{\tau}_{\mathcal{N},N} )
\,=\,
\sum_{i = 1}^N \frac{1}{z-z_i}
- \frac{N}{z}
\ee
where $z_i \in \mathcal{P}_{N,\mathcal{N}}$ are the zeros of $P_{N,\mathcal{N}}(z)$, 
which are non-trivial functions of $\eta$ 
whose explicit expressions depend on $\mathcal{N}$.

The Abel-Ruffini theorem states that
the roots of a polynomial of degree five or higher cannot be written through radicals.
In our analysis, 
this implies that approximations corresponding to $\mathcal{N} \geqslant 25$
can be studied only numerically because $N \geqslant  5$ is required in those cases. 
This leads us to consider $N \leqslant 4$.

Plugging the finite sum \eqref{der-log-tau-approx} into \eqref{entropies-tau} and exploiting the fact that $s_\alpha(0) = 0$, 
one obtains the following approximate result for the entanglement entropies
\be
\label{approx-entropies-small-eta}
\widetilde{S}_{A;\,\mathcal{N}}^{(\alpha)} 
\,\equiv\,
\sum_{j} s_\alpha (\tilde{z}_j)
\;\;\;\qquad\;\;\;
\tilde{z}_j \in \mathcal{P}_{N,\mathcal{N}} \cap [0,1]
\ee
where only the zeros of $P_{N,\mathcal{N}}(z)$ belonging to $[0,1]$ contribute to this finite sum. 
In the Appendix\;\ref{app_small_eta_entropies} we report the analytic expressions for the zeros
of $\mathcal{P}_{N,\mathcal{N}} $ in terms of $\eta$, for various values of $\mathcal{N} \leqslant 24$.
By introducing the $\mathcal{N}$ dependent parameter $\eta_\ast$ through the condition that 
at least one zero does not lie in $(0,1)$ or it has a non-vanishing imaginary part,
for $\eta \in (0, \eta_\ast)$ all the zeros belong to $(0,1)$
and therefore contribute to $\widetilde{S}_{A;\,\mathcal{N}}^{(\alpha)}$.

\begin{figure}[t!]
\subfigure
{\hspace{-1.6cm} \includegraphics[width=.58\textwidth]{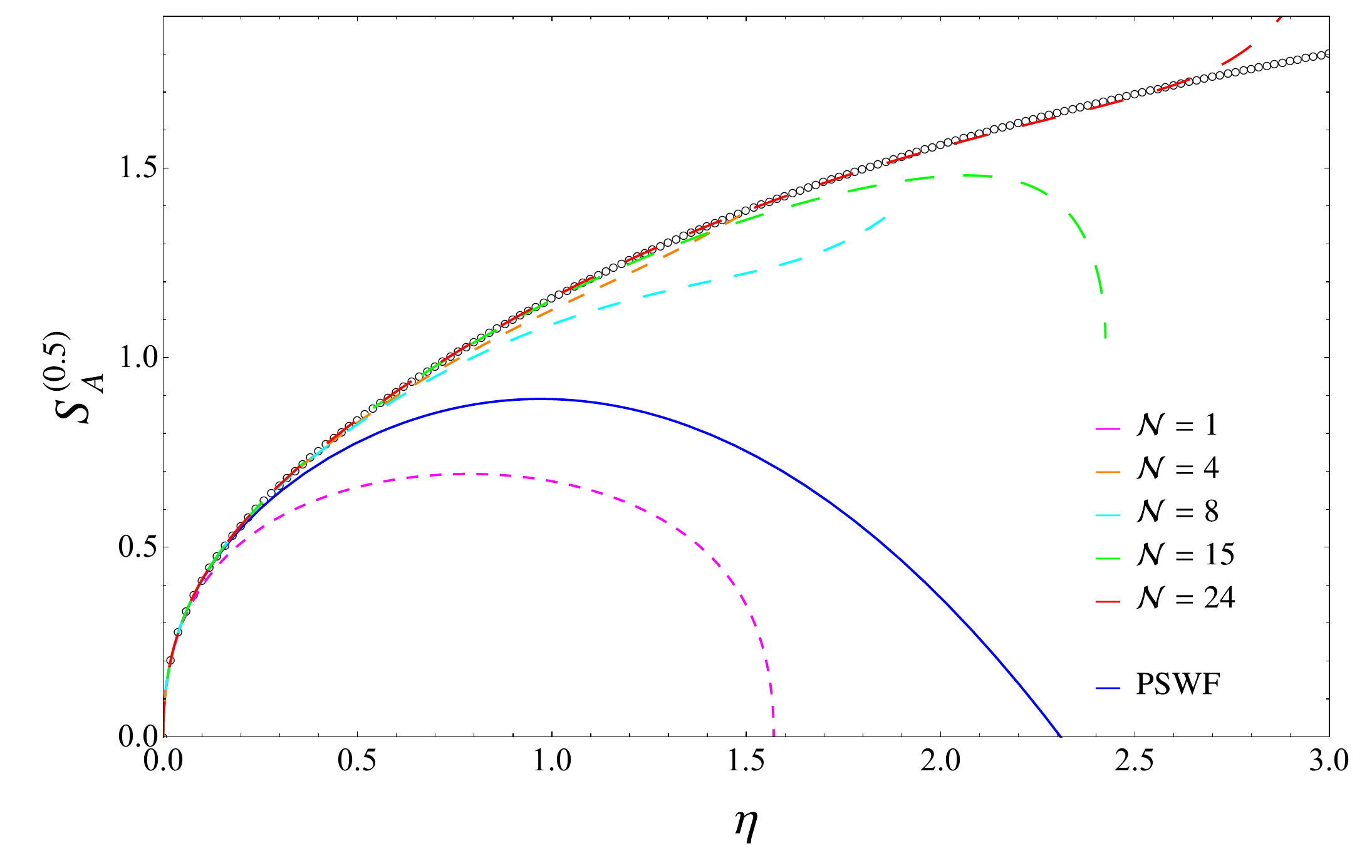}}
\subfigure
{\hspace{.1cm}\includegraphics[width=.58\textwidth]{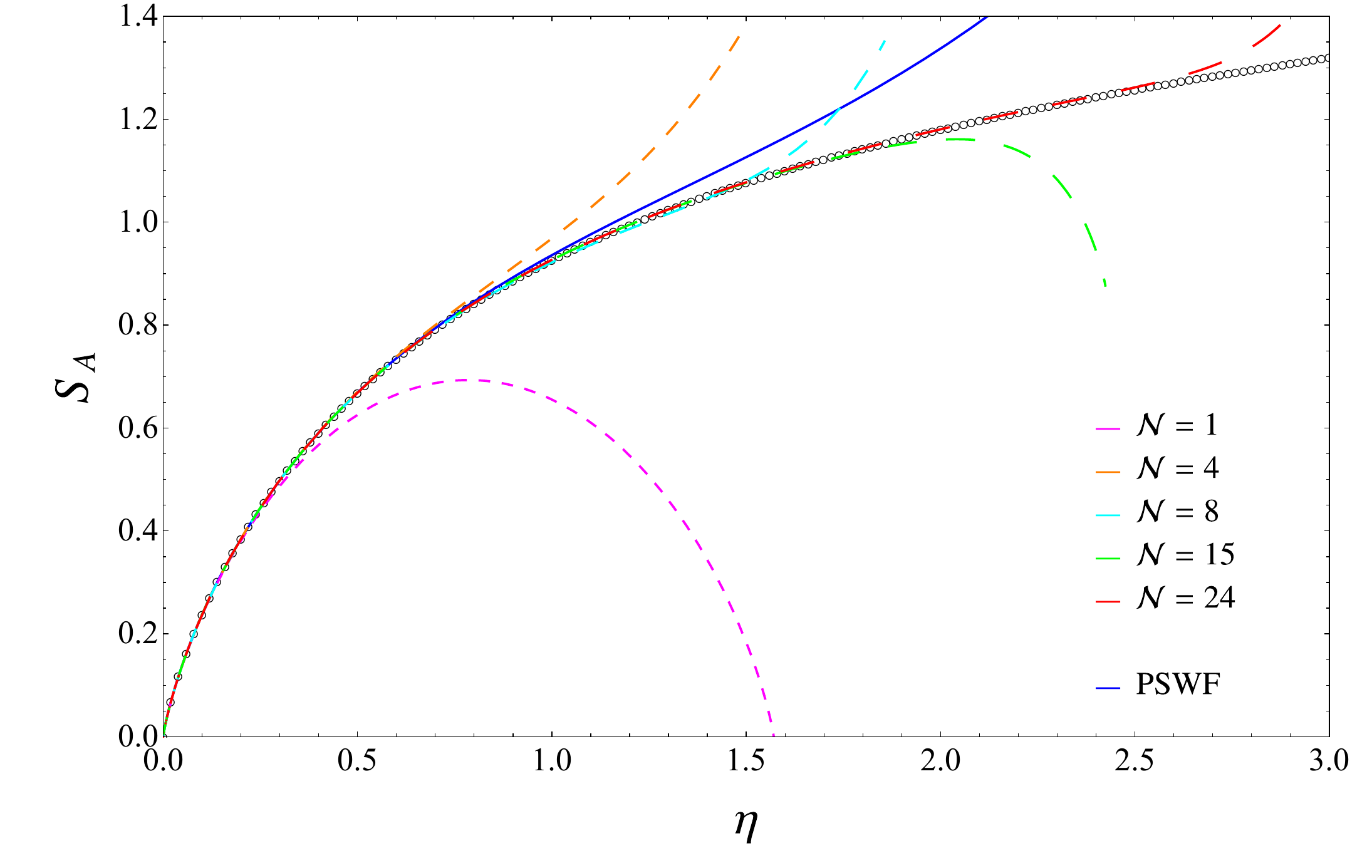}}
\subfigure
{\hspace{-1.45cm}\includegraphics[width=.58\textwidth]{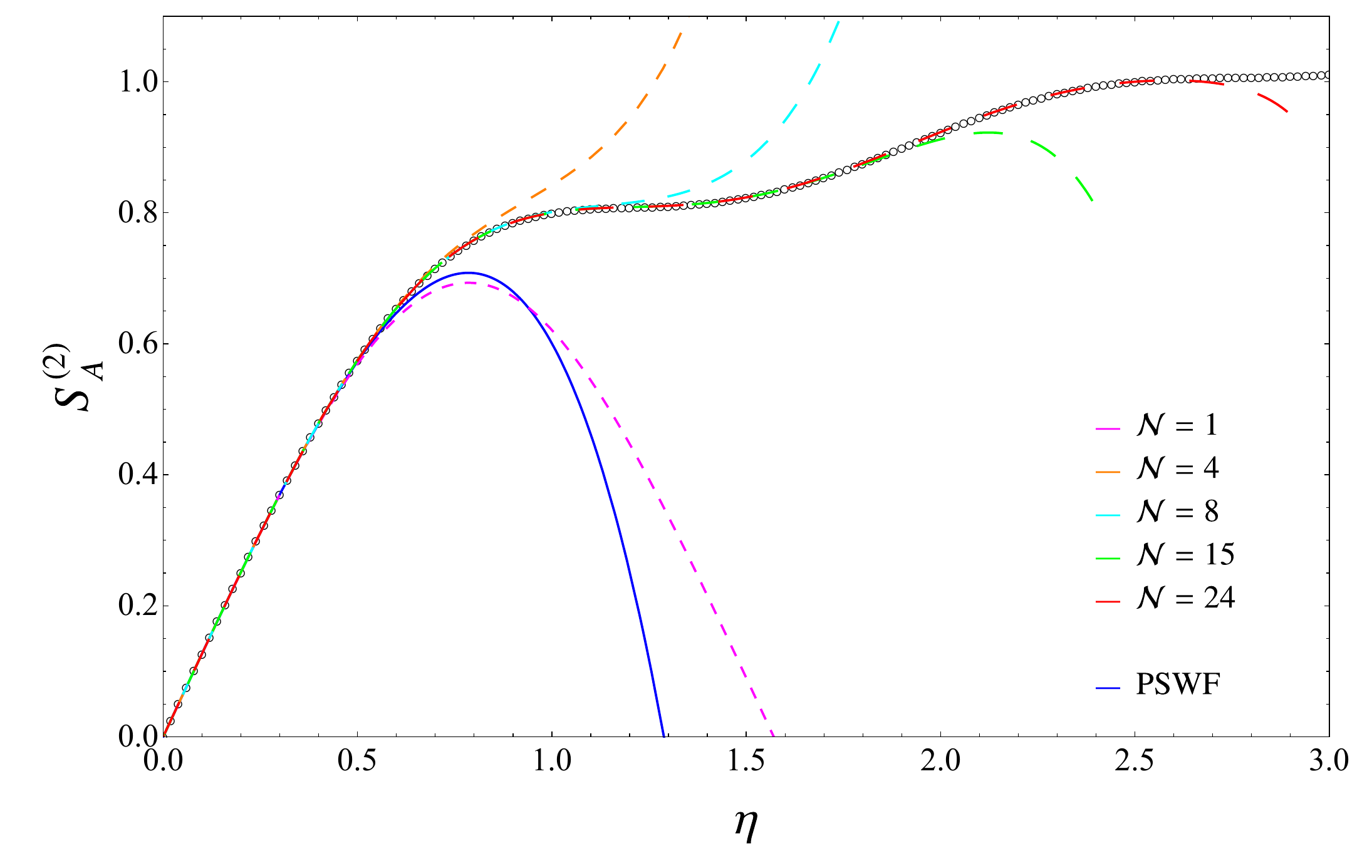}}
\subfigure
{\hspace{-.05cm} \includegraphics[width=.58\textwidth]{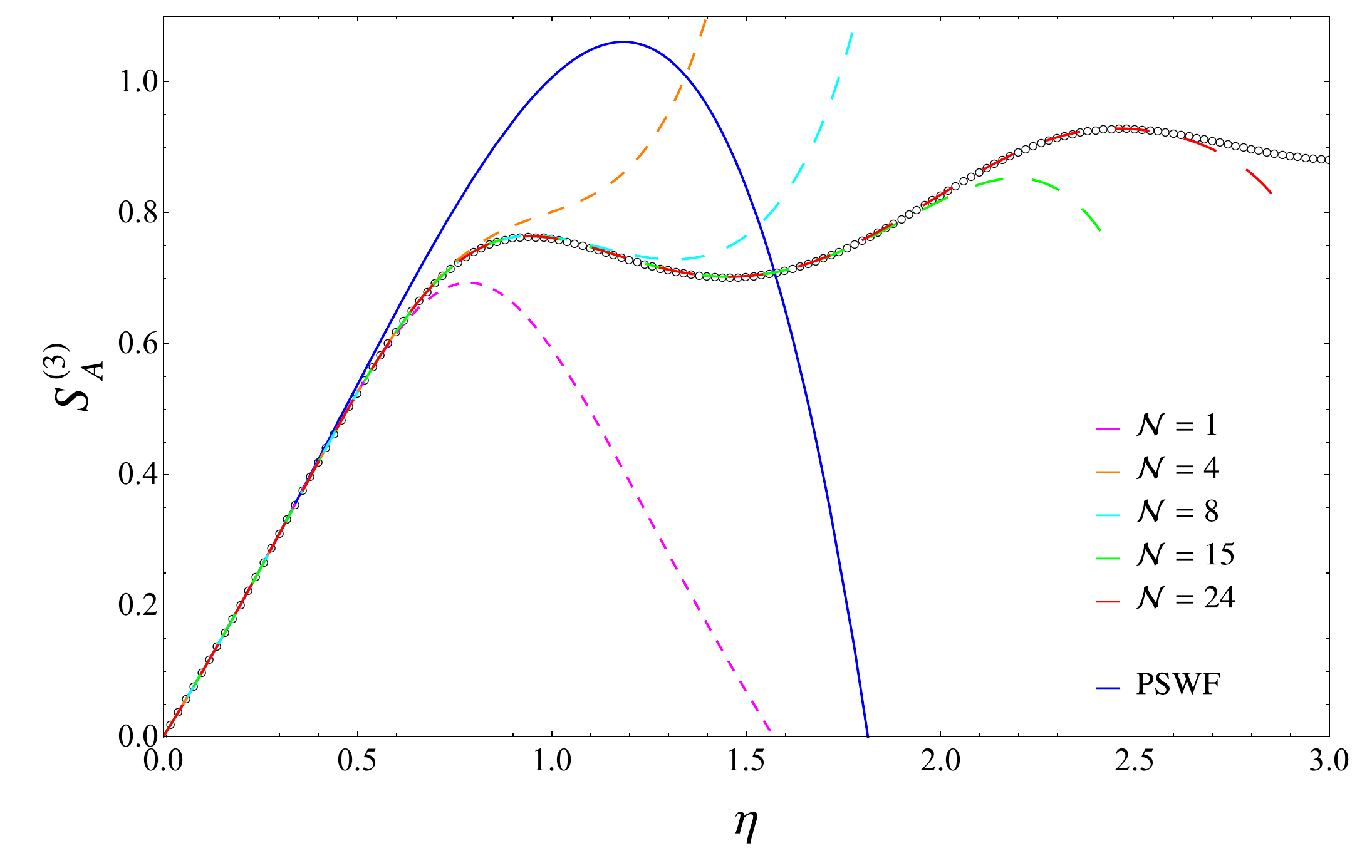}}
\subfigure
{\hspace{-1.55cm} \includegraphics[width=.58\textwidth]{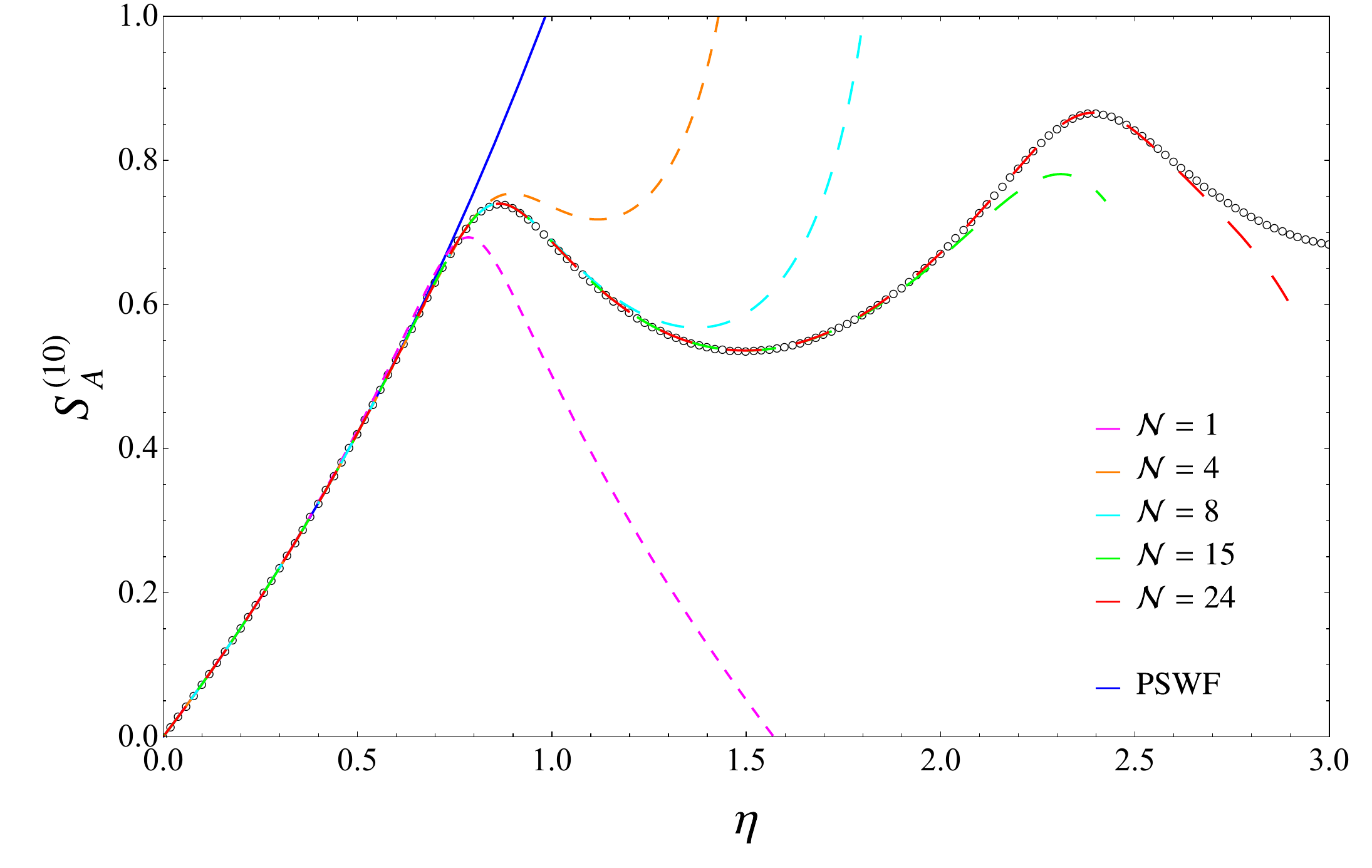}}
\subfigure
{\hspace{-.08cm} \includegraphics[width=.58\textwidth]{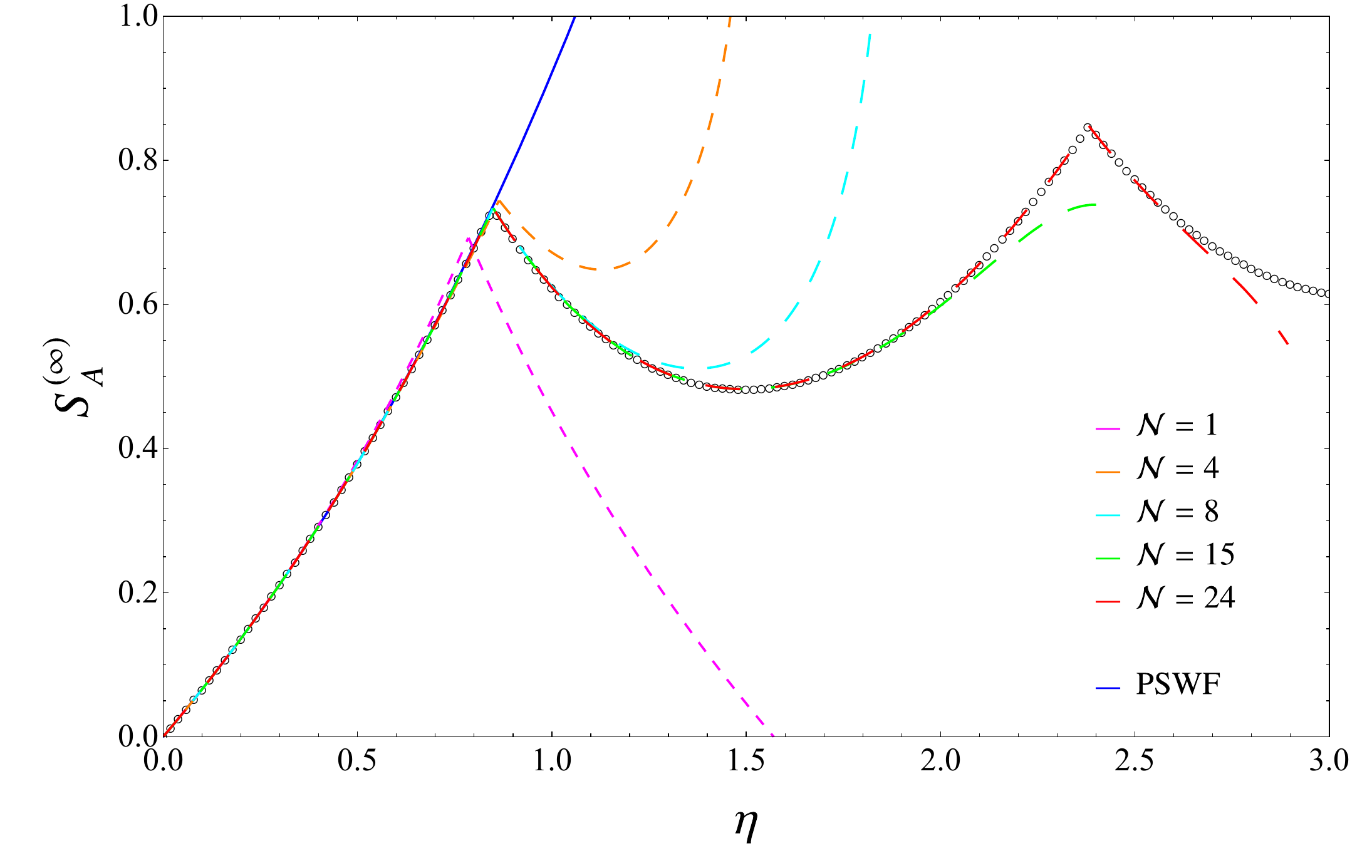}}
\caption{
Entanglement entropies $S_A^{(\alpha)}$ in the small $\eta$ regime.
Different values of $\alpha$ are considered in the different panels. 
The numerical results correspond to the black empty markers
and the coloured curves to the analytic approximate expressions as follows:
the dashed curves are obtained from (\ref{approx-entropies-small-eta}) 
for increasing values of  $\mathcal{N}$
(see Sec.\,\ref{sec_small_distance_P5})
and the solid blue lines from
\eqref{small_c_o5}, \eqref{renyi_small}  and \eqref{pswf_inf}
(see Sec.\,\ref{sec_small_distance_PSWF}).
} 
\vspace{0.4cm}
\label{fig:small-eta-panels}
\end{figure}

From (\ref{approx-entropies-small-eta}) and (\ref{c-function-alpha}),
it is straightforward to introduce 
\be
\label{approx-ch-function-small-eta}
\widetilde{C}_{A;\,\mathcal{N}}^{(\alpha)} 
\,\equiv \,
\eta \, \partial_\eta \widetilde{S}_{A;\,\mathcal{N}}^{(\alpha)}
\ee
which provides analytic expressions for the expansion of (\ref{c-function-alpha})  as $\eta \to 0$.

The cases $\mathcal{N} \leqslant 3$ are the simplest one to explore because  $N=1$.
For $\mathcal{N} = 3$,
by using \eqref{B0-B1-expansion} and the fact that $G(1) = G(2) = G(3) =1$,
one finds $\tilde{\tau}_{3,1} = 1- a_1 \eta / z$ with $a_1 \equiv 2/\pi$;
hence we can consider only one root $z_1 = a_1 \eta= 2\eta/\pi$, 
which belongs to $[0,1]$ when $\eta\leqslant \pi/2$.
In this case \eqref{approx-entropies-small-eta} simplifies to $S_{A;3}^{(\alpha)} = s_\alpha(a_1 \eta)$.
By expanding this result for $\eta \to 0$, we obtain a leading term $O\big(\eta \log (\eta)\big)$.
This is confirmed by the numerical results reported in Fig.\,\ref{fig:small-eta-panels}
and also by the expansion obtained through the PSWF in Sec.\,\ref{sec_small_distance_PSWF}.
Notice that, instead, expanding $\log (\tilde{\tau}_{3,1})$ as $\eta \to 0$ first
and then employing the resulting expansion in \eqref{entropies-tau} 
leads to a wrong leading term $O(\eta)$.

Improved approximations corresponding to $\mathcal{N}>3$ require $N>1$;
hence two or more terms can occur in \eqref{approx-entropies-small-eta}.
All the improved approximations characterised by $\mathcal{N} \leqslant  24$ are discussed
in the Appendix \ref{app_small_eta_entropies}.

In Fig.\,\ref{fig:small-eta-panels}
the numerical results for some entanglement entropies (black data points)
are compared with the corresponding approximate analytic expressions
$\widetilde{S}_{A;\,\mathcal{N}}^{(\alpha)}$ in 
\eqref{approx-entropies-small-eta} (coloured dashed curves)
obtained in the Appendix \ref{app_small_eta_entropies},
which hold in the small $\eta$ regime. 
The domain of $\eta$ where $\widetilde{S}_{A;\,\mathcal{N}}^{(\alpha)}$
reproduce the numerical data points 
becomes wider as $\mathcal{N}$ increases. 
The coloured dashed curves in Fig.\,\ref{fig:small-eta-panels}
ends at some finite value $\eta=\eta_\ast$
where at least a zero  of $P_{N,\mathcal{N}}(z)$ lies outside the interval $[0,1]$.
The value $\eta_\ast$ depends on $\mathcal{N}$ and on the coefficients of the polynomial 
$P_{N,\mathcal{N}}(z)$.
In the simplest case we have $P_{1,3}(z) = z -2\eta/\pi$, hence $\tilde{z}_1 = 2\eta/\pi$,
which leads to $\eta_\ast = \pi/2$.
For higher values of $\mathcal{N}$ we have determined $\eta_\ast$ numerically. 
The best approximation considered in this manuscript corresponds 
to $\mathcal{N}=24$ (red dashed curves).
In this case, $\widetilde{S}_{A;\,24}^{(\alpha)}$
perfectly agree with the corresponding numerical data for $\eta \lesssim 2.5$,
capturing also the second local maximum of $S_A^{(\alpha)}$.

\begin{figure}[t!]
\vspace{-.2cm}
\hspace{-.8cm}
\includegraphics[width=1.05\textwidth]{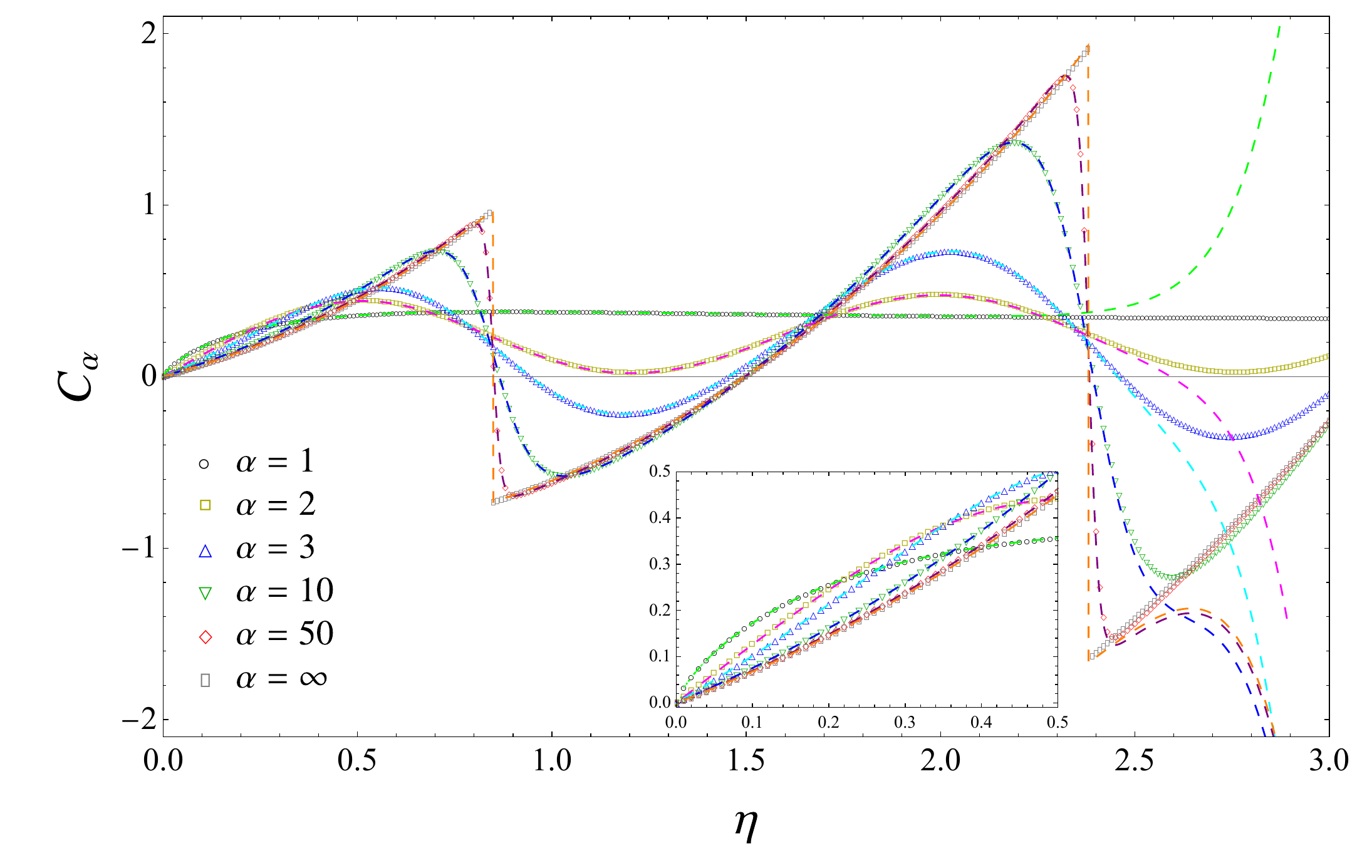}
\vspace{-.7cm}
\caption{
The quantities $C_A^{(\alpha)}$ in the regime of small $\eta$.
The data points are obtained numerically through \eqref{C-function-def},
while the coloured dashed curves represent $\widetilde{C}_{A;\,24}^{(\alpha)}\,$
(see (\ref{approx-ch-function-small-eta})).
}
\label{fig:c-functions-alpha-small}
\end{figure}

In Fig.\,\ref{fig:c-functions-alpha-small} we compare 
the numerical data points for $C_A^{(\alpha)}$
obtained numerically from \eqref{C-functions-gamma-prime}
with the approximate analytic expressions $\widetilde{C}_{A;\,\mathcal{N}}^{(\alpha)}$
defined in (\ref{approx-ch-function-small-eta}),
which hold for small $\eta$.
The coloured dashed curves represent $\widetilde{C}_{A;\,24}^{(\alpha)}\,$,
which is the best approximation considered in our analysis. 
The numerical data agree with the analytic results for the expansions as $\eta \to 0^+$.
At leading order we have 
$C=O(\eta \log(\eta))$ when $\alpha =1$ and 
$C_{\alpha}=O(\eta)$ when $\alpha >1$
(see also (\ref{small_c_o5}) and (\ref{renyi_small_leading}) respectively).
Notice that $C_\alpha$ with $\alpha>1$ displays 
an oscillatory behaviour whose amplitude grows with $\eta$.
When $\alpha \to \infty$, this curve becomes sawtoothed
(see also Fig.\,\ref{fig:c-functions-alpha}).

\subsection{PSWF approach}
\label{sec_small_distance_PSWF}

The expansions of the entanglement entropies for small $\eta$
can be studied also from (\ref{entropies-def-sums}) and (\ref{single-copy-ent-sum})
by employing the expansion of the eigenvalues as $\eta \to 0$.

For the generic eigenvalue $\gamma_n$ of the spectral problem \eqref{spectral-problem-v2}, it has been found that 
\cite{Slepian-expansions}
\be
\label{gamma-n-small-c}
\gamma_n  = \, \tilde{\gamma}_n \, \exp  \left\{ -\frac{(2n+1)\, \eta^2}{(2n-1)^2 \, (2n+3)^2}\, 
+ O(\eta^4)\right\}
\ee
where $\tilde{\gamma}_n $ has been defined in \eqref{gamma-tilde-def}. 
This tells us that
$\gamma_n \to 0$ as $\eta \to 0$, for any $n \geqslant 0$.

A non-trivial approximate analytic expression for the entanglement entropy 
can be obtained by considering the approximation 
$\gamma_n \simeq \tilde{\gamma}_n $ (from \eqref{gamma-n-small-c})
and $s(x) = -\,x \log(x) + x +O(x^2)$ as $x \to 0^+$ (from \eqref{entropies_func}).
This result reads
\be
\label{EE-small-c}
S_A
=
\sum_{n=0}^{\infty} \tilde{\gamma}_n  \big(\! -   \log \tilde{\gamma}_n +  1\, \big)  + \dots
\, =\,
\sum_{n=0}^{\infty} \, \tilde{g}_n  \, \eta^{2n+1} \big[  -(2n+1) \log \eta - \log \tilde{g}_n + 1\,\big] + \dots
\ee
where \eqref{gamma-tilde-def} has been employed. 
The leading term comes from the summand corresponding to 
$n=0$ in the series occurring in the last expression of (\ref{EE-small-c})
and it is given by
\be
\label{EE-small-c-regime}
S_A
=
- \frac{2}{\pi}  \, \eta  \log \eta \,
+ 
\frac{2}{\pi}   \big[ 1- \log (2/\pi) \,\big] \, \eta + O\big(\eta ^2 \big) 
\,=\,
- \,\frac{2\eta}{\pi}  \,   \log(2 \eta/\pi) 
+ \frac{2\eta}{\pi}
+ O\big(\eta ^2 \big) 
\ee
where we used that $\tilde{g}_0 = 2/\pi$ (from \eqref{gamma-tilde-def}).

Higher order terms in the expansion (\ref{EE-small-c-regime}) can be written
by including more terms in the expansion of
$s(x) = -\,x \log(x) + x -x^2/2 -x^3/6 -x^4/12+ O(x^5)$
and also taking into account the exponential correction 
occurring in the r.h.s. of \eqref{gamma-n-small-c}.
This leads to
\bea
 \label{small_c_o5}
S_A
&=&
 -\frac{2}{\pi}\, \eta\log(\eta)
 +\frac{2}{\pi}\left[1-\log\left(2/\pi\right)\right]\eta
 -\frac{2}{\pi^{2}}\,\eta^{2}
 -\frac{4}{9\pi}\,\eta^{3}\log(\eta)
 \\
 \rule{0pt}{.7cm}
& & 
 + \, \frac{2}{3\pi}  \left(\frac{1}{3}+\frac{2}{3}\log(3)-\frac{2}{\pi^{2}}\right)\eta^{3}
 + \frac{4}{3\pi^2} \left(\frac{1}{3}-\frac{1}{\pi^{2}}\right)\eta^{4}
 +O\big(\eta^{5}\log(\eta)\big) 
 \nn
\eea
which has been obtained by considering only the terms coming from $\gamma_{0}$ and $\gamma_{1}$.
The expansion (\ref{small_c_o5})
is the best approximation allowed by \eqref{gamma-n-small-c}.
Indeed, 
$s(\gamma_{n})=O\big(\eta^{5}\log(\eta)\big)$ for $n\geqslant 2$
and  $s(\gamma_{0})$ contain a term of order $O\big(\eta^{5} \log(\eta)\big)$ 
that cannot be evaluated because the $O(\eta^{4})$ term 
in the exponent of \eqref{gamma-n-small-c} has been neglected.

In order to study  the expansion of the R\'enyi entropies with finite index $\alpha \neq 1$ as $\eta \to 0$,
first we rewrite the function $s_\alpha(x)$ in  \eqref{entropies_func} as 
\be
s_{\alpha}(x)=
\frac{\alpha}{1-\alpha}\log(1-x)
+
\frac{1}{1-\alpha}\,\log\!\big[ 1+x^{\alpha}(1-x)^{-\alpha}\big] 
\ee
and then expand these two terms separately, obtaining
\be
\label{renyif_small}
s_{\alpha}(x) 
=
\frac{\alpha}{\alpha-1}
\left[\,x+\frac{x^{2}}{2}+\frac{x^{3}}{3}+\frac{x^{4}}{4}+O(x^{5})\right]
+
\frac{1}{1-\alpha}\left[\,
t^{\alpha}-\frac{t^{2\alpha}}{2}+\frac{t^{3\alpha}}{3}-\frac{t^{4\alpha}}{4}+O(t^{5\alpha})
\right]
\hspace{.5cm}
\ee
where
\be
\label{renyif_small_2}
t^{\beta} 
\equiv
\frac{x^{\beta}}{(1-x)^{\beta}}
\,=\,
x^{\beta}\left[1+\beta \,x+\frac{\beta(\beta+1)}{2}\,x^{2}
+\frac{\beta(\beta+1)(\beta+2)}{6}\,x^{3}
+O(x^4)
\right] 
\ee
By employing \eqref{gamma-n-small-c}, \eqref{renyif_small} and \eqref{renyif_small_2}, 
one finds that $s_{\alpha}(\gamma_{n})=O\big(\eta^{\min\{5\alpha,5\}}\big)$ for $n\geqslant 2$. 
Thus, by considering only the contributions coming from 
$\gamma_{0}$ and $\gamma_{1}$ in \eqref{entropies-def-sums}, 
for the R\'enyi entropies with finite $\alpha \neq 1$ we obtain 
\bea
\label{renyi_small}
S_{A}^{(\alpha)}
&=&
\frac{\alpha}{\alpha-1}  \;
\Bigg\{
\bigg[\,
\frac{2}{\pi}\eta+\frac{2}{\pi^{2}}\,\eta^{2}+\frac{8}{3\pi^{3}}\,\eta^{3}+\left(\frac{4}{\pi^{4}}-\frac{4}{9\pi^{2}}\right)\eta^{4}
\, \bigg]
\nonumber
\\
\rule{0pt}{.7cm}
& & 
-\; \eta^{\alpha}\left(\frac{2}{\pi}\right)^{\alpha}
\bigg[\,
\frac{1}{\alpha}+\frac{2}{\pi}\,\eta
+\left(\frac{2(\alpha+1)}{\pi^{2}}-\frac{1}{9}\right)\eta^{2}
+\left(\frac{4(\alpha+1)(\alpha+2)}{3\pi^{3}}-\frac{2(\alpha+1)}{9\pi}\right)\eta^{3}
\, \bigg]
\nn 
\\
\rule{0pt}{.7cm}
& &
+\;\eta^{2\alpha}\left(\frac{2}{\pi}\right)^{2\alpha}
\bigg[\,
\frac{1}{2\alpha}+\frac{2}{\pi}\eta+\left(\frac{2(2\alpha+1)}{\pi^{2}}-\frac{1}{9}\right)\eta^{2}
\, \bigg]
\nn
\\
\rule{0pt}{.7cm}
& &
-\;\eta^{3\alpha}\left(\frac{2}{\pi}\right)^{3\alpha}
\bigg[\,
\frac{1+3 \,(\pi / 6)^{2\alpha}}{3\alpha}+\frac{2}{\pi}\,\eta
\, \bigg]
+
\eta^{4\alpha}\left(\frac{2}{\pi}\right)^{4\alpha} \frac{1}{4\alpha}
\Bigg\} 
+O\big(\eta^{\min\{5\alpha,5\}}\big)
\eea
where the ordering of the terms based on their relevance 
depends on $\alpha$. 
For instance, we have
\be
\label{renyi_small_leading}
S_{A}^{(\alpha)}
=
\left\{
\begin{array}{ll}
\displaystyle
\frac{1}{1-\alpha}\left(\frac{2}{\pi}\right)^{\alpha}\eta^{\alpha}
+O\big(\eta^{\min\{2\alpha,1\}}\big)
\hspace{1.2cm}
& 
\alpha<1
\\
\rule{0pt}{.8cm}
\displaystyle
\frac{\alpha}{\alpha-1}\;\frac{2}{\pi}\,\eta
+O\big(\eta^{\min\{\alpha,2\}}\big)
& 
\alpha>1 \,.
\end{array}
\right.
\ee
Some terms reported in \eqref{renyi_small}  
could be of order $O\big(\eta^{\min\{5\alpha,5\}}\big)$,
depending on $\alpha$.

For the single-copy entanglement \eqref{single-copy-ent-sum},
by using 
$s_{\infty}(x)=x+x^2/2+x^3/3+x^4/4+O(x^{5})$
we obtain
\be
\label{pswf_inf}
S_{A}^{(\infty)}
=
\frac{2}{\pi}\,\eta+\frac{2}{\pi^{2}}\,\eta^{2}
+\frac{8}{3\pi^{3}}\,\eta^{3}
+\frac{4}{\pi^2}
\left(\frac{1}{\pi^{2}}-\frac{1}{9}\right)\eta^{4}
+O(\eta^{5}) \,.
\ee

In Fig.\,\ref{fig:small-eta-panels}, the solid blue lines correspond to 
\eqref{small_c_o5}, \eqref{renyi_small}, and \eqref{pswf_inf}.
The width of the range of $\eta$ where these curves agree
with the numerical data (black data points) increases with $\alpha$.
We remark that better approximate expressions are obtained through the approach 
based on the sine kernel tau function. 
Indeed, the solid blue lines in Fig.\,\ref{fig:small-eta-panels}
do not capture the first local maximum of the numerical data corresponding to $\alpha > 1$,
while the dashed lines nicely reproduce it when $\mathcal{N}$ is large enough. 
The curves corresponding to the analytic expressions obtained by applying the
differential operator $\eta\,\partial_\eta$ to 
\eqref{small_c_o5}, \eqref{renyi_small}, and \eqref{pswf_inf}
have not been included in Fig.\,\ref{fig:c-functions-alpha-small} 
to make this figure readable; 
but the ranges of $\eta$ where they reproduce the numerical data 
are the same ones of Fig.\,\ref{fig:small-eta-panels}, for any given value of $\alpha$.

The results presented above can be compared with the ones discussed in Sec.\,\ref{sec_small_distance_P5}.
Considering e.g. the entanglement entropy, 
by expanding \eqref{approx-entropies-small-eta} with $\alpha =1 $ as $\eta \to 0$, 
one obtains a series that coincides with \eqref{small_c_o5} up to a certain order
which depends on the value of $\mathcal{N}$ chosen in \eqref{approx-entropies-small-eta}. 
In order to obtain all the terms reported in \eqref{small_c_o5},
the expansion of \eqref{approx-entropies-small-eta} with $\mathcal{N} \geqslant 4$
must be considered.
A similar analysis can be performed for the other entanglement entropies.

\section{Large $\eta$ expansion}
\label{sec_large_distance}

The form \eqref{entropies-tau} of 
the entanglement entropies (\ref{entropies-def-sums}) and (\ref{single-copy-ent-sum})
allows to study their expansions in the regime of large $\eta$
by employing the expansion of the sine kernel tau function (\ref{tau-function-sine}) in this regime.

\subsection{Tau function}

The asymptotic expansion of the sine kernel tau function for large $\eta$ 
is a special case of the large distance expansion of the tau function for the general Painlev\'e V
found in \cite{Lisovyy:2018mnj, Bonelli:2016qwg}.
The details of this analysis are discussed 
in the Appendix\;\ref{app_tau_large_eta_LNR} and the result reads
\bea
\label{tau-expansion-large-eta}
\tau 
&=&
\frac{\textrm{e}^{4 \textrm{i} \nu_\star \eta} }{(4 \eta)^{2\nu_\star^2}}\,
\big[ G(1-\nu_\star)\, G(1+ \nu_\star) \big]^2\,
\\
& &
\times 
\sum_{n \in \mathbb{Z}}
\frac{1}{(2\pi z)^{2n}}
\left[ \frac{G(1+\nu_\star + n)}{G(1+\nu_\star)} \right]^4
\frac{\textrm{e}^{4\textrm{i}  n \eta}}{(4\textrm{i} \eta)^{2n(n+2\nu_\star)}}
\sum_{k = 0}^{\infty} \frac{\mathcal{D}_k(\nu_\star + n)}{(4\textrm{i} \eta)^k}
\nonumber
\eea
where 
\be
\label{nu-star-main}
\nu_\star = \frac{1}{2\pi \textrm{i}}  \log (1-1/z) \,.
\ee
The functions $\mathcal{D}_k(\nu_\star) $ available in the literature are \cite{Lisovyy:2018mnj, Bonelli:2016qwg}
\be
\label{D-012-def}
\mathcal{D}_0(\nu_\star) = 1
\;\;\;\;\qquad\;\;\;\;
\mathcal{D}_1(\nu_\star) = 4\, \nu_\star^3
\;\;\;\;\qquad\;\;\;\;
\mathcal{D}_2(\nu_\star) = 8\, \nu_\star^6 +10 \, \nu_\star^4 \,.
\ee
We report also\footnote{The functions (\ref{D-3-def}) and (\ref{D-4-def}) have been obtained 
by Oleg Lisovyy.
We are very grateful to him for having shared with us a Mathematica code 
to generate the functions $\mathcal{D}_k(\nu_\star)$.}
\bea
\label{D-3-def}
\mathcal{D}_3(\nu_\star) 
&=&
\frac{4}{3} \, 
\big(  8\,\nu_\star^9 + 30\,\nu_\star^7 + 33\,\nu_\star^5 + \nu_\star^3    \,\big)
\\
\rule{0pt}{.65cm}
\label{D-4-def}
\mathcal{D}_4(\nu_\star) 
&=&
\frac{2}{3} \, 
\big(  16\,\nu_\star^{12} + 120\,\nu_\star^{10} + 339\,\nu_\star^{8} + 386\,\nu_\star^6 + 39 \,\nu_\star^4  \,\big) \,.
\eea
In our analyses we employ $\mathcal{D}_k(\nu_\star) $ with $k \in \{0,1,2,3\}$. 

We remark that the area $\mathsf{a}$ of the limited phase space 
(see (\ref{a-parameters-def}) and (\ref{def-red-phae-space}))
is a natural variable for the expansion (\ref{tau-expansion-large-eta}).
We also stress that, differently from the small $\eta$ expansion of the tau function 
 given by the convergent series (\ref{tau-small-distance-expansion}), 
 the large $\eta$ expansion (\ref{tau-expansion-large-eta}) is asymptotic.

We find it useful to write (\ref{tau-expansion-large-eta}) as the following product
\be
\label{tau-infty-dec}
\tau \,=\, \tilde{\tau}_\infty \, \mathcal{T}_\infty
\ee
where 
\be
\label{tau-tilde-infty-def}
\tilde{\tau}_\infty 
\equiv 
\frac{\textrm{e}^{4 \textrm{i} \nu_\star \eta} }{(4 \eta)^{2\nu_\star^2}}\,
\big[ G(1-\nu_\star)\, G(1+ \nu_\star) \big]^2\;
\ee
which has been first obtained in \cite{Budylin1996} 
(see also \cite{Cheianov-04, Abanov-2011, Susstrunk_2012, bothner2015asymptotic}),
and
\bea
\label{cal-T-infty-def}
 \mathcal{T}_\infty 
& \equiv &
 \sum_{n \in \mathbb{Z}}
\frac{1}{(2\pi z)^{2n}}
\left[ \frac{G(1+\nu_\star + n)}{G(1+\nu_\star)} \right]^4
\frac{\textrm{e}^{4\textrm{i}  n \eta}}{(4\textrm{i} \eta)^{2n(n+2\nu_\star)}}
\,\sum_{k = 0}^{\infty} \frac{\mathcal{D}_k(\nu_\star + n)}{(4\textrm{i} \eta)^k} 
\\
\label{cal-T-infty-def-b}
\rule{0pt}{.9cm}
&=&
 \sum_{k = 0}^{\infty} \frac{\mathcal{D}_k(\nu_\star)}{(4\textrm{i} \eta)^k} 
 +
\sum_{n \geqslant 1}
\frac{\textrm{e}^{4\textrm{i}  n \eta} \, \prod_{j=0}^{n-1} \Gamma(1+\nu_\star +j)^4}{(2\pi z)^{2n} \,(4\textrm{i} \eta)^{2n(n+2\nu_\star)}}
\,\sum_{k = 0}^{\infty} \frac{\mathcal{D}_k(\nu_\star + n)}{(4\textrm{i} \eta)^k} 
\\
& & \hspace{2cm}
+ 
\sum_{n \leqslant -1}
\frac{\textrm{e}^{4\textrm{i}  n \eta} }{(2\pi z)^{2n} \,(4\textrm{i} \eta)^{2n(n+2\nu_\star)} \, \prod_{j=-1}^{n} \Gamma(1+\nu_\star +j)^4}
\,\sum_{k = 0}^{\infty} \frac{\mathcal{D}_k(\nu_\star + n)}{(4\textrm{i} \eta)^k} 
\nonumber
\eea
where the identity $G(z+1) = \Gamma(z) \, G(z)$ has been used.

An approximate expression for (\ref{cal-T-infty-def-b}) is obtained 
by considering only the summands corresponding to $n=1$ and $n=-1$  in the second and third terms respectively.
This gives
\be
\label{cal-T-infty-split-101}
\mathcal{T}_\infty 
\simeq
 \sum_{k = 0}^{\infty} \frac{\mathcal{D}_k(\nu_\star)}{(4\textrm{i} \eta)^k} 
 +
\frac{\textrm{e}^{4\textrm{i}  \eta} \, \Gamma(1+\nu_\star)^4}{(2\pi z)^{2} \,(4\textrm{i} \eta)^{2(2\nu_\star+1)}}
\sum_{k = 0}^{\infty} \frac{\mathcal{D}_k(\nu_\star + 1)}{(4\textrm{i} \eta)^k} 
+ 
\frac{ (2\pi z)^2\, (4\textrm{i} \eta)^{2(2\nu_\star-1)} }{\textrm{e}^{4\textrm{i}  \eta} \, \Gamma(\nu_\star )^4}
\sum_{k = 0}^{\infty} \frac{\mathcal{D}_k(\nu_\star -1)}{(4\textrm{i} \eta)^k} \,.
\ee

Since the complex parameter $\nu_\star$ takes the values (\ref{nu_star_app_y_def})
in the computation of the entanglement entropies,
the approximation (\ref{cal-T-infty-split-101}) allows to obtain their expansion up to $O(1/\eta^4)$,
as discussed in Sec.\,\ref{subsec-ee-large-eta} and in the Appendix\;\ref{app_large_eta_Ttilde}
(see (\ref{entropies-tau-dec-large-eta}) and (\ref{exp-S-infty-123})).

By using the identities $(2\pi z)^2\,\textrm{i}^{4\nu_\star} = -[\pi / \sin(\pi \nu_\star)]^2$ 
and $\tfrac{\pi}{\sin(\pi w)} = \Gamma(1-w)\, \Gamma(w)$,
the factors multiplying the last two series in (\ref{cal-T-infty-split-101})
can be written as follows
\bea
\frac{\textrm{e}^{4\textrm{i}  \eta} \, \Gamma(1+\nu_\star)^4}{(2\pi z)^{2} \,(4\textrm{i} \eta)^{2(2\nu_\star+1)}}
&=&
  \frac{4\,\textrm{e}^{4\textrm{i}  \eta} \, \Gamma(1+\nu_\star)^4\, [\sin(\pi \nu_\star)]^2}{ (2\pi)^2\,(4 \eta)^{2(2\nu_\star+1)}}
   \,=\,
  \frac{\textrm{e}^{4\textrm{i}  \eta} \, \Gamma(1+\nu_\star)^2}{ (4 \eta)^{2(2\nu_\star+1)} \, \Gamma(-\nu_\star)^2}
\\
\rule{0pt}{.8cm}
\frac{(2\pi z)^2\, (4\textrm{i} \eta)^{2(2\nu_\star-1)} }{\textrm{e}^{4\textrm{i}  \eta} \, \Gamma(\nu_\star )^4}
&=&
\frac{(2\pi)^2 (4\eta)^{2(2\nu_\star-1)} }{ 4\,\textrm{e}^{4\textrm{i}  \eta} \, \Gamma(\nu_\star )^4\, [\sin(\pi \nu_\star)]^2}
\,=\,
\frac{(4\eta)^{2(2\nu_\star-1)} \, \Gamma(1-\nu_\star)^2}{ \textrm{e}^{4\textrm{i}  \eta} \, \Gamma(\nu_\star )^2}
\eea
hence (\ref{cal-T-infty-split-101}) becomes
\bea
\label{cal-T-infty-split-101-bis}
\mathcal{T}_\infty 
&\simeq&
 \sum_{k = 0}^{\infty} \frac{\mathcal{D}_k(\nu_\star)}{(4\textrm{i} \eta)^k} 
 +
  \frac{\textrm{e}^{4\textrm{i}  \eta} \, \Gamma(1+\nu_\star)^2}{ (4 \eta)^{2(2\nu_\star+1)} \, 
  \Gamma(-\nu_\star)^2}
\sum_{k = 0}^{\infty} \frac{\mathcal{D}_k(\nu_\star + 1)}{(4\textrm{i} \eta)^k} 
\nonumber
\\
\label{cal-T-infty-split-101-bis-1}
\rule{0pt}{.8cm}
& & \hspace{1.95cm}
+ \;
\frac{(4\eta)^{2(2\nu_\star-1)} \, \Gamma(1-\nu_\star)^2}{ \textrm{e}^{4\textrm{i}  \eta} \, \Gamma(\nu_\star )^2}
\sum_{k = 0}^{\infty} \frac{\mathcal{D}_k(\nu_\star -1)}{(4\textrm{i} \eta)^k} 
\,\equiv\,
1 + \widetilde{\mathcal{T}}_\infty
\eea
where the term $1$ in the last expression corresponds to $k=0$ term of the first series.

\subsection{Entanglement entropies}
\label{subsec-ee-large-eta}

The expansion of the entanglement entropies for large $\eta$
can be studied by employing (\ref{tau-infty-dec}) and (\ref{cal-T-infty-split-101-bis-1}) 
into (\ref{entropies-tau}).
This leads to the  decomposition
\be
\label{entropies-tau-dec-large-eta} 
S_A^{(\alpha)} 
= 
S_{A,\infty}^{(\alpha)} 
+
 \widetilde{S}_{A,\infty}^{(\alpha)} 
\ee
where 
\be
\label{SA_infty_def}
S_{A,\infty}^{(\alpha)} 
\equiv
\lim_{\epsilon, \delta \to 0}\,
 \frac{1}{2\pi \textrm{i}} \oint_{\mathfrak{C}}
 s_\alpha(z) \,
 \partial_z \log( \tilde{\tau}_\infty )
 \, \textrm{d}z
 \;\;\qquad\;\;
  \widetilde{S}_{A,\infty}^{(\alpha)} 
\equiv
\lim_{\epsilon, \delta \to 0}\,
  \frac{1}{2\pi \textrm{i}} \oint_{\mathfrak{C}}
 s_\alpha(z) \,
  \partial_z \log \! \big( 1 + \widetilde{\mathcal{T}}_\infty \big)
 \, \textrm{d}z \,.
\ee
In the following we show that the leading term $S_{A,\infty}^{(\alpha)} $ gives a logarithmic growth,
while the subleading corrections in $ \widetilde{S}_{A,\infty}^{(\alpha)}$ 
provide the oscillatory terms observed e.g. in Fig.\,\ref{fig:ee-large-100}.

The term $S_{A,\infty}^{(\alpha)}$ in (\ref{entropies-tau-dec-large-eta}) 
can be computed by adapting 
to the case that we are considering in the continuum
the analysis of \cite{Jin_2004, Keating_04}
for the entanglement entropies of a block made by consecutive sites 
in the infinite one-dimensional spin-$\tfrac{1}{2}$ Heisenberg XX chain in a magnetic field,
which is based on the Fisher-Hartwig conjecture \cite{FHc,Basor-91,Basor-94}.
The result of this calculation,
whose details are reported in the Appendix\;\ref{app_large_eta_tildetau},
is
\be
\label{S_alpha_main_large_eta_0} 
S_{A,\infty}^{(\alpha)}  
\,=\,
\frac{1}{6} \left(1 + \frac{1}{\alpha} \right) \log(4\eta) + E_\alpha
\ee
where the argument of the logarithm 
is the area of the limited phase space (\ref{def-red-phae-space}) when $\hbar = 1$
and the constant term $E_\alpha$ is defined as follows
\cite{Jin_2004, Calabrese-Essler-10}
\be
\label{E_alpha-def}
E_\alpha \equiv
\left(1 + \frac{1}{\alpha} \right)
\int_0^\infty
\!\left( 
\frac{\alpha\, \textrm{csch}(t)}{\alpha^2 - 1} \, 
\big( \textrm{csch}(t/\alpha) -\alpha\, \textrm{csch}(t)  \big)- \frac{\textrm{e}^{-2t}}{6} 
\,\right)
\frac{\textrm{d} t}{t}
\ee
In the limits $\alpha \to 1$ and $\alpha \to +\infty$, 
this constant becomes respectively
\bea
E_1 
&=&
\int_0^\infty\!
\left( 
 [\textrm{csch}(t)]^2
 \big[ t  \coth(t) - 1\big]
- \frac{\textrm{e}^{-2t}}{3} 
\,\right)
\frac{\textrm{d} t}{t}
\\
\rule{0pt}{.7cm}
E_\infty 
&=&
\int_0^\infty\!
\left( 
 \textrm{csch}(t)
 \left[\, \frac{1}{t} - \textrm{csch}(t)\right]
- \frac{\textrm{e}^{-2t}}{6} 
\,\right)
\frac{\textrm{d} t}{t} \,.
\eea
The result \eqref{S_alpha_main_large_eta_0} can be obtained also 
by employing a result of Slepian \cite{Slepian-expansions} 
in this asymptotic regime, as shown in \cite{EislerPeschelProlate, Susstrunk_2012}.
A rigorous derivation of the leading logarithm term in \eqref{S_alpha_main_large_eta_0} has been provided in \cite{Spitzer-14}.

In order to study the expansion of the subleading term $\widetilde{S}_{A,\infty}^{(\alpha)}$ 
in (\ref{entropies-tau-dec-large-eta}) as $\eta \to \infty$,
we adapt to our case in the continuum 
the analysis performed \cite{Calabrese-Essler-10} 
in the spin-$\tfrac{1}{2}$ Heisenberg XX chain in a magnetic field,
which provides the terms subleading to the ones found in \cite{Jin_2004, Keating_04}
by employing the generalised Fisher-Hartwig conjecture \cite{Basor-91,Basor-94,Deift-11}. 
This analysis, described in the Appendix\;\ref{app-large-eta-renyi-ee},
gives the following expansion 
for $\widetilde{S}_{A,\infty}^{(\alpha)}$ in (\ref{entropies-tau-dec-large-eta})
as $\eta \to \infty$
\be
\label{exp-S-infty-123} 
 \widetilde{S}_{A,\infty}^{(\alpha)} 
 =
  \widetilde{S}_{A,\infty,0}^{(\alpha)} 
  +
  \frac{ \widetilde{S}_{A,\infty,1}^{(\alpha)} }{\eta}
    +
  \frac{ \widetilde{S}_{A,\infty,2}^{(\alpha)} }{\eta^2}
    +
  \frac{ \widetilde{S}_{A,\infty,3}^{(\alpha)} }{\eta^3}
    +
O\big(1/\eta^4\big)
\ee
where $\widetilde{S}_{A,\infty,N}^{(\alpha)}$ for $N \in \{0,1,2,3\}$
are functions of $\eta$ which can be conveniently written as 
\be
\label{S-infty-tilde-ab-dec}
\widetilde{S}_{A,\infty,N}^{(\alpha)}
\,=\,   
\widetilde{S}_{A,\infty,N,a}^{(\alpha)}
+
\widetilde{S}_{A,\infty,N,b}^{(\alpha)}
\;\;\;\qquad\;\;\;
N\in \big\{ 0,1,2,3 \big\} \,.
\ee
The constant term $\widetilde{S}_{A,\infty,N,a}^{(\alpha)}$ reads
\be
\label{S-infty-tilde-ab-dec-a-part}
\widetilde{S}_{A,\infty,N,a}^{(\alpha)} =0 
\qquad 
N\in \big\{ 0,1,3 \big\}
\;\;\;\qquad\;\;\;
\widetilde{S}_{A,\infty,2,a}^{(\alpha)} 
= 
\frac{(1+\alpha)\, (3\,\alpha^2 - 7)}{384\, \alpha^3} \,.
\ee
The non-constant contribution $\widetilde{S}_{A,\infty,N,b}^{(\alpha)}$ 
to (\ref{S-infty-tilde-ab-dec}) takes the following form
\be
\label{EE-large-eta-oscillation-b-main}
\widetilde{S}_{A,\infty,N,b}^{(\alpha)}  
\,=\,
\frac{\kappa_N}{(\alpha-1)\, 2^{2N-1}}
\sum_{\substack{j=1 \\ k=0}}^{\infty}
(-1)^j\,
\frac{\big[ \Omega(-\textrm{i} y-1/2)^{j} \,  \mathcal{\widetilde{D}}_{N,j-1}^{-}\big]\!\big|_{\tilde{y}_k}
}{
(4\eta)^{2j(2k+1)/\alpha}}
\, \times
\Bigg\{
\begin{array}{ll}
\cos(4\eta j) \hspace{.8cm} & N\textrm{ even}
\\
\rule{0pt}{.6cm}
\sin(4\eta j)  & N\textrm{ odd} 
\end{array}
\ee
where $ \Omega(z) \equiv  \Gamma(1+z)^2 / \Gamma(-z)^2$,
the polynomials $\mathcal{\widetilde{D}}_{N,j-1}^{-}$ are 
\bea
\mathcal{\widetilde{D}}_{0,j-1}^{-}
&\equiv&
\frac{1}{j}
\\
\rule{0pt}{.5cm}
\mathcal{\widetilde{D}}_{1,j-1}^{-}
&\equiv&
1 - 12 y^2
\\
\rule{0pt}{.5cm}
\mathcal{\widetilde{D}}_{2,j-1}^{-}
&\equiv&
\frac{j}{2} \big( 12y^2 -1 \big)^{2} - 10 \,\textrm{i}\,y \,\big(4y^2-1\big)
\\
\rule{0pt}{.5cm}
\mathcal{\widetilde{D}}_{3,j-1}^{-}
&\equiv&
-\,\frac{j^2}{6} \big( 12y^2 -1 \big)^{3} 
- 10 \,\textrm{i}\,j\, y \,\big(48 y^4 -16 y^2 +1\big)
+
\left( 220 y^4 - 114 y^2 + \frac{37}{12}\, \right)
\hspace{1cm}
\eea
which must be evaluated on the following points belonging to the imaginary axis
\be
 \label{y_k-tilde-def-main}
 \tilde{y}_k \equiv \frac{\textrm{i}}{\alpha}\left(k+\frac{1}{2}\right)
 \ee 
and the constant $ \kappa_{N}  \equiv  \sin(\pi N / 2) + \cos(\pi N / 2) $,
which is equal to $\kappa_{N}  = + 1$ when $N=4J$ or $N=4J+1$
and $\kappa_{N}  = - 1$ when $N=4J+2$ or $N=4J+3$,
for $J\in\mathbb{N}_{0}$.
We remark that only a finite number of terms of the series in (\ref{EE-large-eta-oscillation-b-main})
are involved in the expansion of $S_A^{(\alpha)}$ at a given order as $\eta \to \infty$.
Notice that it is natural to write 
also (\ref{EE-large-eta-oscillation-b-main})
in terms of the area $\mathsf{a}$ of the limited phase space 
(see (\ref{a-parameters-def}) and (\ref{def-red-phae-space})).

The subleading contribution to the entanglement entropy corresponds to the limit $\alpha \to 1$
of $ \widetilde{S}_{A,\infty}^{(\alpha)} $ given by 
(\ref{exp-S-infty-123}), (\ref{S-infty-tilde-ab-dec}), 
(\ref{S-infty-tilde-ab-dec-a-part}) and (\ref{EE-large-eta-oscillation-b-main}).
It is remarkable that 
the oscillating quantity $\widetilde{S}_{A,\infty,N,b}^{(\alpha)} $ in (\ref{EE-large-eta-oscillation-b-main})
vanishes in the limit $\alpha \to 1$
because $\mathcal{\widetilde{D}}_{N,j-1}^{-}\big|_{\tilde{y}_k}$ 
and $\Omega(-\textrm{i} y-1/2) = O((\alpha -1)^2)$.
Thus,  in the regime of large $\eta$,
the subleading corrections to the entanglement entropies do not oscillate. 
%
Further subleading terms in the entanglement entropy have been evaluated in \cite{Susstrunk_2012}
and they do not oscillate.

\begin{figure}[t!]
\vspace{-.5cm}
\hspace{-.8cm}
\includegraphics[width=1.05\textwidth]{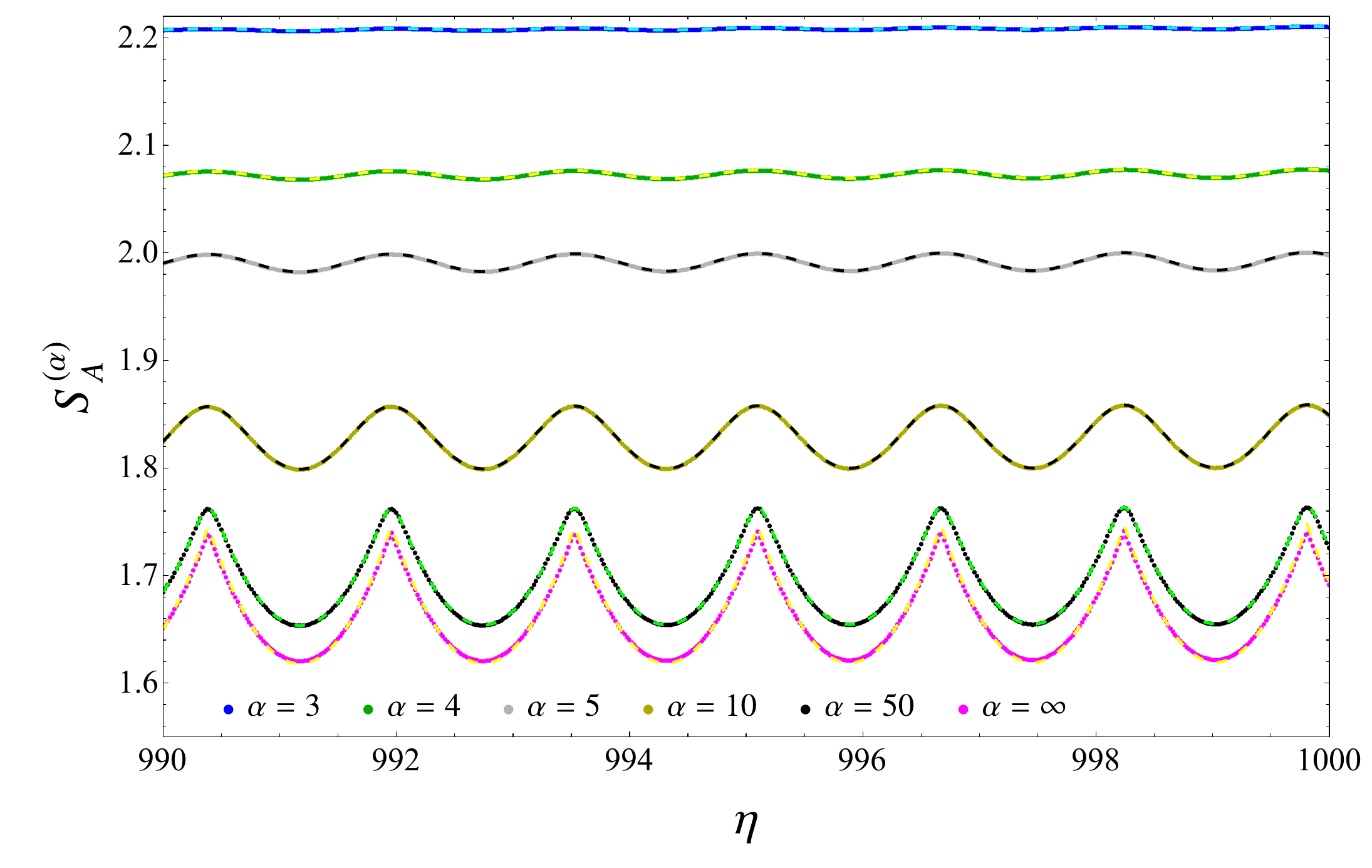}
\vspace{-.8cm}
\caption{
Oscillatory behaviour of the entanglement entropies 
with $\alpha > 1$ in the regime of large $\eta$.
The dashed curves are obtained from \eqref{entropies-tau-dec-large-eta}, 
\eqref{S_alpha_main_large_eta_0} and \eqref{exp-S-infty-123}.
}
\label{fig:ee-large-900}
\end{figure}

In the Appendix\;\ref{app-large-eta-single-copy} 
the expansion of the single copy entanglement $  \widetilde{S}_{A,\infty}^{(\infty)} $ for large $\eta$ has been studied,
finding 
\bea
\label{expansion-infty-app-34-order}
  \widetilde{S}_{A,\infty}^{(\infty)} 
&=&
      \frac{1}{4 \log(4\eta)}\;
      \Bigg\{
       \Big[ \textrm{Li}_{2}(-e^{4\textrm{i} \eta}) + \textrm{Li}_{2}(-e^{-4\textrm{i} \eta})  \Big]
       \left( 1 +\frac{\psi(1/2)}{4 \log(4\eta)} 
       +  \frac{\psi(1/2)^2}{[ 4 \log(4\eta)]^2}
       +  \frac{\psi(1/2)^3}{[ 4 \log(4\eta)]^3}
       \right)
             \nonumber
      \\
            \rule{0pt}{.8cm}
      & &\hspace{2.2cm}
      -\, \Big[ \textrm{Li}_{4}(-e^{4\textrm{i} \eta}) + \textrm{Li}_{4}(-e^{-4\textrm{i} \eta})  \Big]\,
      \frac{7\, \zeta(3)}{32\,  [\,\log(4\eta)]^3}
            \Bigg\}
            +
            O\big(1/[\log(\eta)]^5\big)
\eea
which can be written also in terms of 
the area $\mathsf{a}$ of the limited phase space 
(see (\ref{a-parameters-def}) and (\ref{def-red-phae-space}))
and improves the expansion obtained in \cite{Calabrese-Essler-10} in the lattice model
up to $O\big(1/[\log(\eta)]^3\big)$ term.
We remark that in \eqref{S-single-copy-app-final} all terms up to order $O(1/\eta)$ are reported.

For the sake of completeness, in the appendix\;\ref{sec_large_distance_lattice} 
we briefly discuss the lattice model 
where the Fisher-Hartwig formula and its generalisation
have been applied and the double scaling limit 
providing the results in the continuum.

In Fig.\,\ref{fig:ee-large-100} 
the entanglement entropies $S_A^{(\alpha)}$ 
are shown in a large domain $\eta \in (0\,,100)$
which includes the one considered in Fig.\,\ref{fig:ee-functions-gen}.
The filled circles are the data points obtained numerically,
as discussed in Sec.\,\ref{sec_entropies},
while the dashed lines correspond to the approximate
analytic expressions derived in the regime of large $\eta$ and given by 
\eqref{entropies-tau-dec-large-eta}, 
\eqref{S_alpha_main_large_eta_0}
and \eqref{exp-S-infty-123},
which perfectly agree with the 
numerical results for $\eta > 10$.
In this figure it is evident 
the logarithmic growth of $S^{(\alpha)}_A$
described by \eqref{S_alpha_main_large_eta_0},
the oscillatory behaviour for $S^{(\alpha)}_A$ when $\alpha > 1$
(see \eqref{EE-large-eta-oscillation-b-main})
and also the lack of such oscillations for the entanglement entropy.
The approximate analytic expressions for large $\eta$ mentioned above
have been employed also in Fig.\,\ref{fig:ee-functions-gen}
and they nicely agree with the numerical data points for $\eta > 1$,
which is quite remarkable.

\begin{figure}[t!]
\vspace{.3cm}
\hspace{-.8cm}
\includegraphics[width=1.05\textwidth]{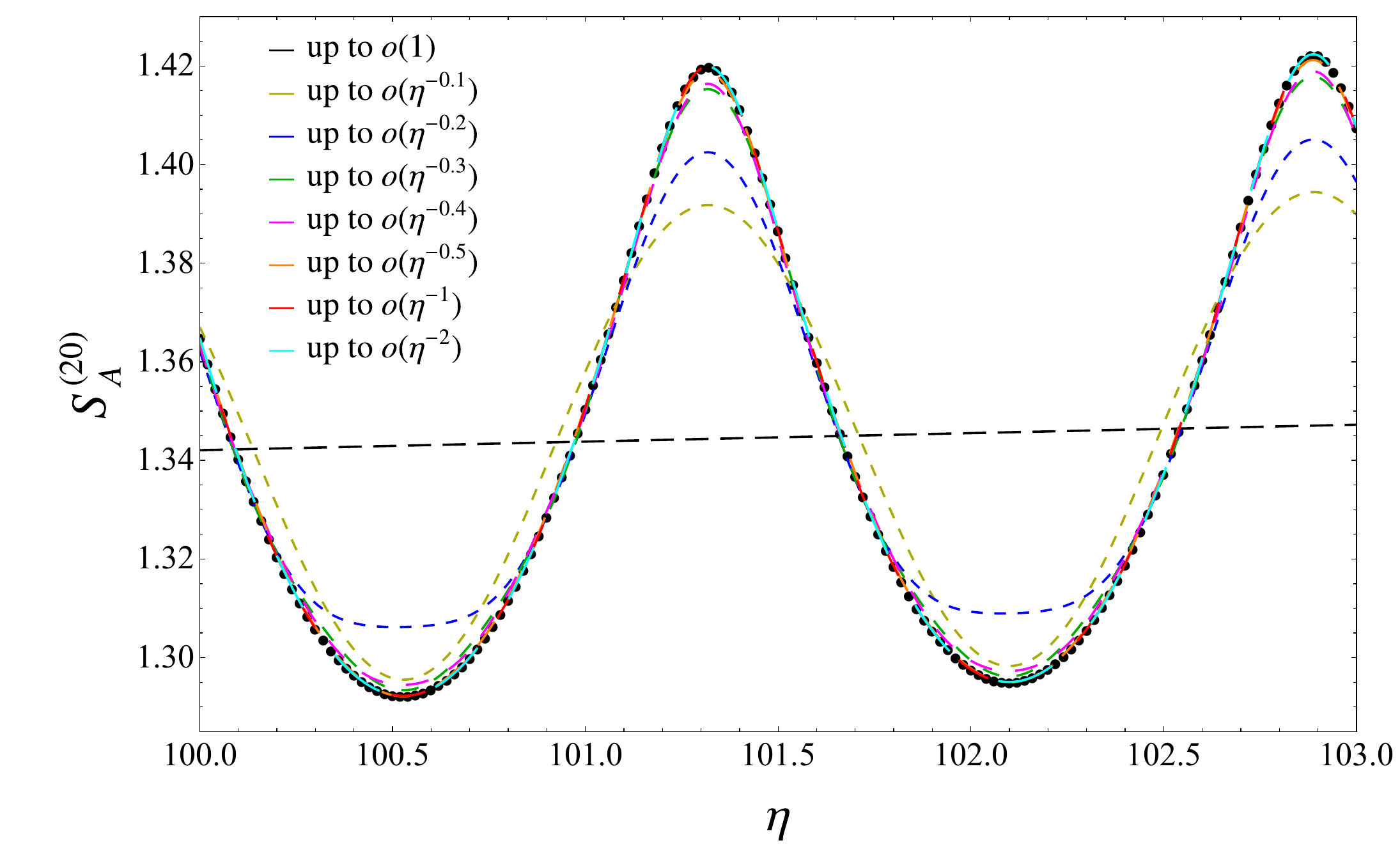}
\vspace{-.8cm}
\caption{
The R\'enyi entropy of order $\alpha=20$ in the regime of large $\eta$. 
}
\label{fig:reny20-large-100}
\end{figure}

The oscillations of the entanglement entropies $S_A^{(\alpha)}$ with index $\alpha \neq 1$
occurring in Fig.\,\ref{fig:ee-large-100} have the same period equal to $\pi/2$,
as one can observe from \eqref{EE-large-eta-oscillation-b-main}.
In Fig.\,\ref{fig:ee-large-900} 
we focus on the range $\eta \in (990\,, 1000)$
and show the change of these oscillations for different values of $\alpha$
in the regime of large $\eta$.
The numerical data points in Fig.\,\ref{fig:ee-large-900}  (filled circles) 
are nicely reproduced  by the dashed curves, obtained from \eqref{entropies-tau-dec-large-eta}, 
\eqref{S_alpha_main_large_eta_0} and \eqref{exp-S-infty-123}
with $o(1/\eta^2)$ terms neglected.
For small values of $\alpha > 1$ harmonic oscillations are observed.
As $\alpha$ increases, this behaviour changes 
and the local maxima of these oscillations become singular 
in the extreme case of the single-copy entanglement. 

\begin{figure}[t!]
\vspace{-.2cm}
\hspace{-.8cm}
\includegraphics[width=1.05\textwidth]{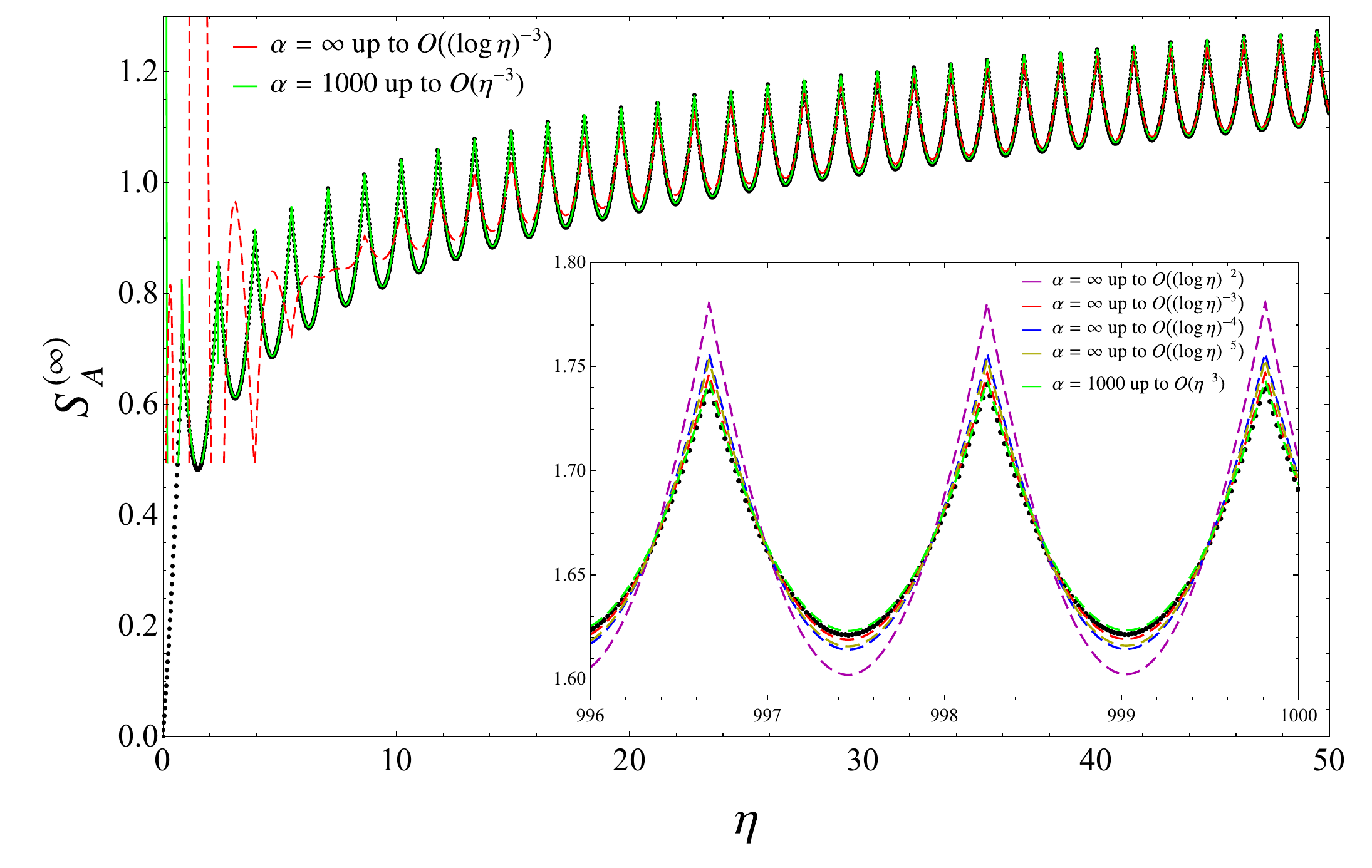}
\vspace{-.7cm}
\caption{
Single copy entanglement for large values of $\eta$.
}
\label{fig:E-infty-large-eta}
\end{figure}

Including more terms in the analytic expressions for the large $\eta$ 
expansion leads to improved approximations for the entanglement entropies, as expected.
This is shown in Fig.\,\ref{fig:reny20-large-100} for the case of $\alpha =20$,
where we compare the numerical data (filled circles) 
with the coloured dashed curves 
obtained from \eqref{entropies-tau-dec-large-eta}, \eqref{S_alpha_main_large_eta_0}, \eqref{S-infty-tilde-ab-dec}
and with the sum in \eqref{EE-large-eta-oscillation-b-main} 
truncated at different orders by neglecting the  $o(1/\eta^{r})$ terms
(in Fig.\,\ref{fig:reny20-large-100} and Fig.\,\ref{fig:E-infty-large-eta},
``up to $o(1/\eta^r)$" means that all the terms proportional to $1/\eta^{b}$ with $b \leqslant r$ have been included,
while ``up to $O(1/\eta^r)$" indicates that all the terms proportional to $1/\eta^{b}$ with $b < r$ have been considered).
For instance, the almost horizontal grey curve is found by neglecting the $o(1)$ terms; 
hence it corresponds to  \eqref{S_alpha_main_large_eta_0},
which is the contribution given by the logarithmic and the constant term. 
Similarly the dashed cyan curve has been obtained by 
neglecting in \eqref{EE-large-eta-oscillation-b-main} all the $o(1/\eta^2)$ terms,
finding a result which is almost indistinguishable 
from the numerical data in the range $\eta \in (100\, , 103)$.
In our numerical analyses, 
we have also observed that, for a given range of large values of $\eta$,
the agreement between the numerical data points and
the curves obtained from the analytic results 
corresponding to a certain approximation
improves as $\alpha$ decreases.  

In Fig.\,\ref{fig:E-infty-large-eta} we show the single copy entanglement $S^{(\infty)}_A$ for $\eta \in [0,50]$. 
The black filled circles correspond to the numerical data points obtained from \eqref{single-copy-ent-sum}.
In the main plot, the dashed red curve represents 
the expansion \eqref{expansion-infty-app-34-order} truncated to $O(1/(\log \eta)^{3})$,
while the solid green curve corresponds to the large $\eta$ expansion of $S^{(1000)}_A$,
found from \eqref{entropies-tau-dec-large-eta}, \eqref{S_alpha_main_large_eta_0} and  \eqref{S-infty-tilde-ab-dec}.
The large $\eta$ expansion of $S^{(1000)}_A$ approximates the numerical data points better than
the expansion of the single copy entanglement.
This may happen because \eqref{S-infty-tilde-ab-dec} contains higher order terms as $\eta \to \infty$
with respect to \eqref{expansion-infty-app-34-order}.
In the inset of Fig.\,\ref{fig:E-infty-large-eta}, we consider the domain $\eta \in [996, 1000]$ and show that
adding more terms in the expansion (\ref{expansion-infty-app-34-order}) 
does not necessarily improves the approximation of the numerical data point. 
Indeed, the best approximation corresponds to the truncation of  (\ref{expansion-infty-app-34-order}) 
up to the term proportional to $1/(\log \eta)^{2}$ included  (dashed red curve).

\begin{figure}[t!]
\vspace{-.2cm}
\hspace{-.8cm}
\includegraphics[width=1.05\textwidth]{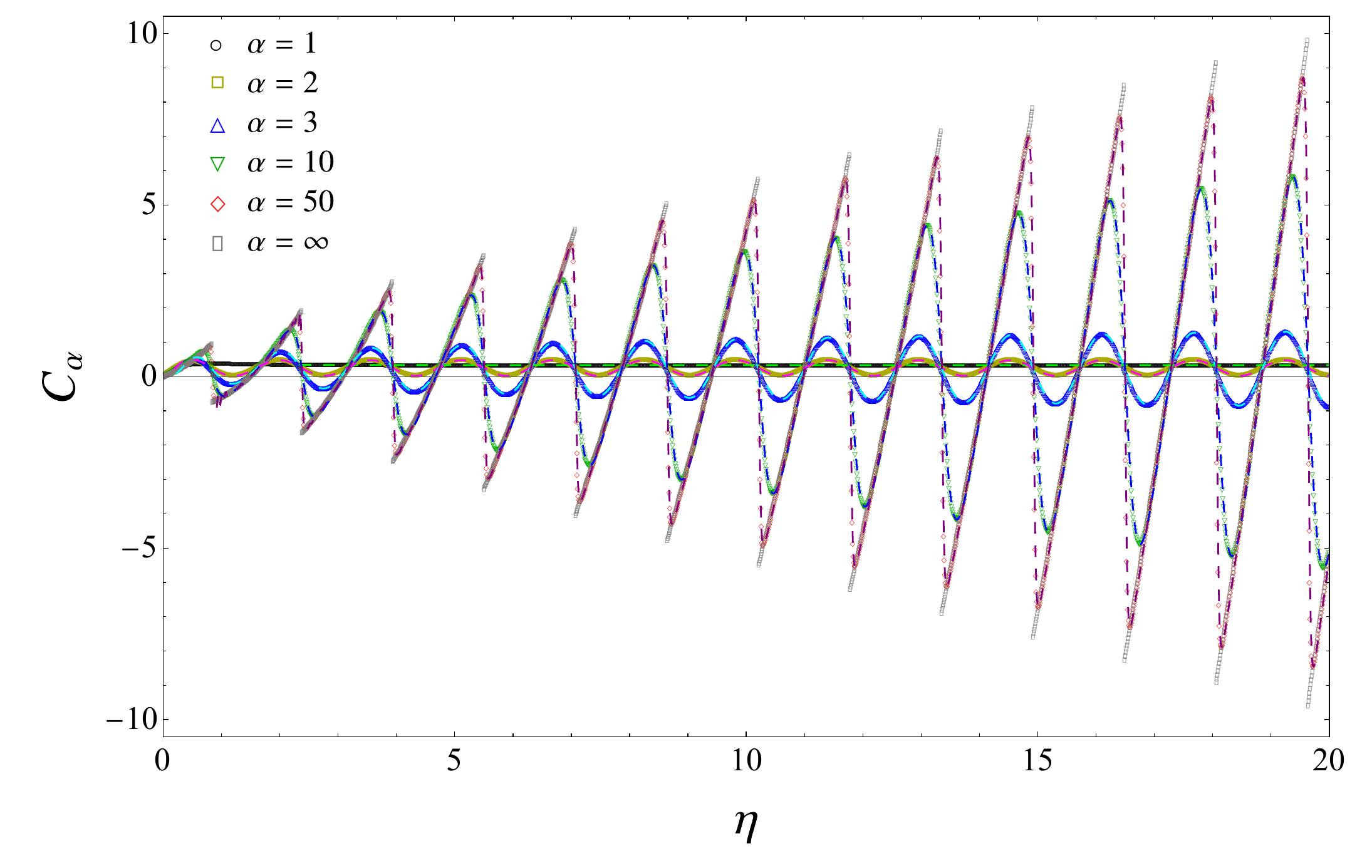}
\vspace{-.7cm}
\caption{
The quantities $C_\alpha$ in  \eqref{c-function-alpha} for different values of $\alpha$.
}
\label{fig:c-20-large-100}
\end{figure}

By applying the differential operator $\eta\, \partial_\eta\;$
to the analytic expressions of the expansion obtained 
from \eqref{entropies-tau-dec-large-eta}, \eqref{S_alpha_main_large_eta_0} and \eqref{S-infty-tilde-ab-dec}
and from \eqref{expansion-infty-app-34-order},
we obtain the large $\eta$ asymptotics for the quantities $C_\alpha$ introduced in  \eqref{c-function-alpha}.
Although we do not report the explicit expressions here, 
we have employed the resulting analytic expressions 
(where $O(1/\eta^{2})$ terms have been discarded)
to draw the dashed coloured curves in Fig.\,\ref{fig:c-20-large-100},
which are compared with the numerical data points obtained from (\ref{C-function-def})
(empty markers).

For the single copy entanglement, 
only the numerical data have been reported
because in this range of $\eta$ the analytic expression coming from
the expansion at large $\eta$ does not agree with the data points, 
as shown by the dashed red curve in Fig.\,\ref{fig:E-infty-large-eta}.
The oscillatory behaviour of $C_\alpha$ with $\alpha \neq 1$
about  the constant value $\tfrac{1}{6} \big( 1 + \tfrac{1}{\alpha}\big)$
is due to the term $\eta \,\partial_\eta \, \widetilde{S}_{A,\infty,N,b}^{(\alpha)}$ 
coming from \eqref{EE-large-eta-oscillation-b-main}. 
The growing amplitudes of the oscillations of  $C_\alpha$  with $\alpha > 2$ can be easily explained. 
Indeed, $ \widetilde{S}_{A,\infty}^{(\alpha)}$ contains a leading oscillating term $\propto \eta^{-2/\alpha} \cos(4\eta)$
from \eqref{EE-large-eta-oscillation-b-main},
which leads to the oscillating term $ \propto \eta^{1-2/\alpha} \sin(4\eta) $ in $C_\alpha$, 
whose amplitude is an increasing function of $\eta$ when $\alpha > 2$. 
This oscillatory behaviour becomes a sawtoothed curve when $\alpha\to \infty$. 
Instead, when $\alpha =1 $ the quantity $C$ in  \eqref{c-function-alpha} does not oscillate for large values of $\eta$
and $C \to (1/3)^+$ (see (\ref{large_eta_c_fun})), 
as shown also in Fig.\,\ref{fig:c-functions-gen} and Fig.\,\ref{fig:c-functions-alpha}.
This lack of oscillations is due to the fact that $\widetilde{S}_{A,\infty,N,b}^{(\alpha)} \to 0$ in \eqref{EE-large-eta-oscillation-b-main}
as $\alpha \to 1$, as already discussed below (\ref{y_k-tilde-def-main}).


\section{Schatten norms}
\label{sec_schatten}

In this section we study
the $p$-th power of the Schatten $p$-norm
$\parallel\! \!K \!\!\parallel_p$ of the sine kernel operator
for integer $p \geqslant 1$, which is defined as follows
\be
\label{powers-corr-def}
\mathcal{P}_A^{(p)}
\equiv
\big(\!
\parallel\!\! K \!\!\parallel_p
\!\big)^p
\equiv
\sum_{n=0}^\infty \gamma_n^p
\;\;\;\;\qquad\;\;\;\;
p > 0 \,.
\ee

These series are convergent; hence they are well defined functions of $\eta$.
For instance, for $p=1$ we have $\mathcal{P}_A^{(1)} = 2\eta /\pi $ (see (\ref{sum-gamma-n})).
In the other cases, the convergence of (\ref{powers-corr-def}) can be proved as follows. 
When $p \geqslant 1$, we have that $0<\mathcal{P}_A^{(p)} \leqslant \mathcal{P}_A^{(1)}$
because the inequalities in (\ref{spectrum-01}) tell us that $0<\gamma_n^p \leqslant \gamma_n$.
For $0 < p <1$, we observe that (\ref{gamma-upper-bound}) provides an upper bound 
which is a convergent series,
as one can show by using (\ref{gamma-tilde-infty}) and the ratio test,
as done in (\ref{ineq-SA-ren}) and (\ref{ratio-test}) for the entanglement entropies.

Following \cite{Klich-Levitov-09, Ivanov-2013, Klich-Laflorencie-12},
a time independent operator $Q_A$ can be introduced such that
\be
\label{QA-def}
\log\! \big[  \langle \e^{\textrm{i} \zeta Q_A} \rangle \big]
=
\textrm{Tr} \big[ \log  \big( I + (\e^{\textrm{i} \zeta} - 1) \, K \big) \,\big]
=
\textrm{Tr} \big[ \log  \big( I - z^{-1}\, K \big) \,\big]
=\,
\log(\tau)
\ee
where $I$ is the identity operator, $K$ is the sine kernel (\ref{sine-kernel def})
and $\zeta = 2\pi \nu_\star$, with $\nu_\star=\nu_\star(z)$ being defined in (\ref{nu-star-main}).
The cumulants of $Q_A$ are 
$\mathcal{C}_A^{(k)} \equiv \big[\partial^k_{\ri \zeta} 
\log \!\big(\langle \e^{\textrm{i} \zeta Q_A} \rangle \big)
\big]\big|_{\zeta=0}\,$, where $k \geqslant 1$ is an integer parameter.
From (\ref{QA-def}), the cumulants $\mathcal{C}_A^{(k)}$ for $k \in \{1,2,3\}$ read
\bea
\label{mean-QA}
\langle Q_A \rangle
&=&
\big[\partial_{\ri \zeta} 
\log \!\big(\langle \e^{\textrm{i} \zeta Q_A} \rangle \big)
\big]\big|_{\zeta=0}
=\,
\textrm{Tr}(K)
\\
\label{var-QA}
\rule{0pt}{.5cm}
\langle \big(Q_A - \langle Q_A \rangle \big)^2  \rangle
&=&
\big[\partial^2_{\ri \zeta} 
\log \!\big(\langle \e^{\textrm{i} \zeta Q_A} \rangle \big)
\big]\big|_{\zeta=0}
=\,
\textrm{Tr}\big(K - K^2\big)
\\
\label{3rd-cum-QA}
\rule{0pt}{.5cm}
\langle \big(Q_A - \langle Q_A \rangle \big)^3  \rangle
&=&
\big[\partial^3_{\ri \zeta} 
\log \!\big(\langle \e^{\textrm{i} \zeta Q_A} \rangle \big)
\big]\big|_{\zeta=0}
=\,
\textrm{Tr}\big(K - 3\,K^2 + 2\,K^3\big) \,.
\eea
These cumulants can be evaluated from (\ref{powers-corr-def}) with $p \in \{1,2,3\}$.
Notice that (\ref{mean-QA}) is equal to (\ref{powers-corr-def}) for $p=1$.

We find it useful to write
the $p$-th power of the Schatten $p$-norm of the sine kernel 
in terms of the sine kernel tau function (\ref{tau-function-sine}).
This can be done by adapting to  (\ref{powers-corr-def}) 
the procedure to obtain the contour integral (\ref{entropies-tau}) for the entanglement entropies,
discussed in Sec.\,\ref{sec_entropies},
and the result reads
\be
\label{powers-corr}
\mathcal{P}_A^{(p)}
\,=
  \lim_{\epsilon, \delta \to 0}\,
   \frac{1}{2\pi \textrm{i}} 
  \oint_{\mathfrak{C}}
 z^p \, \partial_z  \log( \tau )\, \textrm{d}z
 \;\;\;\qquad\;\;\;
 p \in \mathbb{N}
\ee
where we focus on integer values $p \geqslant 1$, for simplicity. 

In order to study (\ref{powers-corr-def}) in the small $\eta$ regime through (\ref{powers-corr}),
let us write the small $\eta$ expansion  
of the sine kernel tau function given in (\ref{tau-small-distance-expansion}) as follows
\be
\label{tau-0-repacked}
\tau\,=\, 1 - \mathcal{T}_0
\;\;\;\;\qquad\;\;\;\;
\mathcal{T}_0 \equiv \sum_{n=1}^{\infty} (-1)^{n-1}\, \frac{\widetilde{\mathcal{B}}_n(\eta) }{z^n}
\ee
where
\be
\label{B-tilde-def}
\widetilde{\mathcal{B}}_n(\eta) \equiv \frac{G(1+n)^6}{(2\pi)^n\,G(1+2n)^2}\;(4\eta)^{n^2}  \mathcal{B}_n(\eta) \,.
\ee
The expansions of the functions $\mathcal{B}_n(\eta)$ as $\eta \to 0$ 
have been obtained in the Appendix\;B of \cite{Gamayun:2013auu} for $n \leqslant 5$ 
and up to a certain order in $\eta$.
We have reported them in the Appendix\;\ref{app_small_eta_entropies}
(see (\ref{B0-B1-expansion})-(\ref{B4-expansion})),
truncated to a certain order as discussed in Sec.\,\ref{sec_small_distance_P5}.

\begin{figure}[t!]
\vspace{-.2cm}
\hspace{-.8cm}
\includegraphics[width=1.05\textwidth]{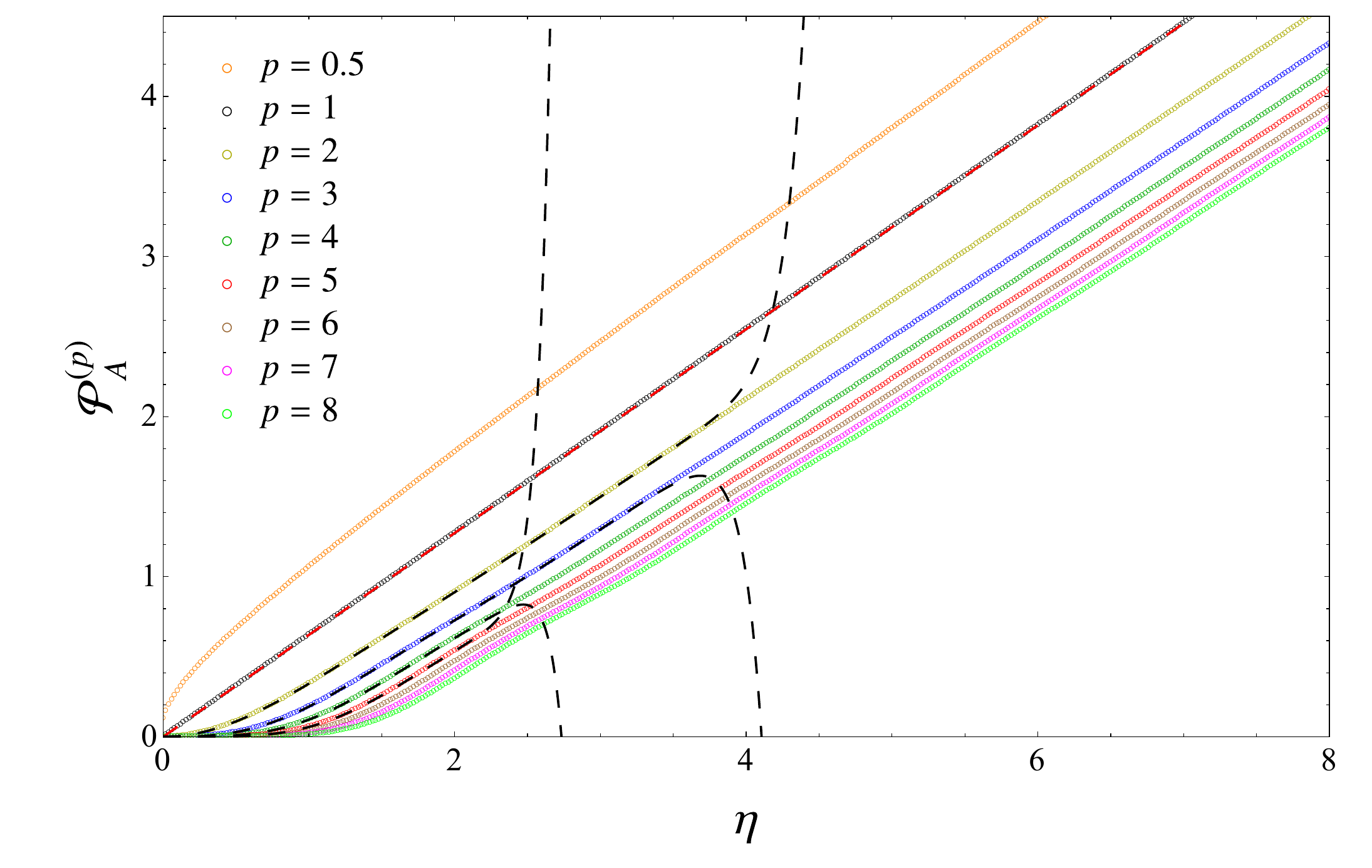}
\vspace{-.7cm}
\caption{
The $p$-th power of the Schatten $p$-norm of the sine kernel operator (see \eqref{powers-corr-def}) for some values of $p>0$, in the regime of small $\eta$.
When $p=1$, the exact result is given by \eqref{sum-gamma-n} for every $\eta$.
The dashed red line corresponds to \eqref{sum-gamma-n} (see also \eqref{P1-ris})
and the dashed black lines to (\ref{P2-ris}-\ref{P5-ris}).
}
\label{fig:app-traces}
\end{figure}

By employing \eqref{tau-0-repacked} into \eqref{powers-corr}, the resulting integral can be evaluated by first expanding the integrand $z^p \, \partial_z  \log( \tau )$ as $z\to \infty$ and then applying the residue theorem, which tells us to select the coefficient of the term corresponding to $1/z$ in the expansion of the integrand. 
For the first values of $p$, we obtain
\bea
\label{P1-ris}
\mathcal{P}_A^{(1)} &=& \widetilde{\mathcal{B}}_1(\eta) \,=\, \frac{2\eta}{\pi}
\\
\rule{0pt}{.5cm}
\label{P2-ris}
\mathcal{P}_A^{(2)} &=& \widetilde{\mathcal{B}}_1(\eta)^2 - 2\,\widetilde{\mathcal{B}}_2(\eta)
\\
\rule{0pt}{.6cm}
\label{P3-ris}
\mathcal{P}_A^{(3)} 
&=& 
\widetilde{\mathcal{B}}_1(\eta)^3 
- 3\,\widetilde{\mathcal{B}}_1(\eta)\, \widetilde{\mathcal{B}}_2(\eta)
+ 3\,\widetilde{\mathcal{B}}_3(\eta)
\\
\rule{0pt}{.6cm}
\label{P4-ris}
\mathcal{P}_A^{(4)} 
&=& 
\widetilde{\mathcal{B}}_1(\eta)^4
- 4\,\widetilde{\mathcal{B}}_1(\eta)^2\, \widetilde{\mathcal{B}}_2(\eta)
+ 4\,\widetilde{\mathcal{B}}_1(\eta)\, \widetilde{\mathcal{B}}_3(\eta)
+ 2\,\widetilde{\mathcal{B}}_2(\eta)^2
- 4\,\widetilde{\mathcal{B}}_4(\eta)
\\
\rule{0pt}{.6cm}
\label{P5-ris}
\mathcal{P}_A^{(5)} 
&=&
\widetilde{\mathcal{B}}_1(\eta)^5
- 5\,\widetilde{\mathcal{B}}_1(\eta)^3\, \widetilde{\mathcal{B}}_2(\eta)
+ 5\,\widetilde{\mathcal{B}}_1(\eta)\, \widetilde{\mathcal{B}}_2(\eta)^2
+ 5\,\widetilde{\mathcal{B}}_1(\eta)^2\, \widetilde{\mathcal{B}}_3(\eta)
- 5\,\widetilde{\mathcal{B}}_1(\eta)\, \widetilde{\mathcal{B}}_4(\eta)
\hspace{1.4cm}
\nonumber
\\
& &
- \,5\,\widetilde{\mathcal{B}}_2(\eta)\, \widetilde{\mathcal{B}}_3(\eta)
+ 5\, \widetilde{\mathcal{B}}_5(\eta)
\eea
in terms of the functions introduced in (\ref{B-tilde-def}).
Notice that (\ref{P1-ris}) is equal to 
the mean value of the particle number operator given in (\ref{N_A-segment}) 
and to the trace of the sine kernel given in (\ref{sum-gamma-n}), 
as expected. 
We observe that the leading behaviour of $\mathcal{P}_A^{(p)}$ as $\eta \to 0$ is $O(\eta^p)$,
which is determined by the term $\widetilde{\mathcal{B}}_1(\eta)^p$.

In Fig.\,\ref{fig:app-traces} 
we show (\ref{powers-corr-def}) in the regime of small $\eta$.
The red dashed line is the exact result (\ref{sum-gamma-n}),
while the black dashed lines correspond to the  analytic expressions in (\ref{P2-ris})-(\ref{P5-ris}),
which hold as $\eta \to 0$ and are polynomials up to $O(\eta^{31})$.
Notice that 
the expansion of $\mathcal{B}_5(\eta)$ in Appendix\;B of \cite{Gamayun:2013auu} 
has been employed for $\mathcal{P}_A^{(5)} $
and that $\widetilde{\mathcal{B}}_p(\eta)$ for $p>5$ are not available in the literature.

\begin{figure}[t!]
\vspace{-.2cm}
\hspace{-.8cm}
\includegraphics[width=1.05\textwidth]{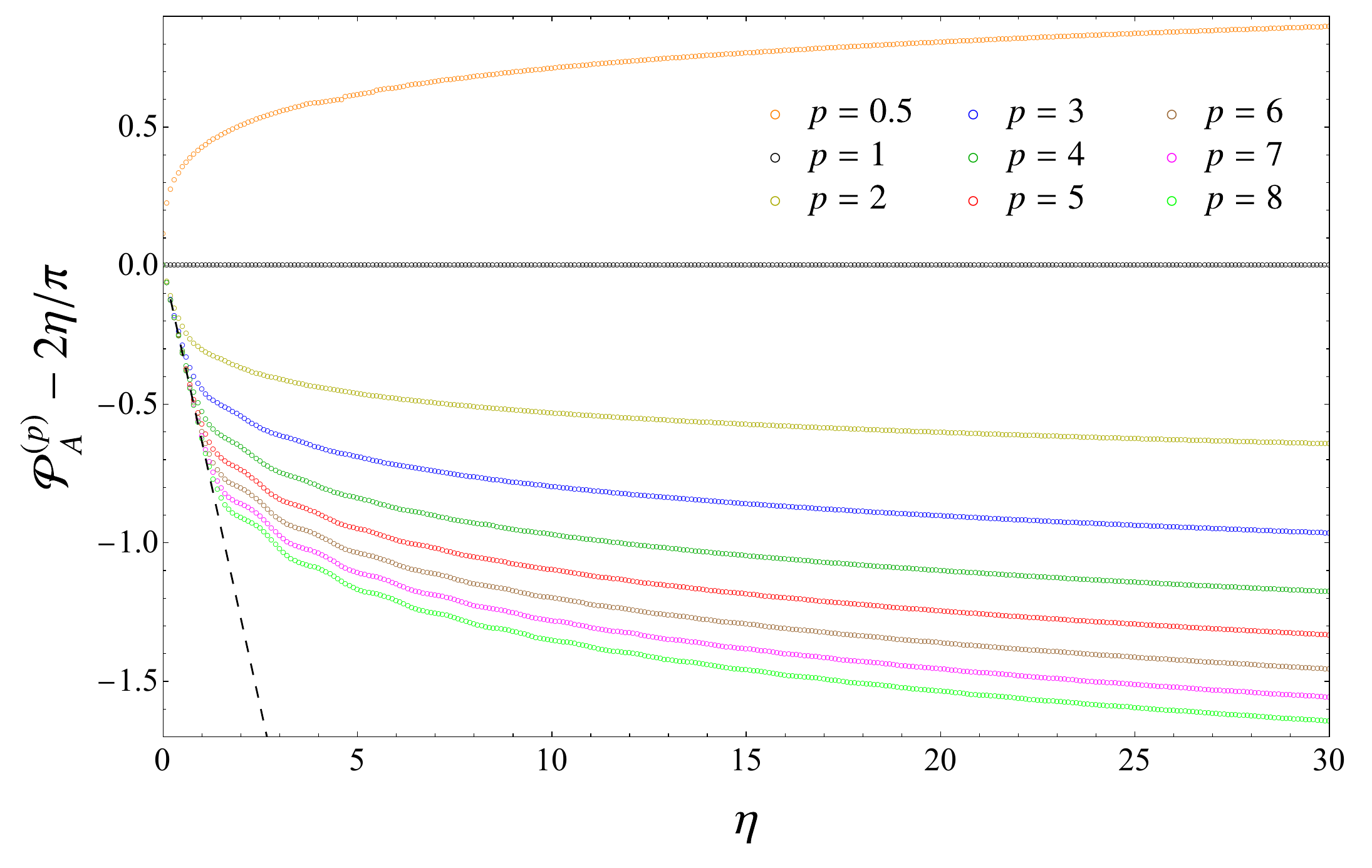}
\vspace{-.7cm}
\caption{
The $p$-th power of the Schatten $p$-norm of the sine kernel operator (see (\ref{powers-corr-def}))
for some values of $p>0$,
where the linear divergence (\ref{Pp-large-eta-linear-term}) for large $\eta$ has been subtracted
to highlight the subleading correction (see (\ref{Pp-large-eta-linear-log-term})).
The black dashed straight line is  $-2\eta/\pi$.
}
\label{fig:app-traces-sub}
\end{figure}

In the large $\eta$ regime, we can employ the factorised form (\ref{tau-infty-dec}),
where the leading terms are contained in $\tilde{\tau}_{\infty} $, defined in (\ref{tau-tilde-infty-def}).
Plugging this factorisation into (\ref{powers-corr}), 
we find that the leading terms of (\ref{powers-corr-def}) as $\eta \to \infty$ are given by 
\be
\label{PA-p-large-eta-contour}
\mathcal{P}_A^{(p)}
\,=
  \lim_{\epsilon, \delta \to 0}\,
   \frac{1}{2\pi \textrm{i}} 
  \oint_{\mathfrak{C}}
 z^p \, \partial_z \log(\tilde{\tau}_{\infty} )\, \textrm{d}z
 \,+ o(1)
 \;\;\;\;\qquad\;\;\;\;
 \eta \,\to \, \infty \,.
\ee
By using (\ref{dlog_taut}) with $\nu_\star(z)$ defined in (\ref{nu-star-main}),
the integrand of (\ref{PA-p-large-eta-contour}) reads
\be
\label{tau-infty-exp-schatten}
z^p\,\partial_z \log(\tilde{\tau}_{\infty} )
\,=\,
\frac{z^{p-1} }{2\pi \textrm{i}\, (z-1)}\,
\Big[\,  4\eta\, \textrm{i}
-4\nu_{\star}\log(4\eta)
+2\nu_{\star} \Big( \psi(1-\nu_{\star}) +\psi(1+\nu_{\star}) - 2\Big)
\Big]
\ee
in terms of the digamma function  $\psi(z)$ (see (\ref{digamma_fun})).

The leading term of (\ref{PA-p-large-eta-contour}) comes
from the linear term in $\eta$ occurring in (\ref{tau-infty-exp-schatten})
and it can be easily evaluated by using the residues theorem, finding 
\be
\label{Pp-large-eta-linear-term}
\mathcal{P}_A^{(p)}
\,=\,
\frac{2}{\pi}\, \eta + O( \log \eta)
 \;\;\;\;\qquad\;\;\;\;
 \eta \,\to \, \infty
\;\;\;\;\qquad\;\;\;\;
p \in \mathbb{N}\,.
\ee
This linear divergence can be observed already in the range of $\eta$ considering in Fig.\,\ref{fig:app-traces}.
\\
When $p=1$, the exact result in (\ref{N_A-segment}) and (\ref{sum-gamma-n}) 
tells us that further corrections do not occur.

The linear divergence (\ref{Pp-large-eta-linear-term}) can be understood also by observing that 
the largest contribution as $\eta \to \infty$ 
comes from the eigenvalues $\gamma_n$ having $n \leqslant n_0$,
where $n_0$ is the critical index (\ref{critical-n0-def}).
Furthermore, in the regime of large $\eta$, an upper bound for the contribution coming 
from the regions I and II of the spectrum (see Sec.\,\ref{sec_spectral}) 
can be obtained by using $\gamma_n^p < 1$ and 
 the Landau-Widom counting formula (\ref{LW-counting}),
 where the coefficient of $\log(\eta)$ is positive.
 
 Notice that the linear divergence (\ref{Pp-large-eta-linear-term}) simplifies in the combinations 
 (\ref{mean-QA}),  (\ref{var-QA}) and  (\ref{3rd-cum-QA}).

The evaluation of the subleading term in (\ref{PA-p-large-eta-contour}),
which corresponds to the term containing $\log(4\eta)$
in (\ref{tau-infty-exp-schatten}), is less straightforward. 
Indeed, splitting the contour $\mathfrak{C}$ as discussed in Sec.\,\ref{sec_entropies},
one finds that, while the integral over $\mathfrak{C}_{0}$ vanishes as $\epsilon \to 0$,
the integral over $\mathfrak{C}_{1}$ diverges like $\log(\epsilon) + o(1)$.
This logarithmic divergence cancels with the same divergence coming from 
the integral over $\mathfrak{C} _{+} \cup \mathfrak{C} _{-}$,
which gives a result proportional to $- \log(\epsilon) - \big[ \gamma_{\textrm{\tiny E}} + \psi(p) \big]$.
Combining these contributions, we obtain
\be
\label{Pp-large-eta-linear-log-term}
\mathcal{P}_A^{(p)}
\,=\,
\frac{2}{\pi}\, \eta -\frac{\log(4\eta)}{\pi^{2}} \, \big[ \gamma_{\textrm{\tiny E}} + \psi(p) \big]+ O(1)
 \;\;\qquad\;\;
 \eta \,\to \, \infty
 \;\;\qquad\;\;
p \in \mathbb{N}
\ee
where $\gamma_{\textrm{\tiny E}} \simeq 0.577$ is the Euler-Mascheroni constant.
Notice that $\gamma_{\textrm{\tiny E}} + \psi(1)  = 0$, as expected. 
Furthermore,  $\gamma_{\textrm{\tiny E}} + \psi(p) $ is a rational number when $p \in \mathbb{N}$.

In order to highlight the subleading corrections for large $\eta$,
in Fig.\,\ref{fig:app-traces-sub} we show $\mathcal{P}_A^{(p)} - 2\eta/\pi$.
In this figure, oscillations are visible for the data corresponding to high values of $p$.
They are expected from the subleading corrections
due to the contribution of $\mathcal{T}_\infty$ (see (\ref{tau-infty-dec})) in (\ref{powers-corr}).
The fact that the numerical data points for small values of $\eta$ in Fig.\,\ref{fig:app-traces-sub}
follow the straight line $-2\eta/\pi$
supports the above observation about the leading behaviour of $\mathcal{P}_A^{(p)}$ as $\eta \to 0$.
The range of $\eta$ considered in Fig.\,\ref{fig:app-traces-sub} is too small 
to compare the coefficient of $\log(\eta)$ in (\ref{Pp-large-eta-linear-log-term})
with the numerical data points. 
%

\section{Conclusions}
\label{sec_conclusions}

We have investigated the entanglement entropies $S_A^{(\alpha)}$ of an interval $A$
in the free fermionic spinless Schr\"odinger field theory on the line, 
at finite density $\mu$ and zero temperature.

This problem can be studied without introducing approximations
because the spectral problem (\ref{spectral-problem-R})
associated to the sine kernel in the interval $A$
has been solved long ago by Slepian, Pollak and Landau in the seminal papers
\cite{Slepian-part-1, Slepian-part-2, Slepian-part-3, Slepian-part-4, Slepian-83},
which have generated a vast literature afterwards, in various directions \cite{Rokhlin-book}.
The eigenvalues $\gamma_n$ of this spectral problem (see (\ref{eigenvalues})) 
depend only on the dimensionless parameter $\eta$ defined in (\ref{eta-def}),
which is proportional to the area of the limited phase space (\ref{def-red-phae-space}).
By employing the fact that $\gamma_n$ vanish super-exponentially as $n \to \infty$ 
(see (\ref{gamma-upper-bound}) and (\ref{gamma-tilde-infty})),
we proved that the entanglement entropies $S_A^{(\alpha)}$ are finite functions of $\eta$.
These functions are displayed e.g. in Fig.\,\ref{fig:ee-functions-gen} and Fig.\,\ref{fig:ee-large-100} for some values of $\alpha$,
where the data points have been generated
through an efficient code optimised to evaluate numerically the PSWFs
(kindly given to us by Vladimir Rokhlin).

In Sec.\,\ref{subsec_ent_loss} it is shown that 
$S_A$ is a function that monotonically increases with the area of the limited phase space
(see also the numerical results for $S_A^{(\alpha)}$ in Fig.\,\ref{fig:ee-functions-gen} and Fig.\,\ref{fig:ee-large-100}).
This property does not hold for the entanglement entropies with index $\alpha \neq 1$.
We proved that  the analogue of the entropic $C$ function for the $d=1$ relativistic models  \cite{Casini:2004bw},
defined in (\ref{c-function-alpha}), is not a monotonous function of $\eta$ (see Fig.\,\ref{fig:c-functions-gen}).
Notice that, in the context of the gauge/gravity correspondence, 
non-monotonic holographic entropic $C$ functions \cite{Myers:2010xs, Myers:2010tj}
have been also found when Lorentz invariance is broken \cite{Hoyos:2020zeg}.

We have studied also the entanglement entropies of an interval 
for a class of free fermionic massless Lifshitz models on the line (see (\ref{e1L}))
at zero temperature and finite density,
which are labelled by their integer Lifshitz exponent $z\geqslant 1$.
This class includes the relativistic massless chiral fermion ($z=1$)
and the above mentioned free fermionic spinless Schr\"odinger field theory ($z=2$).
Important qualitative differences in the entanglement entropies are observed, depending on the parity of $z$.
For instance, 
the models with even $z$ have finite and $\mu$-dependent $S_A^{(\alpha)}$,
while the models with odd $z$ have UV divergent and $\mu$-independent $S_A^{(\alpha)}$.
For the subclass of models having odd Lifshitz exponents, 
we have computed the entanglement Hamiltonian (\ref{K_A-local-def}),
the modular flow of the field (\ref{mod-flow-mu-main})
and the corresponding correlation function (\ref{mod-corr-1}),
finding that, instead, these entanglement quantifiers 
explicitly depend both on $\mu$ and on the Lifshitz exponent
through the Fermi momentum (\ref{kF-def-L}).

Finally, we have employed the method of \cite{Jin_2004, Keating_04}
and the asymptotic expansions for small and large $\eta$ of the sine kernel tau function
(see (\ref{tau-small-distance-expansion}) and (\ref{tau-expansion-large-eta}) respectively)
found in \cite{Gamayun:2013auu, Lisovyy:2018mnj, Bonelli:2016qwg}
to write the expansions of the entanglement entropies in these limiting regimes
(see Sec.\,\ref{sec_small_distance} and Sec.\,\ref{sec_large_distance} respectively).
The analytic expressions approximating the entanglement entropies
obtained from these expansions give the dashed curves 
in all the figures from Fig.\,\ref{fig:ee-functions-gen} to Fig.\,\ref{fig:c-20-large-100},
which display a remarkable agreement with the numerical data points. 
These analytic results have also allowed to prove 
the non-monotonicity of the function $C$  (see Sec.\,\ref{sec_flow}).
Thus, our analysis provides a new application of the results 
for the general solution of the Painlev\'e V equation
obtained in \cite{Gamayun:2013auu, Lisovyy:2018mnj, Bonelli:2016qwg},
specialised to the simple case of the sine kernel tau function. 

Our results can be extended in many interesting directions.
Since we have described an explicit example where the entanglement entropy is finite 
and monotonically increasing along the $\eta$-flow, 
it would be insightful to find whether this interesting feature occurs also in other models
and whether it provides some new insights about the RG flows in non-relativistic field theories. 
For instance, it would be instructive to find a RG flow involving 
non-relativistic models where the entanglement entropy 
plays the role of the entropic $C$ function.

It would be interesting to study the
entanglement entropies of an interval on the line for the fermionic spinless Schr\"odinger model
in a general Gibbs state, where both the density and the temperature occur \cite{Spitzer-16,Spitzer-17,Spitzer-22}.
The most important generalisation to explore 
is the spinfull model with a quartic interaction  \cite{Benfatto-book, Gallavotti_01, Gentile:2001gb}.
It would be insightful to study also
the entanglement entropies corresponding to bipartitions of the line where the subsystem is the union of disjoint intervals \cite{Calabrese:2009ez,Calabrese:2010he,Casini:2009vk,Coser:2013qda,Coser:2015dvp,DeNobili:2015dla,Arias:2018tmw} (see also \cite{Brightmore_2020} for lattice computations), or other entanglement measures, like e.g. the logarithmic negativity
\cite{Peres_96,Werner_02,Plenio_05, Calabrese:2012ew, Calabrese:2012nk, Calabrese:2014yza, Zimboras_15, Coser:2015mta, Coser:2015eba, Grava:2021yjp, Zimboras_16, DeNobili:2016nmj}.
Other interesting extensions involve non-relativistic models for bosonic fields  
and higher dimensions.

\vskip 20pt 
\centerline{\bf Acknowledgments} 
\vskip 5pt 

We are deeply grateful to Vladimir Rokhlin for having shared with us his code, 
that has been employed to generate the numerical data points reported in this manuscript. 
We thank Oleg Lisovyy for having shared with us his code to find the functions $\mathcal{D}_k$
and Wolfgang Spitzer for having allowed us to report his argument about the monotonicity of $S_A$.
We are grateful to Ingo Peschel for useful comments on the draft.
We acknowledge Filiberto Ares, Marco Bertola, Giulio Bonelli,  Viktor Eisler, Valentina Giangreco, Tamara Grava, 
Jelle Hartong, Vieri Mastropietro, Giuseppe Mussardo, Domenico Seminara and Alessandro Tanzini
for insightful discussions.
%
%
ET’s research has been conducted within the framework of the Trieste Institute for Theoretical Quantum Technologies (TQT).

\vskip 30pt

\appendix


\section{Bounding the entanglement entropy} 
\label{app_bounds}

In order to bound the entanglement entropies,
let us introduce a parameter $q \in \big( 0, 1/2 \big)$ 
and the corresponding partition of $\mathbb{N}_0$ as follows
\bea
\Lambda_{1,q}
&\equiv &
\big\{\, n \in \mathbb{N}_0\; ; \; 1-q \leqslant  \gamma \leqslant 1\, \big\}
\\
\rule{0pt}{.6cm}
\Lambda_{\frac{1}{2},q}
&\equiv &
\big\{\, n \in \mathbb{N}_0\; ; \; q < \gamma < 1-q\, \big\}
\\
\rule{0pt}{.4cm}
\Lambda_{0,q}
&\equiv &
\big\{\, n \in \mathbb{N}_0\; ; \; 0\leqslant  \gamma \leqslant q\, \big\} \,.
\eea

Considering e.g. the entanglement entropy (\ref{entropies-def-sums}) 
(the discussion can be easily adapted to the other entanglement entropies), 
the above partition of $\mathbb{N}_0$ naturally provides the decomposition
\be
\label{ee-dec-0-1/2-1}
S_{A}  =
S_{A,1}+S_{A,\frac{1}{2}}+S_{A,0}
\;\;\qquad \;\;
S_{A,r}\equiv\sum_{n\in\Lambda_{r,q}}s(\gamma_{n})
\qquad
r \in \big\{0, \tfrac{1}{2}, 1\big\}\,.
\ee

Upper and lower bounds depending on $\eta$
for $S_{A,r}$ in (\ref{ee-dec-0-1/2-1}) can be studied
by using approximating formulas for $\gamma_n$
whose validity depends on which $\Lambda_{r,q}$
the label $n$ belongs to.

A numerical inspection shows that  $S_{A,r}=O(\log(\eta))$ as $\eta\to \infty$;
hence we expect
\be
L_{r}^{\textrm{\tiny low}}
\log(\eta)+C_{r}^{\textrm{\tiny low}}
\leqslant 
\, S_{A,r} \,
\leqslant 
L_{r}^{\textrm{\tiny up}}\log(\eta)+C_{r}^{\textrm{\tiny up}}
\;\;\qquad\;\;
\eta \gg 1
\qquad
r \in \big\{0, \tfrac{1}{2}, 1\big\}
\ee
where the constants $L_{r}^{\textrm{\tiny low}}$, $C_{r}^{\textrm{\tiny low}}$, 
$L_{r}^{\textrm{\tiny up}}$ and $C_{r}^{\textrm{\tiny up}}$ do not depend on $\eta$. 

The simplest term to consider is $S_{A,\frac{1}{2}}$.
For this quantity, it is straightforward to write 
\be
\label{SA1/2}
s(q) \, \big|\Lambda_{\frac{1}{2},q}\big|
\leqslant S_{A,\frac{1}{2}}\leqslant 
s(1/2) \, \big|\Lambda_{\frac{1}{2},q}\big|
\ee
where  $ |Q |$ denotes the cardinality of $Q$.
The asymptotic behaviour of $\big|\Lambda_{\frac{1}{2},q}\big|$ for large $\eta$ is 
\cite{Landau-Widom}
\be
\label{size_plunge}
\big|\Lambda_{\frac{1}{2},q}\big|
=
\frac{2}{\pi^{2}} \log(1/q-1)\, \log(\eta) +o(\log(\eta))
\;\;\;\qquad\;\;\;
\eta\rightarrow \infty 
\ee
The upper bound in \eqref{SA1/2} can be studied also by
employing the following result \cite{KARNIK202197}
\be
\big|\Lambda_{\frac{1}{2},q}\big|
\,\leqslant\,
\frac{2}{\pi^{2}}\log\!\left(\frac{5}{q(1-q)}\right)
\log\!\left(\frac{100\,\eta}{\pi}+25\right)+7
\ee
which gives
\bea
L_{\frac{1}{2}}^{\textrm{\tiny up}} 
& \equiv & 
\frac{2\log(2)}{\pi^{2}}\log\!\left(\frac{5}{\delta(1-\delta)}\right)
\\
\rule{0pt}{.7cm}
C_{\frac{1}{2}}^{\textrm{\tiny up}} 
& \equiv & 
7\log(2)
+\frac{2\log(2)}{\pi^{2}}\
\log\!\left(\frac{5}{\delta(1-\delta)}\right)
\big[\log(100/\pi )+\log(1+\pi/4)\big] \,.
\eea
We have not found a lower bound for $\big|\Lambda_{\frac{1}{2},q}\big|$.

By applying the procedure described above for $S_{A,\frac{1}{2}}$ 
to $S_{A,1}$ and $S_{A,0}$ in (\ref{ee-dec-0-1/2-1}),
one obtains 
\be
\label{SA10}
0  \leqslant S_{A,r}\leqslant s(q)\,
\big|\Lambda_{r,q}\big|
\;\;\qquad \;\;
r\in \big\{ 0,1\big\}\,.
\ee

The Landau-Widom counting formula (see \cite{Landau-Widom} and theorem 3.14 of \cite{Rokhlin-book})
\be
\label{LW-counting}
\big| \Lambda_{1,q}\big|  + \big| \Lambda_{\frac{1}{2},q}\big|
\,=\,
\frac{2\eta}{\pi} 
+ \frac{1}{\pi^2} \,\log\!\left(\frac{1-q}{q} \right) \, \log (\eta) 
+ o\big(\log (\eta) \big)
\ee
and \eqref{size_plunge} imply that $\left|\Lambda_{1,q}\right|=O(\eta)$  as $\eta\rightarrow+\infty$.
Moreover, $\left|\Lambda_{0,q}\right|$ is infinite for any $\eta > 0$. 
Hence, \eqref{SA10} are not useful to show that $S_A$ grows logarithmically as $\eta \to \infty$.

\section{Cumulants of the entanglement spectrum}
\label{app_cumulants}

The reduced density matrix $\rho_A$ introduced in Sec.\,\ref{sec_intro}
(normalised by the condition $\textrm{Tr} \rho_A = 1$)
naturally leads to define the entanglement Hamiltonian $K_A$
as $\rho_A = e^{-K_A}$.
The spectrum of $K_A$ is called entanglement spectrum
and its relevance has been discussed e.g. in \cite{EislerPeschel:2009review, Li:2008kda}.
Important information about the entanglement spectrum 
can be obtained by considering 
the moments of $K_A$, i.e. $\textrm{Tr}(\rho_A^n)$,
or, equivalently,  the cumulants of the entanglement spectrum. 
In this appendix we evaluate the first cumulants 
of the entanglement spectrum corresponding to the interval $A =[-R,R] \subset \RR$ 
for the free fermionic Schr\"odinger field theory at zero temperature and finite density
considered in this manuscript.

The moments of the reduced density matrix are 
\be
\label{M-alpha-def}
\mathcal{M}_A^{(\alpha)} 
\equiv 
\textrm{Tr}\big( \rho_A^\alpha \big)
\,=\,
\textrm{exp}\big[ (1-\alpha) S_A^{(\alpha)}\big]\,.
\ee
The moments of $K_A$ can be obtained through the following analytic continuation 
\be
\label{p-moment}
\big\langle K_A^p \big\rangle_{A}
=
(-1)^p  \lim_{\alpha \to 1}
\partial_\alpha^p \,\mathcal{M}_A^{(\alpha)} \,.
\ee

The logarithm of the generating function $\mathcal{M}_A^{(\alpha)} $ 
in  (\ref{M-alpha-def}), i.e.  $(1-\alpha) S_A^{(\alpha)}$, 
provides the generating function of the 
cumulants of the entanglement spectrum
\be
\label{p-cumulant}
\widetilde{\mathcal{C}}_A^{\,(p)}
\equiv
\big\langle K_A^p \big\rangle_{_A,\,\textrm{\tiny con}}
=
(-1)^p \lim_{\alpha \to 1}\partial_\alpha^p \Big[(1-\alpha)\, S_A^{(\alpha)}\Big]
\ee
which are the connected correlators of $K_A$.
The entanglement entropy is the expectation value of $K_A$;
hence it corresponds to $p=1$, both in (\ref{p-moment}) and in (\ref{p-cumulant}).
The second cumulant, which is (\ref{p-cumulant}) for $p=2$ and gives the variance of $K_A$,
is called capacity of entanglement \cite{yao-10,Schliemann-11,DeBoer:2018kvc}.

\begin{figure}[t!]
\vspace{-.2cm}
\hspace{-.8cm}
\includegraphics[width=1.05\textwidth]{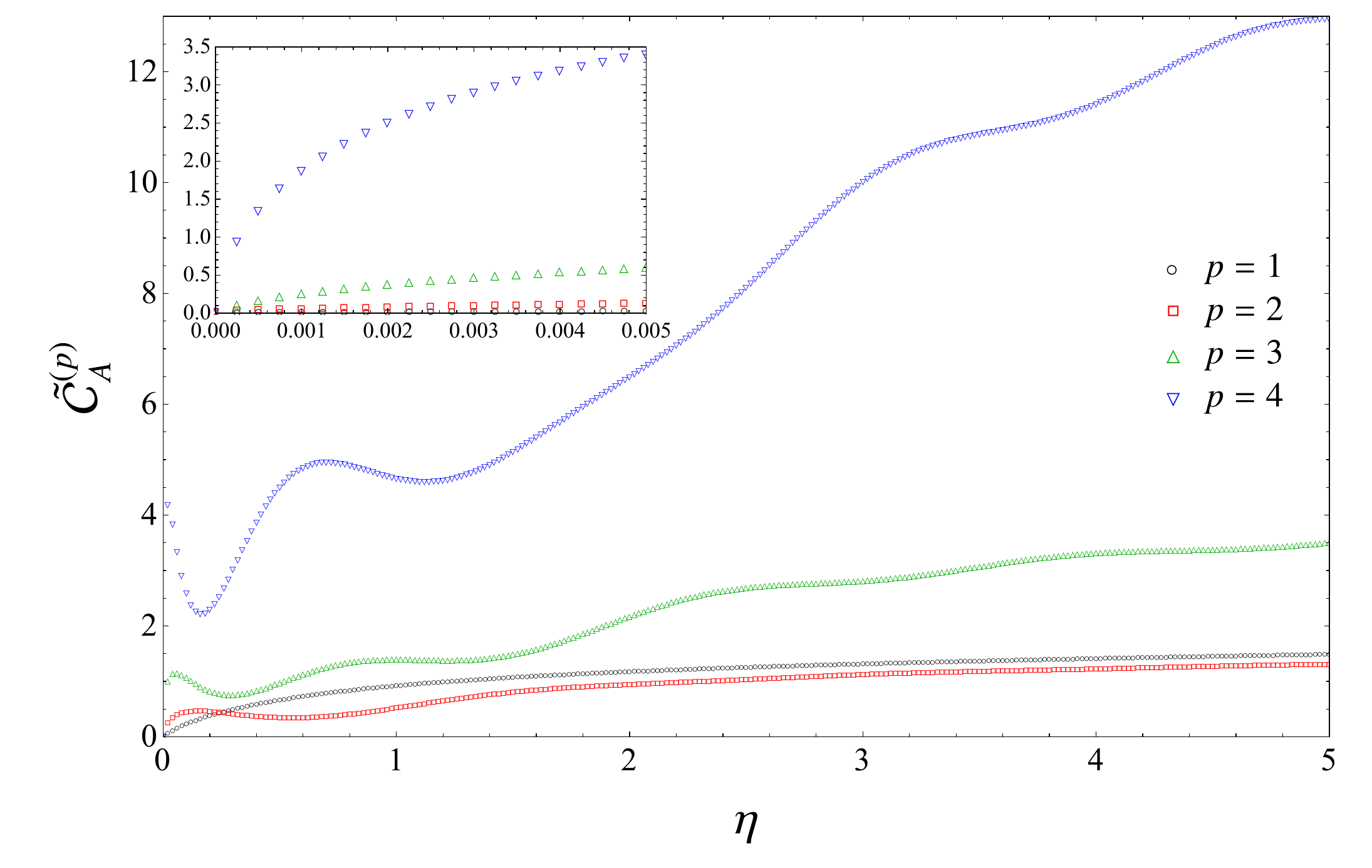}
\vspace{-.7cm}
\caption{
Cumulants $\widetilde{\mathcal{C}}_A^{\,(p)}$ for $p \in \{1,2,3,4\}$,
obtained numerically from  (\ref{capacity-def}).
For $p=1$ we have $\widetilde{\mathcal{C}}_A^{\,(1)}= S_A$.
The inset zooms in on $\eta \sim 0^+$
and shows that $\widetilde{\mathcal{C}}_A^{\,(p)}\to 0$ vanish as $\eta \to 0$.
}
\label{fig:app-cumulants}
\end{figure}

In the free fermionic Schr\"odinger field theory that we are exploring, 
$S_A^{(\alpha)}$ are (\ref{entropies-def-sums}) and (\ref{entropies_func});
hence the cumulants of the entanglement spectrum (\ref{p-cumulant}) read
\be
\label{capacity-def}
\widetilde{\mathcal{C}}_A^{\,(p)} = \sum_{n = 0}^{\infty} c_p(\gamma_n)
\;\;\qquad\;\;
c_p(x) \equiv
(-1)^p \lim_{\alpha \to 1}\partial_\alpha^p \, \Big\{\! \log\!\big[ x^\alpha +(1-x)^\alpha\big] \Big\}\,.
\ee
For the first integer values of $p$, we obtain
\bea
\label{app-c2-def}
c_2(x) 
&=&
(1-x)\, x\, \big[ \log(1-x) - \log(x)\big]^2
\\
\label{app-c3-def}
c_3(x) 
&=&
\big(1-3x+2x^2\big)\, x\, \big[ \log(1-x) - \log(x)\big]^3
\\
\label{app-c4-def}
c_4(x) 
&=&
\big(1-7x+12 x^2 - 6 x^3\big)\, x\, \big[ \log(1-x) - \log(x)\big]^4
\eea
which have the form
$c_p(x) = x\,P(x)\, \big[ \log(1-x) - \log(x)\big]^p$,
where  $P(x)$ is a polynomial of order $p-1$ such that $P(0) = 1$.

By adapting to (\ref{capacity-def})
the analysis described in Sec.\,\ref{sec_entropies} to prove the finiteness of $S_A^{(n)}$,
we can easily find that the series (\ref{capacity-def}) is convergent;
hence $\mathcal{C}_A^{(p)}$ are well defined functions of $\eta$.

The method of \cite{Jin_2004, Keating_04} described in 
Sec.\,\ref{sec_entropies} for the entanglement entropies can be also adapted 
to the cumulants (\ref{capacity-def}), finding
\be
\label{cumulants-tau}
\widetilde{\mathcal{C}}_A^{\,(p)}
= 
  \lim_{\epsilon, \delta \to 0}\,
   \frac{1}{2\pi \textrm{i}} 
  \oint_{\mathfrak{C}}
 c_p(z) \;
 \partial_z  \log( \tau )\, \textrm{d}z
\ee
where $\tau$ is the sine kernel tau function (\ref{tau-function-sine})
and $\mathfrak{C}$ is the closed path in the complex plane introduced in Sec.\,\ref{sec_entropies}.
Thus, by adapting the analyses performed in 
Sec.\,\ref{sec_small_distance} and Sec.\,\ref{sec_large_distance},
analytic expressions for the expansions of the cumulants
either as $\eta \to 0$ or as $\eta \to \infty$ can be found.

In Fig.\,\ref{fig:app-cumulants} we show the cumulants $\widetilde{\mathcal{C}}_A^{\,(p)}$ for $p \in \{1,2,3,4\}$,
obtained numerically from (\ref{capacity-def}), where the infinite sum is truncated to $n_{\textrm{\tiny max}}$
introduced in Sec.\,\ref{sec_entropies}.
While $\widetilde{\mathcal{C}}_A^{\,(1)} = S_A$ does not oscillate
(see Sec.\,\ref{subsec_ent_loss} and in Sec.\,\ref{subsec-ee-large-eta}),
$\widetilde{\mathcal{C}}_A^{\,(p)}$ with $p \in \{2,3,4\}$ display an oscillatory behaviour.

From Fig.\,\ref{fig:app-cumulants}  we also observe that $\widetilde{\mathcal{C}}_A^{\,(p)}\geqslant0$ 
and that $\widetilde{\mathcal{C}}_A^{\,(p)} \rightarrow 0 $ as $\eta \rightarrow 0 $ (see the inset).
The former property for $p=1,2,3$ follows from the fact that the corresponding $c_p(x)$
(see also (\ref{app-c2-def}) and (\ref{app-c3-def}))
are  positive when $x\in[0,1]$. 
The inequality $\widetilde{\mathcal{C}}_A^{\,(4)} \geqslant 0$ is less trivial because 
$c_4(x)$ is negative for a finite region around $x=1/2$ and positive otherwise. 
However, this property holds because
the number of the eigenvalues lying in such region is finite 
and the negative contribution that they provide to \eqref{capacity-def} 
is smaller than the positive contribution coming from the remaining eigenvalues lying the positive part of $c_4(x)$.
The limit $\widetilde{\mathcal{C}}_A^{\,(p)} \rightarrow 0 $ as $\eta \rightarrow 0 $ 
can be obtained by combining the fact that
$\gamma_n(\eta)\rightarrow 0 $ as $\eta \rightarrow 0 $ uniformly in $n\in \mathbb{N}$
and that $c_p(x)\rightarrow 0$ as $x \rightarrow 0 $.


\section{Finiteness of $C_\alpha$}
\label{app_finiteness-C-alpha}

In this appendix we show that the functions (\ref{c-function-alpha}) are well defined functions of $\eta$,
i.e. that the series (\ref{C-function-def}) are convergent.

Consider the integer value $n_c \in\mathbb{N}_{0}$ 
such that $\gamma_{n}\leqslant a$ for all $n>n_c$,
where $a$ is such that the function $x\mapsto x\log(1/x -1)$ 
is positive and increasing for $x \in [0,a]$. 
Here we choose $a=\tfrac{1}{5}$.
Then, one can split the infinite sum defining $C$ in (\ref{C-function-def}) 
into the finite sum over $n \in [0,n_c]$ and the remaining infinite sum.
Since the function $x\mapsto x\log(1/x -1)$ is positive and increasing when $x\in[0,\tfrac{1}{5}]$
and $f_n(\eta;1)^2<(n+1/2)$ for all $n \in\mathbb{N}_{0}$ \cite{Rokhlin-book},
by employing also (\ref{gamma-upper-bound}) 
we obtain the following upper bound
\be
\label{C-upper-bound}
C(\eta) 
\, \leqslant   \,
 \sum_{n=0}^{n_c}
 f_{n}(\eta;1)^{2} \, \gamma_{n}\log(1/\gamma_{n}-1)
 +
 \sum_{n>n_c} (n+1/2)\,\tilde{\gamma}_{n}\log(1/\tilde{\gamma}_{n} -1)
\ee
where $\tilde{\gamma}_{n}$ are defined in (\ref{gamma-tilde-def}).
The infinite sum over $n > n_c$ in the r.h.s. of (\ref{C-upper-bound}) is convergent;
indeed, by applying the ratio test, one obtains
\be
\lim_{n\rightarrow\infty}
\frac{\left(n+1+\frac{1}{2}\right)\tilde{\gamma}_{n+1}\log(1/\tilde{\gamma}_{n+1}-1)}{\left(n+\frac{1}{2}\right)\tilde{\gamma}_{n}\log(1/\tilde{\gamma}_{n}-1)}
\,=\,0\,.
\ee

Similarly, we can shown that $C_\alpha$ in (\ref{C-function-def}) is finite for any $\eta>0$.
First notice that
\be
\frac{2\alpha}{\alpha-1}\;
\frac{x\,\big[(1-x)^{\alpha-1}-x^{\alpha-1}\big]}{(1-x)^{\alpha}+x{}^{\alpha}}
\,=\,
\left\{\begin{array}{ll}
\displaystyle
\frac{2\alpha}{1-\alpha}\;x^{\alpha}+\dots
 \hspace{.7cm} & 0<\alpha<1
\\
\rule{0pt}{.8cm}
\displaystyle
\frac{2\alpha}{\alpha-1}\;x+\dots
 & \alpha>1 
\end{array}\right.
\hspace{1.5cm}
x \to 0^+\,.
\ee
By employing the limit comparison test for the convergence of series, 
the finiteness of $C_\alpha$ is guaranteed 
by the convergence of the following series
\bea
\label{ineq-c-alpha-1}
&& \frac{2\alpha}{1-\alpha}
\sum_{n=0}^{\infty} \gamma_{n}^{\alpha} \,f_{n}(\eta;1)^{2} 
\,\leqslant\,
\frac{2\alpha}{1-\alpha}\sum_{n=0}^{\infty} (n+1/2)\, 
\tilde{\gamma}_{n}^{\alpha}
\hspace{2cm}
0<\alpha<1
\\
\label{ineq-c-alpha-2}
\rule{0pt}{.8cm}
&& 
\frac{2\alpha}{\alpha-1}
\sum_{n=0}^{\infty} \gamma_{n} \,f_{n}(\eta;1)^{2}
\,\leqslant \,
\frac{2\alpha}{1-\alpha}\sum_{n=0}^{\infty}(n+1/2)\, \tilde{\gamma}_{n}
\hspace{2cm}
\alpha>1
\eea
where (\ref{gamma-upper-bound}) and the upper bound reported in Eq.\,(3.117)
of \cite{Rokhlin-book} have been used.
The infinite sums in the r.h.s.'s of (\ref{ineq-c-alpha-1}) and (\ref{ineq-c-alpha-2}) are convergent
because, by applying the ratio test, we have that
\be
\lim_{n\rightarrow\infty}
\frac{\tilde{\gamma}_{n+1}^{\alpha}\,(n+3/2)}{\tilde{\gamma}_{n}^{\alpha}\,(n+1/2)}
=0
\;\;\;\qquad\;\;\;
\lim_{n\rightarrow\infty}
\frac{\tilde{\gamma}_{n+1}\,(n+3/2)}{\tilde{\gamma}_{n}\,(n+1/2)}
=0\,.
\ee

The finiteness of $C_\infty$ can be shown by adapting the analysis discussed above.
Data points for $C_\infty$ are shown in Fig.\,\ref{fig:c-functions-alpha-small} and Fig.\,\ref{fig:c-20-large-100}.

\section{Modular Hamiltonian and modular flow for odd Lifshitz exponents}
\label{app_mod-ham-flow}

In Sec.\,\ref{sec_lifshitz_exponent} we have observed that 
the entanglement entropies of an interval 
for the Lifshitz fermion fields with odd Lifshitz exponents
are UV divergent and  independent of $k_{\textrm{\tiny F},z}$, 
which contains both the density and the Fermi momentum.
In this appendix we show that, instead,
the modular Hamiltonian \eqref{K_A-local-def}
and the corresponding modular flow of the field, defined in (\ref{mod-evolution-lambda}),
depend on $k_{\textrm{\tiny F},z}$.

The free fermionic Lifshitz models that we are considering are quadratic field theories; 
hence  the modular Hamiltonian $K_A$ of the interval $A=[-R,R] \subset \RR$ 
can be written in the following quadratic form
\be
\label{K_A lambda}
K_A \,= 
\int_A \int_A 
:\! \psi^* (0,x) \,
H_A(x,y)\, 
\psi(0,y)\!:
\rd x  \,\rd y 
\ee
where $:\cdots :$ denotes the normal product in the oscillator algebra ${\cal A}$ 
introduced in Sec.\,\ref{sec_model}.
The kernel $H_A(x,y)$ in (\ref{K_A lambda}) can be found through the Peschel's formula 
\cite{Peschel:2003rdm, EislerPeschel:2009review,Casini:2009sr}
\be
\label{eh-matrix-peschel}
H_A(x,y) \,=\, \log \! \big(C_A(x,y)^{-1} - 1\,\big) 
\;\;\qquad\;\;
x,y \in A
\ee
where $C_A$ is the kernel defined by the two point function (\ref{gLL}) restricted to $A$.

When the Lifshitz exponent $z$ is odd, 
the spectral problem is (\ref{Lif2}) and
from (\ref{Lif3}), its eigenfunctions  can be easily obtained 
from the eigenfunctions of the spectral problem (\ref{Lif4}),
discussed in \cite{Musk-book, Casini:2009vk}.
They are
\be 
\label{eigenfunction}
\phi_s(x) 
=
\e^{-\ri k_{\textrm{\tiny F},z} x} \,
\tilde{\phi}_s(x) 
\qquad\;\;
\tilde{\phi}_s(x) 
=
\sqrt{\frac{R}{\pi (R^2-x^2)}}\; \e^{\ri s w(x)}
\qquad\;\;
w(x) = \log\!\left( \frac{x+R}{R-x} \right) .
\ee

The Peschel's formula (\ref{eh-matrix-peschel}) tells us that the kernels $C_A$ and $H_A$ have the same eigenfunctions
and that the eigenvalues of $H_A$ are $2\pi s = \log(1/\gamma_s - 1) $ with $s\in \RR$,
where $\gamma_s$ are the eigenvalues of $C_A$, given in  (\ref{Lif5}).
Thus, the spectral representation of the kernel $H_A$ reads
\be
\label{H_A spectral rep}
H_A(x,y)
\,=\,
2\pi \int_{-\infty}^{+\infty} \!\! s \; \phi_s(x) \, \phi_{s}^\ast (y) \, \rd s 
\,=\,
2\pi \,\e^{-\ri k_{\textrm{\tiny F},z}(x-y)}
\int_{-\infty}^{+\infty} \!\! s \; \tilde{\phi}_s(x) \, \tilde{\phi}_{s}^\ast (y) \, \rd s 
\ee
where (\ref{Lif3}) has been employed. 
The integral in the last expression of (\ref{H_A spectral rep}) does not contain $k_{\textrm{\tiny F},z}$;
hence  $H_A(x,y)$ explicitly depends on $k_{\textrm{\tiny F},z}$.
Plugging the explicit expression of $\tilde{\phi}_s(x)$ (given in (\ref{eigenfunction})) 
into (\ref{H_A spectral rep}),
we find
\be
H_A(x,y)
\,=\,
-\, 2\pi  \textrm{i}\, 
\e^{-\ri k_{\textrm{\tiny F},z}(x-y)}\;
\frac{\sqrt{(R^2-x^2)(R^2-y^2)}}{4R}\;
\big(\partial_x - \partial_y\big) \delta(x-y)\,.
\ee
Finally, the modular Hamiltonian (\ref{K_A-local-def}) can be obtained
by inserting this kernel into (\ref{K_A lambda}).

The modular flow of the field has been defined in (\ref{mod-evolution-lambda}).
It is the solution the following partial differential equation
\be
\label{partial differential equation-mod-evolution}
\textrm{i}\,\frac{ d\psi(\tau,x)}{d\tau}  \,=\, \big[\,K_A \,, \psi(\tau,x)\,\big]_{-}
\;\;\;\;\; \qquad \;\;\;\;\;
x\in A
\ee
where $K_A$ is (\ref{K_A-local-def}), for a given initial field configuration $\psi(x)$ when $\tau=0$.
The explicit form of (\ref{partial differential equation-mod-evolution}) reads
\be
\label{pde-mod-flow-1}
\textrm{i}\,\frac{ d\psi(\tau,x)}{d\tau}  
\,=\,
2\pi \, \textrm{i}\; 
\left(  \beta_{\textrm{\tiny loc}}(x)\, \partial_x + \frac{1}{2}\, \partial_x \beta_{\textrm{\tiny loc}}(x) \right)
 \psi(\tau,x)
 -
  2\pi \, k_{\textrm{\tiny F},z}\, \beta_{\textrm{\tiny loc}}(x) \, \psi(\tau,x) \,.
  \ee

This partial differential equation  has the following structure
\be
\label{pde-gen-AB}
\partial_{\mathsf{t}} \psi(\mathsf{t},x)
=
V(x) \, \partial_x \psi(\mathsf{t},x) + Y(x)\, \psi(\mathsf{t},x) 
\;\;\qquad\;\;
\psi(0,x) = \psi(x)
\ee
where $\psi(x)$ corresponds to the initial configuration of the field.
The solution of (\ref{pde-gen-AB}) has been discussed e.g. in the Appendix\,B of \cite{Mintchev:2020uom}
and it can be written as 
\be
\label{gen-solution-mod-flow}
\psi(\mathsf{t},x)  
\,=\,
e^{\Phi(\mathsf{t},x)}\,
\psi \big(\xi(\mathsf{t},x)\big)
\;\;\;\qquad\;\;\;
\Phi(\mathsf{t},x) \equiv \int_{x}^{\xi(\mathsf{t},x)} \frac{Y(y)}{V(y)}\, \rd y
\ee
where $\xi(\mathsf{t},x)$ is defined as follows
\be
\label{w-prime-def}
w'(x) = \frac{1}{V(x)}
\;\;\qquad\;\;
\xi(\mathsf{t},x)\,\equiv\, w^{-1}\big( w(x)+ \mathsf{t} \big)
\ee
which satisfies $\xi(0,x) = x$.

The partial differential equation (\ref{pde-mod-flow-1}) belongs to the class of partial differential equations
defined by (\ref{pde-gen-AB}); indeed,  it corresponds to
\be
\label{our-case-V-Y}
\mathsf{t} = 2\pi \, \tau
\;\;\;\qquad\;\;\;
V(x) \equiv  \beta_{\textrm{\tiny loc}}(x)
\;\;\;\qquad\;\;\;
Y(x) \equiv \frac{1}{2} \, \partial_x\beta_{\textrm{\tiny loc}}(x) - \ri \, k_{\textrm{\tiny F},z}\,\beta_{\textrm{\tiny loc}}(x) \,.
\ee
In this special case, the functions and $ \xi(\tau,x) $ and $\Phi(\tau,x)$, 
defined respectively in (\ref{w-prime-def}) and (\ref{gen-solution-mod-flow}),
become respectively (\ref{xi-interval-line-gs}) and 
\be
\label{Gamma-function-explicit-app}
\Phi(\tau,x)
=
\frac{1}{2}\, \log\!
\left( \frac{\beta_{\textrm{\tiny loc}}(\xi(\tau,x))}{\beta_{\textrm{\tiny loc}}(x)} \right)
-
\ri \, k_{\textrm{\tiny F},z} \big[ \,\xi(\tau,x) - x \,\big] \,.
\ee
Finally, the expression (\ref{mod-flow-mu-main}) for the modular flow of the field 
is (\ref{gen-solution-mod-flow}) specialised 
to the case given by (\ref{xi-interval-line-gs}), (\ref{our-case-V-Y}) and (\ref{Gamma-function-explicit-app}).

From (\ref{pde-mod-flow-1}), it is straightforward to observe that the correlator along the modular flow
$\mathcal{W}_{1,2} \equiv \mathcal{W}(\tau_1,x_1; \tau_2,x_2) 
\equiv \langle \psi^\ast(\tau_1,x_1)  \, \psi(\tau_2,x_2)  \rangle_{\infty, \mu} $ in
(\ref{mod-corr-1})
satisfies the following 
\be
\label{mod-eom-x12}
\textrm{i}\,\frac{d \,\mathcal{W}_{1,2} }{d\tau_2}  
=
\left[\,
2\pi \, \textrm{i}
\left(  \beta_{\textrm{\tiny loc}}(x_2)\, \partial_{x_2} + \frac{1}{2}\, \partial_{x_2} \beta_{\textrm{\tiny loc}}(x_2) \right)
 -
  2\pi \, k_{\textrm{\tiny F},z}\, \beta_{\textrm{\tiny loc}}(x_2) \,
  \right]
\mathcal{W}_{1,2} 
  \ee
  which can be interpreted as a modular equation of motion \cite{Mintchev:2020uom}.


\section{On the small $\eta$ expansion}
\label{app_small_distance}

In this appendix we derive the 
analytic expressions for the expansion of the entanglement entropies as $\eta \to 0$.
They are reported in Sec.\,\ref{sec_small_distance_P5}
and employed in various figures.

\subsection{Sine kernel tau function from the Painlev\'e V tau function}
\label{app_tau_small_eta_GIL}

We find it instructive to report the derivation of (\ref{tau-small-distance-expansion})
as a special case of the expansion of the solution of the general Painlev\'e V
found in \cite{Gamayun:2013auu, Lisovyy:2018mnj}.
%
In the following, a given quantity 
indicated through a certain notation in \cite{Gamayun:2013auu, Lisovyy:2018mnj}
is denoted by the same symbol with the subindex $\star$.

The expansion of the tau function for the general Painlev\'e V as $t_\star \to 0$
is given in Eq.\,(1.11a) of \cite{Lisovyy:2018mnj}\footnote{By setting 
$\mathcal{N}_0 = 1$ and $s_{\textrm{\tiny V}} = \e^{2\pi \textrm{i} \eta_\star}$
in Eq.\,(1.11a) of \cite{Lisovyy:2018mnj}, one obtains 
Eq.\,(4.14) of  \cite{Gamayun:2013auu}.}.
In the notation of  \cite{Gamayun:2013auu, Lisovyy:2018mnj},
the sine kernel tau function corresponds to the special case
characterised by $\theta_0 = \theta_t = \theta_\ast = 0$ 
and $\sigma_{\!\star} \to 0$.
When $\theta_0 = \theta_t = \theta_\ast = 0$,
the general expansion found in \cite{Gamayun:2013auu, Lisovyy:2018mnj}
simplifies to 
\be
\label{expansion-tau-small-eta-general}
\tau
\,=\,
\mathcal{N}_0\,
\sum_{n \in \mathbb{Z}} 
\e^{2\pi \textrm{i} n \eta_\star}\,
\mathcal{C}_0(\sigma_{\!\star} + n)\, 
\mathcal{B}_0(\sigma_{\!\star} + n; t_\star)
\ee
where 
\be
\label{C0-B0-def-LNR}
\mathcal{C}_0(\sigma_{\!\star}) 
\equiv
\frac{G(1+ \sigma_{\!\star})^3 \, G(1 -\sigma_{\!\star})^3}{G(1+ 2\sigma_{\!\star}) \, G(1 -2\sigma_{\!\star})}
\;\;\;\qquad\;\;
\mathcal{B}_0(\sigma_{\!\star}; t_\star)
\equiv
t_\star^{\sigma_{\!\star}^2} \,\widetilde{\mathcal{B}}_0(\sigma_{\!\star}; t_\star)
\ee
with $\widetilde{\mathcal{B}}_0(\sigma_{\!\star}; t_\star)$ 
being the summation over $\mathbb{Y}$ in Eq.\,(1.6) of \cite{Lisovyy:2018mnj}
specialised to $\theta_0 = \theta_t = \theta_\ast = 0$.
To obtain the sine kernel tau function, 
for the parameters $\sigma_{\!\star}$ and $\eta_\star$ in (\ref{expansion-tau-small-eta-general}) 
one imposes \cite{Jimbo-82}
\be
\label{sigma-eta-P5-parameters}
\sigma_{\!\star} \to 0
\;\;\;\;\qquad\;\;\;\;
\e^{2\pi \textrm{i} \eta_\star} 
= 
\frac{4\textrm{i} \, \xi_\star}{\sigma_{\!\star}}\,.
\ee

We remark that
the limit $\sigma_{\!\star} \to 0$ of (\ref{expansion-tau-small-eta-general})
is not straightforward because
$G(m) = 0$ for $m \in \mathbb{Z}$ and $m \leqslant 0$.
%
From (\ref{sigma-eta-P5-parameters}) and
the asymptotic behaviour of the Barnes $G$ function (see e.g. Eq.\,(A.3) of \cite{Gamayun:2013auu}),
we find
\be
\label{lim-sigma0-513-GIL}
\lim_{\sigma_{\!\star} \to 0} 
\e^{2\pi \textrm{i} n \eta_\star}\,
\mathcal{C}_0(\sigma_{\!\star} + n)
=
\left\{
\begin{array}{ll}
0 
&
n < 0
\\
\rule{0pt}{.8cm}
\displaystyle
\frac{G(1+n)^6}{G(1+2n)^2} \; \frac{(- \,\xi_\star)^n}{\textrm{i}^{n^2}}
\hspace{.8cm}
&
n \geqslant 0\,.
\end{array}
\right.
\ee
In order to obtain the tau function for the sine kernel, 
we have to evaluate (\ref{expansion-tau-small-eta-general}) in $\textrm{i} t_\star$ 
(see Eq.\,(5.10) of \cite{Gamayun:2013auu}) and in the limit $\sigma_{\!\star} \to 0$.
The function $\mathcal{B}_0(n; \textrm{i} t_\star)$ provides 
both the factor that cancels $\textrm{i}^{n^2}$ in the denominator of (\ref{lim-sigma0-513-GIL}) 
and $\widetilde{\mathcal{B}}_0(n; \textrm{i} t_\star)$.
In the Appendix\;B of \cite{Gamayun:2013auu}
the function $\widetilde{\mathcal{B}}_0(n; \textrm{i} t_\star)$
is called $\mathcal{B}_{\textrm{\tiny sine}}(n; t_\star)$
and its expansions as $t_\star \to 0$ are reported,
for various values of $n$
(see also (\ref{B0-B1-expansion})-(\ref{B4-expansion}), 
where only the terms employed in our analysis are shown).
By using 
(\ref{lim-sigma0-513-GIL}),
we can take the limit $\sigma_{\!\star} \to 0$ of (\ref{expansion-tau-small-eta-general}) with $\mathcal{N}_0 = 1$.
Finally, we obtain (\ref{tau-small-distance-expansion}) 
by introducing the following change of notation
\be
\label{GIL-to-our-notation}
t_{\star} = 4\,\eta = \mathsf{a}
\;\;\;\qquad\;\;\;
 \xi_{\star} =  \frac{1}{2\pi\,z}
\ee
and by setting $\widetilde{\mathcal{B}}_0(n; \textrm{i} t_\star) \equiv \mathcal{B}_n(\eta)$.


\subsection{Approximate entanglement entropies}
\label{app_small_eta_entropies}

In Sec.\,\ref{sec_small_distance_P5} 
the approximate expression (\ref{approx-entropies-small-eta}) for the entanglement entropies 
as $\eta \to 0$ has been obtained from the small distance expansion 
of the sine kernel tau function given in (\ref{tau-small-distance-expansion}).
In the following we derive these approximate expressions 
up to $O(\eta^{\mathcal{N}})$ included, for $\mathcal{N} \leqslant  24$.
In order to consider terms up to $O(\eta^{\mathcal{N}})$ included
in the expansion of $S_A^{(\alpha)}$ as $\eta \to 0$,
we first write the finite sum $\tilde{\tau}_{\mathcal{N},N}= P_{N,\mathcal{N}}(z) /z^N$
obtained by truncating (\ref{tau-small-distance-expansion}),
where $P_{N, \mathcal{N}}(z)$ is a monic polynomial of order $N$.
Then, the zeros of $P_{N,\mathcal{N}}(z)$ are needed  in (\ref{approx-entropies-small-eta})
and they can be computed analytically through radicals only for $N \leqslant 4$;
hence we consider only $N\leqslant 4$.

To write $\tilde{\tau}_{\mathcal{N},N}$, 
the expansions of the functions $ \mathcal{B}_n(\eta)$ in (\ref{tau-small-distance-expansion}) 
as $\eta \to 0$ must be taken into account up to the proper order. 
In the Appendix\;B of \cite{Gamayun:2013auu},
the expansions of $ \mathcal{B}_n(\eta)$ 
up to a certain order in $\eta$ for any $n \leqslant 5$ have been obtained. 
In the following,
by using also the first expression of (\ref{GIL-to-our-notation}),
we report these expansions truncated to the order employed in this manuscript, 
where $N\leqslant 4$.
We use that
\be
\label{B0-B1-expansion}
\mathcal{B}_0(\eta)=\mathcal{B}_1(\eta)=1
\ee 
identically and also the following expansions
\bea
\label{B2-expansion}
\mathcal{B}_2(\eta)
&=&
1  - \frac{4^2}{75}\, \eta^2 + \frac{4^4}{7840}\, \eta^4
- \frac{4^6}{1134000}\, \eta^6 
+ \frac{4^8}{219542400}\, \eta^8
- \frac{4^{10}}{55091836800}\, \eta^{10}
\\
\rule{0pt}{.7cm}
& &
+\, \frac{4^{12}}{17435658240000}\, \eta^{12} 
- \frac{4^{14}}{6802522062336000}\, \eta^{14} 
+ \frac{4^{16}}{3210079038566400000}\, \eta^{16} 
\nonumber
\\
\rule{0pt}{.7cm}
& &
- \,\frac{4^{18}}{1803084500809912320000}\, \eta^{18} 
+ \frac{4^{20}}{1189192769988708925440000}\, \eta^{20} 
+ O\big(\eta^{22}\big)
\nonumber
\\
\rule{0pt}{1.1cm}
\label{B3-expansion}
\rule{0pt}{1.1cm}
\mathcal{B}_3(\eta)
&=&
1  - \frac{18 \cdot 4^2}{1225}\,\eta^2 
+ \frac{4^4}{8820} \,\eta^4
- \frac{2293\cdot 4^6}{3922033500}\,\eta^6
+ \frac{3581\cdot 4^8}{1616027212800}\,\eta^8
\\
\rule{0pt}{.7cm}
& &
- \,\frac{71\cdot 4^{10}}{10908183686400}\,\eta^{10}
+ \frac{94789\cdot 4^{12}}{6178831567324416000}\,\eta^{12}
\nonumber
\\
\rule{0pt}{.7cm}
& &
- \,\frac{76477 \cdot 4^{14}}{2570452778021883955200} \,\eta^{14}
+ O\big(\eta^{16}\big)
\nonumber
\eea
\bea
\label{B4-expansion}
\rule{0pt}{1.1cm}
\mathcal{B}_4(\eta)
&=&
1 
- \frac{20\cdot 4^2}{1323} \,\eta^2 
+ \frac{83 \cdot 4^4}{711480}\,\eta^4
- \frac{174931  \cdot 4^6}{286339821768}\,\eta^6
+ \frac{9605 \cdot 4^8}{3926946127104} \,\eta^8
+ O\big(\eta^{10}\big)
\nonumber
\\
& &
\eea
where we have highlighted the possibility to express them in terms of the area $\mathsf{a}$ 
of the limited phase space (see (\ref{a-parameters-def})).

The simplest approximate expression is (\ref{approx-entropies-small-eta}) with $N=1$,
and it has been discussed in Sec.\,\ref{sec_small_distance_P5},
by employing (\ref{B0-B1-expansion}).
In the following we derive the improved approximate expressions,
which correspond to (\ref{approx-entropies-small-eta}) with $2\leqslant N \leqslant 4$.

\subsubsection{$N=2$}

When $\mathcal{N} \geqslant 4$, at least $N=2$ is needed.
The simplest case corresponds to $\mathcal{N} = 4$ and $N=2$, where 
from (\ref{tau-small-distance-expansion}), we find
\be
\tilde{\tau}_{2,4} 
\,=\, 
1 - a_1 \,\frac{\eta}{z} + b_4 \,\frac{\eta^4}{z^2} 
+ O(\eta^6)
\,\equiv\,
\frac{P_{2,4}(z)}{z^2} + O(\eta^6)
\ee
being $P_{2,4}(z)$ the following monic polynomial
\be
\label{P-2-4-def}
P_{2,4}(z)  \,\equiv\,
z^3 - a_1 \eta\, z^2 + b_4 \eta^4
\ee
with $a_1$ and $b_4$ given by (here $G(5) = 12$ is needed)
\be
\label{a1-b4-def}
a_1 = \frac{2}{\pi}
\;\;\;\qquad\;\;\;
b_4 = \frac{1}{G(5)^2}\,\frac{4^4}{(2\pi)^2} \,.
\ee

We find it convenient to adopt the notation $z_{N, \mathcal{N}, j}$ for the zeros of $P_{N, \mathcal{N}}(z)$, 
which are denoted simply by $z_j$ in (\ref{approx-entropies-small-eta}).
The zeros of (\ref{P-2-4-def}) read
\be
z_{2,4,\pm} \equiv
\frac{\eta }{2} \left( a_1  \pm \sqrt{a_1^2 - 4\,b_4\, \eta^2}\,\right)
\ee
hence in this case  the approximate expression (\ref{approx-entropies-small-eta}) for the entanglement entropies becomes 
\be
\widetilde{S}_{A;2,4}^{(\alpha)} 
= 
s_\alpha(z_{2,4,+}) + s_\alpha(z_{2,4,-}) 
\ee
which has been employed to obtain the curves
corresponding to $\mathcal{N}=4$ in Fig.\,\ref{fig:small-eta-panels}.

When $N=2$, the highest order that can be considered is $\mathcal{N} = 8$.
For $N=2$ and $\mathcal{N} = 8$, from (\ref{tau-small-distance-expansion}) we obtain 
\be
\label{tau-8-approx}
\tilde{\tau}_{2,4} 
\,=\, 
1 - a_1 \,\frac{\eta}{z} 
+ b_4 \,\frac{\eta^4}{z^2}
- b_6 \,\frac{\eta^6}{z^2}
+ b_8 \,\frac{\eta^8}{z^2}
+O(\eta^9)
\,\equiv\,
\frac{P_{2,8}(z)}{z^2} + O(\eta^9)
\ee
where, 
by using also (\ref{B2-expansion}) up to $O(\eta^4)$ included, beside (\ref{a1-b4-def}) we also have 
\be
\label{b6-b8-def}
b_6 = \frac{1}{G(5)^2}\,\frac{4^4}{(2\pi)^2}\;\frac{4^2}{75}
\;\;\;\qquad\;\;\;
b_8 = \frac{1}{G(5)^2}\,\frac{4^4}{(2\pi)^2}\;\frac{4^4}{7840}\,.
\ee
The zeros of the monic polynomial $P_{2,8}(z)$ introduced in (\ref{tau-8-approx}) read
\be
z_{2,8,\pm} \equiv
\frac{\eta }{2} \left( a_1  \pm 
\sqrt{a_1^2 - 4\, \eta^2 \big( b_4 - b_6 \,\eta^2+ b_8 \,\eta^4 \big)}\;\right)
\ee
which are employed in (\ref{approx-entropies-small-eta}) to obtain the following approximate expression
\be
\widetilde{S}_{A;2,8}^{(\alpha)} 
= 
s_\alpha(z_{2,8,+}) + s_\alpha(z_{2,8,-}) 
\ee
which has been used to find the curves
corresponding to $\mathcal{N}=8$ in Fig.\,\ref{fig:small-eta-panels}.

\subsubsection{$N=3$}

Improved approximate expressions
for the entanglement entropies can be found 
by increasing $N$ in (\ref{approx-entropies-small-eta}).
However, the explicit analytic expressions for the zeros of $P_{N, \mathcal{N}}(z)$ 
become more complicated as $N$ increases.

For $N=3$, we have $\mathcal{N} \leqslant 15$.
Considering the optimal case $\mathcal{N}=15$, 
from (\ref{tau-small-distance-expansion}) we have\footnote{The expansion (\ref{tau-15-approx}) includes more terms with respect to the one reported in Eq.\,(8.114) of \cite{forrester-book}.}
\bea
\label{tau-15-approx}
\tilde{\tau}_{3,15} 
&=&
1 - a_1 \,\frac{\eta}{z} 
+ 
b_4 \,\frac{\eta^4}{z^2}
- b_6 \,\frac{\eta^6}{z^2}
+ b_8 \,\frac{\eta^8}{z^2}
- 
b_{10} \,\frac{\eta^{10}}{z^2}
+ b_{12} \,\frac{\eta^{12}}{z^2}
- b_{14} \,\frac{\eta^{14}}{z^2}
\\
\rule{0pt}{.7cm}
& &\hspace{1.5cm}
-\,
c_9 \,\frac{\eta^9}{z^3}
+ c_{11} \,\frac{\eta^{11}}{z^3}
- c_{13} \,\frac{\eta^{13}}{z^3}
+ c_{15} \,\frac{\eta^{15}}{z^3}
+ O(\eta^{16})
\,\equiv\,
\frac{P_{3,15}(z)}{z^3} + O(\eta^{16})
\nonumber
\eea
with the polynomial $P_{3,15}(z)$ being defined as follows
\be
\label{poly-3-15-def}
P_{3,15}(z)
\,\equiv\,
z^3 - a\, z^2 + b\, z - c
\ee
where
\bea
a 
&\equiv &
a_1\, \eta
\\
b 
&\equiv &
b_4 \, \eta^{4} - b_6 \, \eta^{6} + b_8 \, \eta^{8} 
- b_{10} \, \eta^{10} + b_{12} \, \eta^{12} - b_{14} \, \eta^{14}
\\
c
&\equiv &
 c_{9} \, \eta^{9} -  c_{11} \, \eta^{11} +  c_{13} \, \eta^{13} -  c_{15} \, \eta^{15}   \,.
\eea
Some coefficients have been already introduced in (\ref{a1-b4-def}) and (\ref{b6-b8-def}).
The remaining ones can be written by using that $G(4) = 2$, $G(7) = 34560$,
(\ref{B2-expansion}) up to $O(\eta^{10})$ included
and (\ref{B3-expansion}) up to $O(\eta^6)$ included,
finding 
\bea
\label{b10-b12-def}
& &
b_{10} = \frac{1}{G(5)^2}\,\frac{4^4}{(2\pi)^2}\;\frac{4^6}{1134000}
\qquad
b_{12} = \frac{1}{G(5)^2}\,\frac{4^4}{(2\pi)^2}\;\frac{4^8}{219542400}
\\
\rule{0pt}{.7cm}
\label{b14-def}
& &
b_{14} = \frac{1}{G(5)^2}\,\frac{4^4}{(2\pi)^2}\;\frac{4^{10}}{55091836800}
\eea
and
\bea
\label{c9-c15-def}
& & 
c_9 = \frac{2^6}{G(7)^2}\,\frac{4^9}{(2\pi)^3}
\qquad
c_{11} = \frac{2^6}{G(7)^2}\,\frac{4^9}{(2\pi)^3}\;\frac{18}{1225}\, 4^2
\\
\rule{0pt}{.7cm}
& &
c_{13} = \frac{2^6}{G(7)^2}\,\frac{4^9}{(2\pi)^3}\;\frac{4^4}{8820}
\qquad
c_{15} = \frac{2^6}{G(7)^2}\,\frac{4^9}{(2\pi)^3}\;\frac{2293}{3922033500}\, 4^6  \,.
\nonumber
\eea

The  roots $\{z_{3,15,j} ; j=1,2,3\}$ of the cubic polynomial $P_{3,15}(z)$ defined in (\ref{poly-3-15-def}) are
\bea
z_{3,15,1}
&=&
\frac{1}{3}  \left(
a+ \frac{a^2 - 3b}{\Delta^{1/3}} + \Delta^{1/3}
\right)
\\
\rule{0pt}{.7cm}
z_{3,15,2}
&=&
\frac{1}{3}  \left(
a+ e^{2\textrm{i} \pi/3}\,\frac{a^2 - 3b}{\Delta^{1/3}} + e^{-2\textrm{i} \pi/3}\,\Delta^{1/3}
\right)
\\
\rule{0pt}{.7cm}
z_{3,15,3}
&=&
\frac{1}{3}  \left(
a+ e^{-2\textrm{i} \pi/3}\,\frac{a^2 - 3b}{\Delta^{1/3}} + e^{2\textrm{i} \pi/3}\,\Delta^{1/3}
\right)
\eea
where
\be
\Delta \,\equiv \,
a^3 - \frac{9}{2}\, ab+\frac{27}{2}\, c+ \textrm{i}\,\frac{3\sqrt{3\, \Delta_3}}{2}
\;\;\qquad\;\;
\Delta_3 \equiv  18\, a b c - 4\, a^3 c + a^2 b^2 - 4\, b^3 - 27\, c^2  \,.
\ee

The expression $\Delta_3$ is the discriminant of the cubic equation,
whose three roots are real and distinct whenever $\Delta_3 >0$.
In our case $\Delta_3$ is a polynomial in $\eta$ of high order,
hence its positivity can be checked numerically.

Specialising (\ref{approx-entropies-small-eta}) to this case, 
one obtains the best approximation for the entanglement entropies when $N=3$,
namely
\be
\label{approx-small-eta-N3}
\widetilde{S}_{A;3,15}^{(\alpha)} 
= 
\sum_{j=1}^3 s_\alpha(z_{3,15,j}) 
\ee
which provides  the curves
corresponding to $\mathcal{N}=15$ in Fig.\,\ref{fig:small-eta-panels}
as discussed in Sec.\,\ref{sec_small_distance_P5}.

\subsubsection{$N=4$}

The approximations corresponding to $N=4$ have $16 \leqslant   \mathcal{N} \leqslant  24$.
Also in these cases we can obtain  analytic expressions 
for the approximate entanglement entropies (\ref{approx-entropies-small-eta}).
Instead, for higher order approximations with $\mathcal{N} \geqslant  25$,
where $N \geqslant 5$ is required, 
the zeros of $P_{N, \mathcal{N}}(z)$ cannot be found through analytic expressions involving radicals
(Abel-Ruffini theorem).
Thus, the best approximation accessible  through analytic expressions 
corresponds to $\mathcal{N} =  24$, which requires $N=4$.
In this case, by truncating (\ref{tau-small-distance-expansion}) to these orders we find 
\be
\label{tau-24-approx}
\tilde{\tau}_{4,24} 
\,\equiv\,
1 - \frac{a}{z} + \frac{b}{z^2} - \frac{c}{z^3} + \frac{d}{z^4} + O(\eta^{25})
\,\equiv\,
\frac{P_{4,24}(z)}{z^4} + O(\eta^{25})
\ee
where
\be
\label{poly-4-24-def}
P_{4,24}(z)
\,\equiv\,
z^4 - a\, z^3 + b\, z^2 - c\, z + d
\ee
with
\bea
a 
&\equiv &
a_1\, \eta
\\
\rule{0pt}{.6cm}
b 
&\equiv &
b_4 \, \eta^{4} - b_6 \, \eta^{6} + b_8 \, \eta^{8} 
- b_{10} \, \eta^{10} + b_{12} \, \eta^{12} - b_{14} \, \eta^{14}
\nonumber
\\
& &
+\, b_{16} \, \eta^{16} - b_{18} \, \eta^{18} + b_{20} \, \eta^{20}
- b_{22} \, \eta^{22} + b_{24} \, \eta^{24} 
\eea
\bea
c
&\equiv &
 c_{9} \, \eta^{9} -  c_{11} \, \eta^{11} +  c_{13} \, \eta^{13} -  c_{15} \, \eta^{15} 
 \nonumber
 \\
& &
+\,  c_{17} \, \eta^{17} -  c_{19} \, \eta^{19} +  c_{21} \, \eta^{21} -  c_{23} \, \eta^{23}  
\\
\rule{0pt}{.7cm}
d
&\equiv &
 d_{16} \, \eta^{16} -  d_{18} \, \eta^{18} +  d_{20} \, \eta^{20} -  d_{22} \, \eta^{22} + d_{24} \, \eta^{24} 
\eea
whose coefficients can be found by employing 
all the terms in the expansions (\ref{B0-B1-expansion})-(\ref{B4-expansion}). 
Beside the coefficients already defined in (\ref{a1-b4-def}), (\ref{b6-b8-def}), 
(\ref{b10-b12-def}), (\ref{b14-def}) and  (\ref{c9-c15-def}), 
we also have to employ
\bea
& & \hspace{-.4cm}
b_{16} = \frac{1}{G(5)^2}\,\frac{4^4}{(2\pi)^2}\;\frac{4^{12}}{17435658240000}
\qquad
b_{18} = \frac{1}{G(5)^2}\,\frac{4^4}{(2\pi)^2}\;\frac{4^{14}}{6802522062336000}
\nonumber
\\
\rule{0pt}{.7cm}
& & \hspace{-.4cm}
b_{20} = \frac{1}{G(5)^2}\,\frac{4^4}{(2\pi)^2}\;\frac{4^{16}}{3210079038566400000}
\qquad
b_{22} = \frac{1}{G(5)^2}\,\frac{4^4}{(2\pi)^2}\;\frac{4^{18}}{1803084500809912320000}
\nonumber
\\
\rule{0pt}{.7cm}
& & \hspace{-.4cm}
b_{24} = \frac{1}{G(5)^2}\,\frac{4^4}{(2\pi)^2}\;\frac{4^{20}}{1189192769988708925440000}
\eea
and
\bea
& & 
c_{17} =  \frac{G(4)^6}{G(7)^2}\,\frac{4^9}{(2\pi)^3}\;\frac{3581}{1616027212800}\, 4^8
\qquad
c_{19} =  \frac{G(4)^6}{G(7)^2}\,\frac{4^9}{(2\pi)^3}\;\frac{71}{10908183686400}\, 4^{10}
\nonumber
\\
\rule{0pt}{.7cm}
& &
c_{21} =  \frac{G(4)^6}{G(7)^2}\,\frac{4^9}{(2\pi)^3}\;\frac{94789}{6178831567324416000}\, 4^{12}
\nonumber
\\
\rule{0pt}{.7cm}
& &
c_{23} =  \frac{G(4)^6}{G(7)^2}\,\frac{4^9}{(2\pi)^3}\;\frac{76477}{2570452778021883955200}\, 4^{14}
\eea
and 
\bea
& & 
d_{16} =  \frac{G(5)^6}{G(9)^2}\,\frac{4^{16}}{(2\pi)^4}
\qquad
d_{18}  = \frac{G(5)^6}{G(9)^2}\,\frac{4^{16}}{(2\pi)^4}\; \frac{20}{1323}\,4^2
\nonumber
\\
\rule{0pt}{.7cm}
& &
d_{20} =  \frac{G(5)^6}{G(9)^2}\,\frac{4^{16}}{(2\pi)^4}\; \frac{83}{711480}\, 4^4
\qquad
d_{22}  = \frac{G(5)^6}{G(9)^2}\,\frac{4^{16}}{(2\pi)^4}\; \frac{174931}{286339821768}\, 4^6
\nonumber
\\
\rule{0pt}{.7cm}
& &
d_{24}  = \frac{G(5)^6}{G(9)^2}\,\frac{4^{16}}{(2\pi)^4}\; \frac{9605}{3926946127104}\, 4^8  \,.
\eea

In order to find the  roots $\{z_{4,24,j} ; j=1,2,3,4\}$ of the quartic polynomial $P_{4,24}(z)$  in (\ref{poly-4-24-def}),
one first introduces
\be
\alpha  \,\equiv\,
-\frac{3a^2}{8} + b
\qquad
\beta \,\equiv\,
-\frac{a^3}{8} + \frac{ab}{2} - c
\qquad
\gamma \,\equiv\,
-\frac{3a^4}{256} + \frac{a^2 b}{16} - \frac{ac}{4} + d
\ee
and finds a solution $y$ of the following cubic equation
\be
y^3 + \frac{5\alpha}{2}\, y^2 
+\big( 2\alpha^2 - \gamma \big) y
+\frac{1}{2} \left(\alpha^3 - \alpha \gamma -\frac{\beta^2}{4} \,\right) 
=\, 0  \,.
\ee
Then, the four roots of the quartic polynomial $P_{4,24}(z)$ are 
\be
\label{zeros-4-24-j}
z_{4,24,j}
\,=\,
\frac{a}{4}
+\frac{1}{2}
\Bigg(
\eta_1 \, \sqrt{\alpha + 2 y} 
+ \textrm{i} \,\eta_2\, 
\sqrt{ 3 \alpha + 2 y +\eta_2\, \frac{2 \beta}{\sqrt{\alpha + 2 y} } }
\; \Bigg)
\;\qquad\;
\eta_1\, , \eta_2 \in \big\{ 1, -1\big\}  \,.
\;\;\;
\ee

The optimal approximation for the entanglement entropies when $N=4$, 
which corresponds to $\mathcal{N} = 24$,
is obtained by employing (\ref{zeros-4-24-j}) into (\ref{approx-entropies-small-eta}) specialised to $N=4$.
This gives the best approximation that we consider when $\eta \to 0$
(as discussed in Sec.\,\ref{sec_small_distance_P5}), namely
\be
\label{approx-small-eta-N4}
\widetilde{S}_{A;4,24}^{(\alpha)} 
= 
\sum_{j=1}^4 s_\alpha(z_{4,24,j}) 
\ee
which provides  the curves corresponding to $\mathcal{N}=24$ in Fig.\,\ref{fig:small-eta-panels}.


\section{On the large $\eta$ expansion}
\label{app_large_distance}

In this appendix we discuss the derivation of the
analytic results reported in Sec.\,\ref{sec_large_distance}
for the expansion of the entanglement entropies in the regime of large $\eta$,
which have been employed in various figures.

\subsection{Large $\eta$ expansion of the sine kernel tau function}
\label{app_tau_large_eta_LNR}

The asymptotic expansion (\ref{tau-expansion-large-eta}) for the sine kernel tau function 
is a special case of the large $\eta$ expansion for the tau function of the general Painlev\'e V,
found in \cite{Lisovyy:2018mnj}.
As done in the Appendix\;\ref{app_tau_small_eta_GIL},
also in the following a certain quantity having the subindex $\star$ 
corresponds to the same quantity (without this subindex) in \cite{Lisovyy:2018mnj} and in their notation, 
if not otherwise specified.

When $\theta_0 = \theta_t = \theta_\ast = 0$ 
and in the  limit $\sigma_{\!\star} \to 0$ (see (\ref{sigma-eta-P5-parameters})),
the large distance expansion of the tau function given in Eq.\,(1.12a) of \cite{Lisovyy:2018mnj}  
simplifies to
\be
\label{tau-large-eta-gen}
\tau\,=\,
\mathcal{N}_{\textrm{i}\infty}
\sum_{n \in \mathbb{Z}} \e^{2\pi \textrm{i} n \rho_\star}\,
\mathcal{C}_{\textrm{i}\infty}(\nu_{\star} + n)\, 
\mathcal{D}(\nu_{\star} + n;  \textrm{i} t_\star)
\ee
where
\be
\label{C-D-i-infty}
\mathcal{C}_{\textrm{i}\infty}(\nu_{\star}) 
\,\equiv\,
\frac{G(1+ \nu_{\star})^4}{(2\pi)^{2\nu_\star}}
\;\;\;\qquad\;\;\;
\mathcal{D}(\nu_{\star}; \textrm{i}t_\star)
\,\equiv\,
\frac{\e^{\nu_{\star} \textrm{i} t_\star}}{ (\textrm{i}t_\star)^{2\nu_{\star}^2}}
\sum_{k=0}^{\infty} \frac{\mathcal{D}_k(\nu_{\star})}{(\textrm{i} t_\star)^k}
\ee
in terms of the parameters $\rho_\star$ and $\nu_\star$.

The multiplicative constant $\mathcal{N}_{\textrm{i}\infty} $ in (\ref{tau-large-eta-gen}) can be obtained
from the expression of 
$\Upsilon_{\textrm{i}\infty \to 0}  \equiv \mathcal{N}_{\textrm{i}\infty} / \mathcal{N}_{0}$ 
given in Eq.\,(1.19a) of \cite{Lisovyy:2018mnj},
specialised to $\theta_0 = \theta_t = \theta_\ast = 0$ and 
in the limit $\sigma_{\!\star} \to 0$.
By using also that $ \mathcal{N}_{0} = 1$, 
in this special case one finds
\be
\label{N-i-infty}
\mathcal{N}_{\textrm{i}\infty} 
=
(2\pi)^{2\nu_{\star}} \, \e^{\textrm{i} \pi \nu_{\star}^2}\; \frac{G(1-\nu_{\star})^2}{G(1+\nu_{\star})^2}  \,.
\ee

From Eqs\,(1.14) and (1.15) of \cite{Lisovyy:2018mnj} 
specialised to $\theta_0 = \theta_t = \theta_\ast = 0$ and $\sigma_{\!\star} \to 0$,
one obtains the auxiliary parameters $X_{\pm} = 1 \pm 2\pi \xi_\star$.
Then, the connection formulae in Eq.\,(1.17) of \cite{Lisovyy:2018mnj} 
allow to express the parameters $\rho_\star$ and $\nu_\star$ in (\ref{tau-large-eta-gen})
as follows
\be
\label{connection-0-i-infty}
\e^{2\pi \textrm{i} \nu_\star} = X_- = 1 - 2\pi \xi_\star
\;\;\;\;\qquad\;\;\;\;
\e^{2\pi \textrm{i} \rho_\star} =  1- X_+ X_- = (2\pi \xi_\star)^2  \,.
\ee
By employing the second expression of (\ref{GIL-to-our-notation}), these relations give respectively
\be
\label{nu-rho-LNR-from-z}
\nu_\star = \frac{1}{2\pi \textrm{i}}  \log (1-1/z) 
\;\;\;\;\qquad\;\;\;\;
\e^{2\pi \textrm{i} \rho_\star} =   \frac{1}{z^2}  \,.
\ee

Finally, the expansion (\ref{tau-expansion-large-eta}) is obtained
by plugging (\ref{C-D-i-infty}), (\ref{N-i-infty}) 
and the second expression of (\ref{connection-0-i-infty}) into (\ref{tau-large-eta-gen}) first
and then using (\ref{GIL-to-our-notation}), (\ref{nu-rho-LNR-from-z}) 
and $\textrm{e}^{\textrm{i} \pi \nu_\star^2} = \textrm{i}^{2\nu_\star^2}$.

\subsection{Contribution from $\tilde{\tau}_\infty$}
\label{app_large_eta_tildetau}

Consider the first integral in (\ref{SA_infty_def}).
From the definition of $\tilde{\tau}_{\infty} $ in (\ref{tau-tilde-infty-def}), 
we have
\be
\label{dlog_taut}
\partial_z \log\tilde{\tau}_{\infty} 
\,=\,
\Big[\,  4\eta\, \textrm{i}
-4\nu_{\star}\log(4\eta)
+2\nu_{\star} \Big( \psi(1-\nu_{\star}) +\psi(1+\nu_{\star}) - 2\Big)
\Big]
\frac{\partial\nu_{\star}}{\partial z}
\ee
where $\nu_\star$ is defined in (\ref{nu-star-main}),
which implies $\frac{\partial\nu_{\star}}{\partial z} = \frac{1}{2\pi \textrm{i}\, z(z-1)}$,
and $\psi(z)$ is the digamma function
\be
\label{digamma_fun}
\psi(z)
=\frac{d}{dz}\log\Gamma(z)
=
\int_{0}^{\infty}\left(\frac{e^{-t}}{t}-\frac{e^{-zt}}{1-e^{-t}}\right)dt
\;\;\;\qquad\;\;\;
\mathrm{Re}(z)>0  \,.
\ee
By using \eqref{dlog_taut}, the first integral in (\ref{SA_infty_def})
can be written as the sum of a term proportional to $\eta$,
a term  proportional  to $\log(\eta)$ and a constant term.

The leading term in the first integral in (\ref{SA_infty_def}), proportional to $\eta$,
vanishes because $s_{\alpha}(0)=s_{\alpha}(1)=0$.
Instead, the contribution proportional to $\log (\eta)$ is non-vanishing.
It reads 
\bea
& &
\label{int_log_term_0}
\frac{\log(4\eta)}{\pi^{2}}\lim_{\epsilon\rightarrow0^{+}}\lim_{\delta\rightarrow0^{+}}
\Bigg\{
\int_{\mathfrak{C} _{0}}+\int_{\mathfrak{C} _{1}}
+\int_{\mathfrak{C} _{-}}+\int_{\mathfrak{C} _{+}}
\Bigg\}
\, \frac{\nu_{\star}\,s_{\alpha}(z)}{z(z-1)}\, \textrm{d}z
\\
\rule{0pt}{.8cm}
\label{int_log_term_1}
& &
=\frac{\log(4\eta)}{\pi^{2}}\lim_{\epsilon\rightarrow0^{+}}
\left\{ 
\int_{\epsilon/2}^{1-\epsilon/2}  \frac{\nu_{\star}^- \,s_{\alpha}(x)}{x(x-1)}\, \textrm{d}x
-\int_{\epsilon/2}^{1-\epsilon/2} \frac{\nu_{\star}^+ \,s_{\alpha}(x)}{x(x-1)}\, \textrm{d}x
\right\} 
\\
\rule{0pt}{.8cm}
& &
\label{int_log_term_2}
 =
 \frac{\log(4\eta)}{\pi^{2}}\int_{0}^{1} \frac{(\nu_{\star}^{-}-\nu_{\star}^{+}) \,s_{\alpha}(x)}{x(x-1)}\, \textrm{d}x
 = -\frac{\log(4\eta)}{\pi^{2}}\int_{0}^{1}\frac{s_{\alpha}(x)}{x(x-1)} \, \textrm{d}x
 = \frac{1}{6} \left(1+\frac{1}{\alpha}\right) \log(4\eta)
 \hspace{1.3cm}
\eea
where in (\ref{int_log_term_1}) we used that
the integrals along $\mathfrak{C} _{0}$ and $\mathfrak{C} _{1}$ vanish
and the following functions have been introduced in the remaining two integrals 
\begin{equation}
\label{nu_star_pm_def}
\lim_{\delta\rightarrow0^{+}}\nu_{\star}(x\pm \textrm{i}\delta)
=
\frac{1}{2\pi \textrm{i}}\log\left(1/x-1\right)\pm\frac{1}{2}
\equiv
\nu_{\star}^{\pm}(x)
\;\;\;\qquad\;\;\;
x\in [0,1]  \,.
\end{equation}

The result (\ref{int_log_term_2}) provides
both the logarithmic divergence and a contribution to the constant term.
From (\ref{dlog_taut}), we have that 
the remaining contribution to the constant term 
in the first integral in (\ref{SA_infty_def}) reads
\bea
\label{int_const_term_0}
& &\hspace{-.2cm}
-\,\frac{1}{2\pi^{2}}\lim_{\epsilon\rightarrow0^{+}}\lim_{\delta\rightarrow0^{+}}
\oint_{\mathfrak{C}}
\Big(\psi(1-\nu_{\star})+\psi(1+\nu_{\star})-2\Big) \nu_{\star}\,
\frac{s_{\alpha}(z)}{z(z-1)}\, \textrm{d}z
\\
\rule{0pt}{.8cm}
& & \hspace{-.2cm}
=
\frac{1}{2\pi^{2}}
\int_{0}^{1}
\Big\{ 
\nu_{\star}^{+}\big[\psi(1-\nu_{\star}^{+})+\psi(1+\nu_{\star}^{+})-2\big]
-\nu_{\star}^{-}\big[\psi(1-\nu_{\star}^{-})+\psi(1+\nu_{\star}^{-}) - 2\big]
\Big\} \;
\frac{s_{\alpha}(x)}{x(x-1)}\,\textrm{d}x
\nonumber 
\\
\rule{0pt}{.8cm}
& & \hspace{-.2cm}
=
-\frac{1}{\pi}\int_{\mathbb{R}}
\Big\{ 
\nu_{\star}^{+}\big[\psi(1-\nu_{\star}^{+})+\psi(1+\nu_{\star}^{+})- 2\big]
-\nu_{\star}^{-}\big[\psi(1-\nu_{\star}^{-})+\psi(1+\nu_{\star}^{-}) - 2\big]
\Big\} \;
s_{\alpha}(x(y))\,\textrm{d}y
\nonumber 
\\
\label{int_const_term_1}
\rule{0pt}{.8cm}
& & \hspace{-.2cm}
=
-\frac{1}{\pi}\int_{\mathbb{R}}
\Big\{ 
\psi(1/2 + \textrm{i} y) + \psi(1/2 - \textrm{i} y) 
\Big\} \;
s_{\alpha}(x(y))\,\textrm{d}y
\eea
where the integral over $\mathfrak{C}$ 
has been decomposed as done in (\ref{int_log_term_0}) 
and we used the fact that,
in the limits $\delta\rightarrow0^{+}$ and $\epsilon\rightarrow0^{+}$,
the contributions corresponding to $\mathfrak{C} _{0}$ and $\mathfrak{C} _{1}$ vanish
and (\ref{nu_star_pm_def}) holds. 
In the second step of (\ref{int_const_term_0}),
the integration variable $y\equiv\frac{1}{2\pi}\log(1/x-1)$ has been employed;
hence (\ref{nu_star_pm_def}) and (\ref{entropies_func})
give respectively $\nu_{\star}^{\pm} = -\textrm{i}y\pm\frac{1}{2}$ and 
\bea
s(x(y)) 
&=&
\frac{\log(1+e^{2\pi y})}{1+e^{2\pi y}}+\frac{\log(1+e^{-2\pi y})}{1+e^{-2\pi y}}
\\
\rule{0pt}{.8cm}
s_\alpha(x(y)) 
&=&
\frac{1}{1-\alpha}\,\Big[\log\left(1+e^{\alpha2\pi y}\right)-\alpha\log\left(1+e^{2\pi y}\right)\Big]
\eea
which leads to
\be
\label{s-infty-y-def}
s_{\infty}(x(y)) 
=
 \log\! \big(1+e^{-2\pi |y|}\big)  \,.
\ee
The final expression (\ref{int_const_term_1}) can be found by exploiting the identity
$\psi(1+z)=\psi(z)+1/z$.
Then, by employing in \eqref{int_const_term_1}
the integral representation of the digamma function (see \eqref{digamma_fun})
and exchanging the order of the two integrations, 
the integral in $y$ can be performed,
finding the integrand occurring in (\ref{E_alpha-def}).
Combining this result with (\ref{int_log_term_2}), the expression (\ref{S_alpha_main_large_eta_0}) is obtained.

\subsection{Contribution from $\widetilde{\mathcal{T}}_\infty$}
\label{app_large_eta_Ttilde}

\subsubsection{R\'enyi and entanglement entropies}
\label{app-large-eta-renyi-ee}

The second expression in (\ref{SA_infty_def}) can be treated 
by first decomposing $\mathfrak{C}$ as done in  (\ref{int_log_term_0}) 
and then adopting (\ref{nu-star-main})  as integration variable. 
The remaining non-vanishing contributions, 
which come from the integration along $\mathfrak{C} _{\pm}$,
can be written in terms of the following quantities
\be
\label{nu_star_app_y_def}
 \nu_{\star}^{\pm} \equiv -\textrm{i}y\pm\frac{1}{2}
\; \;\;\qquad\;\;\;
\widetilde{\mathcal{T}}_\infty^{\pm} 
\equiv \widetilde{\mathcal{T}}_\infty\big|_{ \nu_{\star} = \nu_{\star}^{\pm}}
\;\;\;\qquad\;\;\;
s_\alpha(z(\nu_\star^{+})) = s_\alpha(z(\nu_\star^{-})) \equiv \hat{s}_\alpha(y)
\ee
and they read
\bea
\label{Stilde-alpha-app-nustar}
 \widetilde{S}_{A,\infty}^{(\alpha)} 
&=&
   \frac{1}{2\pi \textrm{i}}
  \int_{\mathbb{R}}
  \hat{s}_\alpha(y)\,
  \Big[
 \partial_{y}\! 
  \log \! \big( 1 + \widetilde{\mathcal{T}}_\infty^{+} \big)
  -
 \partial_{y}
  \log \! \big( 1 + \widetilde{\mathcal{T}}_\infty^{-} \big)
  \Big]
 \, \textrm{d}y
 \\
 \label{Stilde-alpha-app-nustar-step2}
 \rule{0pt}{.7cm}
 &=&
   \frac{1}{2\pi \textrm{i}}
  \int_{\mathbb{R}}
 \big( \partial_y\hat{s}_\alpha(y) \big)\,
  \Big[
  \log \! \big( 1 + \widetilde{\mathcal{T}}_\infty^{-} \big)
  -
  \log \! \big( 1 + \widetilde{\mathcal{T}}_\infty^{+} \big)
  \Big]
 \, \textrm{d}y
\eea
where the last expression has been obtained through an integration by parts and
\be
\label{der-s-y-explicit}
\partial_y\hat{s}(y)
=
-\, \frac{\pi^2  y}{[\cosh(\pi y)]^2}
\;\;\;\qquad\;\;\;
\partial_y\hat{s}_\alpha(y)
=
\frac{\pi \,\alpha}{\alpha - 1} \Big( \tanh(\pi y) - \tanh(\alpha\,\pi  y) \Big)  \,.
\ee

Here we consider finite values of $\alpha>0$.
The limiting case $\alpha \to \infty$ 
is discussed in Sec.\,\ref{app-large-eta-single-copy}.

When $\alpha =1$, the function $\partial_y\hat{s}(y)$ has poles of order two 
at $y_k \equiv \textrm{i}(k+1/2)$ with $k \in \mathbb{Z}$.
Instead, for finite $\alpha > 1$, 
the term containing $\tanh(\pi y)$ in $\partial_y\hat{s}_\alpha(y)$
has poles of order one 
at $y_k = \textrm{i}(k+1/2)$ with $k \in \mathbb{Z}$
whose residue is $\alpha /(\alpha - 1)$.
The term containing $\tanh(\alpha \pi y)$ provides poles of order one 
with residue equal to $1 /(1-\alpha )$ at
\be
\label{y_k-tilde-def}
 \tilde{y}_k \equiv \frac{\textrm{i}}{\alpha}\left(k+\frac{1}{2}\right)
 \;\;\;\qquad\;\;\;
 k \in \mathbb{Z}
 \ee

By introducing the following notation
\be
\label{Omega-def}
\Omega(z) \equiv  \left(\frac{\Gamma(1+z)}{\Gamma(-z)} \right)^2
\ee
for $\widetilde{\mathcal{T}}_\infty^{\pm}$ in (\ref{nu_star_app_y_def}),
which are obtained by evaluating 
\eqref{cal-T-infty-split-101-bis} in $\nu_\star=\nu_\star^{\pm}$, 
we find
\bea
\label{cal-T-infty-split-101-bis-app-plus}
\widetilde{\mathcal{T}}_\infty^{+}
&=&
 \sum_{k = 1}^{\infty} \frac{\mathcal{D}_k(\nu_\star^{+})}{(4\textrm{i} \eta)^k} 
 +
  \frac{\textrm{e}^{4\textrm{i}  \eta} \, (4 \eta)^{4\textrm{i}y} \,  \Omega(\nu_\star^{+})}{ (4 \eta)^{4}}\,
\sum_{k = 0}^{\infty} \frac{\mathcal{D}_k(\nu_\star^{+} + 1)}{(4\textrm{i} \eta)^k} 
+ 
\frac{ \Omega(-\nu_\star^{+})}{ \textrm{e}^{4\textrm{i}  \eta} \, (4\eta)^{4\textrm{i}y}}\,
\sum_{k = 0}^{\infty} \frac{\mathcal{D}_k(\nu_\star^{+} -1)}{(4\textrm{i} \eta)^k} 
\nonumber
\\
& &
\\
\label{cal-T-infty-split-101-bis-app-minus}
\widetilde{\mathcal{T}}_\infty^{-}
&=&
 \sum_{k = 1}^{\infty} \frac{\mathcal{D}_k(\nu_\star^{-})}{(4\textrm{i} \eta)^k} 
 +
\textrm{e}^{4\textrm{i}  \eta} \, (4 \eta)^{4\textrm{i}y} \,\Omega(\nu_\star^{-})\,
\sum_{k = 0}^{\infty} \frac{\mathcal{D}_k(\nu_\star^{-} + 1)}{(4\textrm{i} \eta)^k} 
+ 
\frac{ \Omega(-\nu_\star^{-})}{ (4\eta)^{4} \,\textrm{e}^{4\textrm{i}  \eta} \, (4\eta)^{4\textrm{i}y}}\,
\sum_{k = 0}^{\infty} \frac{\mathcal{D}_k(\nu_\star^{-} -1)}{(4\textrm{i} \eta)^k}  \,.
\nonumber
\\
& &
\eea

The expansions of these expressions for large $\eta$ take the following form
\be
\label{R_tilde_infty_pm_def}
\widetilde{\mathcal{T}}_\infty^\pm
\,=\,
\sum_{k =0}^\infty \frac{\mathcal{R}_k^\pm}{\eta^k}
\ee
where $\mathcal{R}_k^\pm$ are functions of $\eta$.
By using (\ref{R_tilde_infty_pm_def}) and 
 introducing 
\be
\label{Bk-def}
     \mathcal{B}_k^\pm \equiv \frac{\mathcal{R}_k^\pm}{1 + \mathcal{R}_0^\pm}
\ee
the expressions occurring in the integrand of (\ref{Stilde-alpha-app-nustar-step2})
can be expanded as 
\bea
\label{log-Ttilde-expansion}  
& &\hspace{-1cm}
  \log \! \big( 1 + \widetilde{\mathcal{T}}_\infty^\pm \big)
\,=\,
    \log \! \big( 1 + \mathcal{R}_0^\pm \big)
    +
     \log \! \Bigg( 
     1 + \sum_{k=1}^{\infty} \frac{\mathcal{B}_k^\pm}{\eta^k} 
     \Bigg)
\\
\rule{0pt}{.8cm}
& & 
\hspace{-1cm}
=\,
      \log \! \big( 1 + \mathcal{R}_0^\pm \big)
    +
    \frac{\mathcal{B}_1^\pm}{\eta}
    +
    \frac{\mathcal{B}_2^\pm - \tfrac{1}{2} \big(\mathcal{B}_1^\pm\big)^2}{\eta^2}
        +
    \frac{\mathcal{B}_3^\pm - \mathcal{B}_2^\pm \, \mathcal{B}_1^\pm
    + \tfrac{1}{3}\big(\mathcal{B}_1^\pm\big)^3}{\eta^3}
    + O(1/\eta^4)
\,\equiv\,
\sum_{k=0}^{\infty} \frac{\mathcal{Y}_k^\pm}{\eta^k}   \,.
\nonumber
\eea

We remark that our analysis is based on the approximate expression \eqref{cal-T-infty-split-101},
obtained by considering only the summands corresponding to $n \in \{-1,0,1\}$ in (\ref{cal-T-infty-def}).
Including also terms corresponding to $|n| \geqslant 2$ 
leads to $O(1/\eta^4)$ terms in the entanglement entropies. 
Hence, in order to be consistent with the approximation made in \eqref{cal-T-infty-split-101},
we truncate (\ref{cal-T-infty-split-101-bis-app-plus}) and (\ref{cal-T-infty-split-101-bis-app-minus})
by considering only the terms up to $O(1/\eta^3)$ included.

Furthermore, the $O(1/\eta^k)$ terms having $k \geqslant 4$ 
in (\ref{cal-T-infty-split-101-bis-app-plus}) and (\ref{cal-T-infty-split-101-bis-app-minus})
are combinations of terms coming from all the three series.
In this approximation,
(\ref{cal-T-infty-split-101-bis-app-plus}) and (\ref{cal-T-infty-split-101-bis-app-minus}) become 
\bea
\label{Ttilde_inf_plus-minus}
\widetilde{\mathcal{T}}_\infty^{\pm}
&=&
 \sum_{k = 1}^{3} \frac{\mathcal{D}_k(\nu_\star^{\pm})}{(4\textrm{i} \eta)^k} 
 +
 \frac{ \Omega(\mp\nu_\star^{\pm})}{ \textrm{e}^{\pm4\textrm{i}  \eta} \, (4\eta)^{\pm4\textrm{i}y}}\,
\sum_{k = 0}^{3} \frac{\mathcal{D}_k(\nu_\star^{\pm} \mp 1)}{(4\textrm{i} \eta)^k} 
+ O(1/\eta^4)
\\
&=&
 \frac{ \Omega(\mp \nu_\star^{\pm})}{ \textrm{e}^{\pm4\textrm{i}  \eta} \, (4\eta)^{\pm4\textrm{i}y}}
 +
 \sum_{k = 1}^{3} \frac{1}{(4\textrm{i} \eta)^k} 
 \left(
 \mathcal{D}_k(\nu_\star^{\pm})
 +
  \frac{ \Omega(\mp\nu_\star^{\pm}) \, \mathcal{D}_k(\nu_\star^{\pm} \mp 1)}{ 
  \textrm{e}^{\pm4\textrm{i}  \eta} \, (4\eta)^{\pm4\textrm{i}y}}\,
 \right)
 + O(1/\eta^4) 
 \nn
\eea
which allows to write 
the explicit expressions of $\mathcal{R}_k^\pm$ 
for $k \in \{0,1,2,3\}$ (see (\ref{R_tilde_infty_pm_def})) as 
\be
\label{R_0_pm_explicit}
\mathcal{R}_0^\pm
=
 \frac{ \Omega(\mp \nu_\star^{\pm})}{ \textrm{e}^{\pm 4\textrm{i}  \eta} \, 
 (4\eta)^{\pm 4\textrm{i}y}}
 =
 \frac{ \Omega( \pm \textrm{i} y - 1/2)}{ \textrm{e}^{\pm 4\textrm{i}  \eta} \, 
 (4\eta)^{\pm 4\textrm{i}y}} 
\ee
and 
\be
\label{R_123_pm_explicit}
 \mathcal{R}_k^\pm
=
\frac{1}{(4\textrm{i})^k}\,
\big[\,
\mathcal{D}_k(\nu_\star^{\pm})
+
\mathcal{R}_0^\pm \, \mathcal{D}_k( \nu_\star^{\pm} \mp 1)
\,\big] 
\;\;\;\qquad\;\;\;
k \in \big\{1,2,3\big\}  \,.
\ee

By using  (\ref{log-Ttilde-expansion}) 
into (\ref{Stilde-alpha-app-nustar-step2}),
the expansion (\ref{exp-S-infty-123}) is obtained with
\be
\label{StildeN}
\widetilde{S}_{A,\infty,N}^{(\alpha)}
=  
\frac{1}{2\pi \textrm{i}}
\int_{-\infty}^{+\infty} \!\!
\big(\partial_{y}\hat{s}_{\alpha}(y)\big)\,
\big( \mathcal{Y}_{N}^{-}-\mathcal{Y}_{N}^{+} \big)
\textrm{d}y
 \;\;\; \qquad \;\;\; 
 N \in \big\{0,1,2,3\big\} 
\ee
In order to write these expressions more explicitly, 
from \eqref{Bk-def}, \eqref{R_0_pm_explicit} and \eqref{R_123_pm_explicit} 
one finds
\be
\label{Bk-decomposition-DP}
\mathcal{B}_k^\pm 
=
\frac{1}{(4\textrm{i})^k}
\Big(
\mathcal{D}_k^{\pm} + \widetilde{\mathcal{R}}_0^\pm \, \mathcal{P}_k^\pm 
\Big)
\;\;\;\qquad\;\;\;
k \in \big\{1,2,3\big\}
\ee
where the following notation has been adopted
\be
\label{R0tilde-D-P-def}
\widetilde{\mathcal{R}}_0^\pm \equiv \frac{\mathcal{R}_0^\pm}{1+\mathcal{R}_0^\pm}
\;\;\;   \qquad \;\;\;
\mathcal{D}_k^\pm \equiv \mathcal{D}_k(\nu_\star^{\pm})
\;\;\;   \qquad \;\;\;
\mathcal{P}_k^\pm \equiv \mathcal{D}_k( \nu_\star^{\pm} \mp 1) - \mathcal{D}_k(\nu_\star^{\pm})  \,.
\ee
By employing the decomposition (\ref{Bk-decomposition-DP}) for $\mathcal{B}_k^\pm $,
for the $O(1/\eta^2)$ and $O(1/\eta^3)$ terms in (\ref{log-Ttilde-expansion}) 
we obtain respectively 
\be
\label{B2-from-P}
\mathcal{Y}_2^\pm 
\equiv 
\mathcal{B}_2^\pm - \frac{1}{2} \big(\mathcal{B}_1^\pm\big)^2
\,=\,
\frac{1}{(4\textrm{i})^2}
\left\{\,
\mathcal{D}_2^\pm - \frac{1}{2} \big(\mathcal{D}_1^\pm\big)^2
+
\Big[
\mathcal{P}_2^\pm - \mathcal{D}_1^\pm\, \mathcal{P}_1^\pm
\Big]
\,\widetilde{\mathcal{R}}_0^\pm 
-
 \frac{1}{2}\,\big(\mathcal{P}_1^\pm\big)^2 \big(\widetilde{\mathcal{R}}_0^\pm\big)^2
 \right\}
\ee
and
\bea
\label{B3-from-P}
\mathcal{Y}_3^\pm 
&\equiv&
\mathcal{B}_3^\pm - \mathcal{B}_2^\pm \, \mathcal{B}_1^\pm + \frac{1}{3}\big(\mathcal{B}_1^\pm\big)^3
\\
\rule{0pt}{.6cm}
& = &
\frac{1}{(4\textrm{i})^3}\,
\bigg\{ 
\mathcal{D}_3^\pm - \mathcal{D}_2^\pm \, \mathcal{D}_1^\pm + \frac{1}{3}\big(\mathcal{D}_1^\pm\big)^3
+
\Big[
\mathcal{P}_3^\pm + \mathcal{P}_1^\pm \big(\mathcal{D}_1^\pm \big)^2 
- \mathcal{P}_2^\pm \, \mathcal{D}_1^\pm - \mathcal{P}_1^\pm \,\mathcal{D}_2^\pm
\Big]
\,\widetilde{\mathcal{R}}_0^\pm
\nonumber
\\
\rule{0pt}{.6cm}
& & \hspace{5.5cm}
-\,
\mathcal{P}_1^\pm
\Big[
\mathcal{P}_2^\pm  - \mathcal{P}_1^\pm \,\mathcal{D}_1^\pm
\Big] \big( \widetilde{\mathcal{R}}_0^\pm \big)^2
+
\frac{ 1}{3}\, \big(\mathcal{P}_1^\pm\big)^3\,\big( \widetilde{\mathcal{R}}_0^\pm \big)^3
\bigg\}  \,.
\nonumber
\eea

At this point, each term of the expansion \eqref{log-Ttilde-expansion} can be written 
as a power series of $\mathcal{R}_0^\pm$.
Indeed,  for the leading term it is straightforward to write
\be
\label{log-expansion-R0}
      \mathcal{Y}_0^\pm \,\equiv \, 
      \log \! \big( 1 + \mathcal{R}_0^\pm \big)
      =
      \sum_{j =1}^\infty \frac{(-1)^{j+1}}{j}\, \big(\mathcal{R}_0^\pm\big)^j 
\ee
In the subleading terms,
 the expansion of $( 1 + \mathcal{R}_0^\pm )^{-p}$ for integer $p\geqslant 1$ is needed
and it can be found through  the following recursion rule
\be
\label{expansions-R0-recursion}
      \frac{1}{1 + \mathcal{R}_0^\pm }
            =
      \sum_{j =0}^\infty (-1)^{j} \,\big(\mathcal{R}_0^\pm\big)^j 
      \;\;\qquad\;\;
      \frac{1}{\big( 1 + \mathcal{R}_0^\pm \big)^p}
            \,=\,
           \frac{1}{1-p}\;
\partial_{\mathcal{R}_0^\pm}\! \left(   \frac{1}{\big( 1 + \mathcal{R}_0^\pm \big)^{p-1}} \right)
\qquad
p\geqslant 2  \,.
\ee
In particular, in (\ref{B2-from-P}) and (\ref{B3-from-P}) we need
$\widetilde{\mathcal{R}}_0^\pm =   \sum_{j =0}^\infty (-1)^{j} \,(\mathcal{R}_0^\pm)^{j+1}$
(which can be easily obtained from (\ref{R0tilde-D-P-def})
and the first expression in (\ref{expansions-R0-recursion}))
and also $( \widetilde{\mathcal{R}}_0^\pm)^2 $ and $( \widetilde{\mathcal{R}}_0^\pm)^3 $,
that can be derived through the recursive formula  in (\ref{expansions-R0-recursion}), 
finding respectively
\be
\label{tildeR0-expansion-23}
\big( \widetilde{\mathcal{R}}_0^\pm \big)^2 =
     - \sum_{j =0}^\infty (-1)^{j}\, j \,\big(\mathcal{R}_0^\pm\big)^{j+1}
\qquad
\big( \widetilde{\mathcal{R}}_0^\pm \big)^3 = 
      \frac{1}{2} \sum_{j =0}^\infty (-1)^{j}\, j(j-1) \,\big(\mathcal{R}_0^\pm\big)^{j+1}  \,.
\ee

The expansion of $\widetilde{\mathcal{R}}_0^\pm$ 
allows us to write $\mathcal{B}_1^\pm $ (defined in (\ref{Bk-decomposition-DP})) as follows
\be
\label{B1-from-P-a-b}
\mathcal{Y}_1^\pm 
\equiv 
\mathcal{B}_1^\pm 
=
\frac{1}{4\textrm{i}}\,
\mathcal{D}_1^{\pm}
+
\frac{\mathcal{R}_0^\pm}{4\textrm{i}}\,
 \sum_{j =0}^\infty (-1)^{j}\,  \mathcal{P}_1^\pm \, \big(\mathcal{R}_0^\pm\big)^{j}
 \,=\,
\mathcal{Y}_{1,a}^\pm + \mathcal{Y}_{1,b}^\pm 
\ee
where $\mathcal{Y}_{1,a}^\pm \equiv  \tfrac{1}{4\textrm{i}}\, \mathcal{D}_1^{\pm}$
and $\mathcal{Y}_{1,a}^\pm$ is a power series in $\mathcal{R}_0^\pm$.
A similar decomposition can be written for the higher order terms. 
In particular, 
by employing also (\ref{tildeR0-expansion-23}),
for $\mathcal{Y}_2^\pm $ in (\ref{B2-from-P}) we find
\be
\label{B2-from-P-expansion}
\mathcal{Y}_2^\pm 
\,=\,
\mathcal{B}_2^\pm - \frac{1}{2} \big(\mathcal{B}_1^\pm\big)^2
\,=\,
\mathcal{Y}_{2,a}^\pm + \mathcal{Y}_{2,b}^\pm 
\ee
where 
\be
\label{B2-from-P-a-b}
\mathcal{Y}_{2,a}^\pm
\,\equiv\,
\frac{1}{(4\textrm{i})^2}
\left(
\mathcal{D}_2^\pm - \frac{1}{2} \big(\mathcal{D}_1^\pm\big)^2
\right)
\;\;\qquad\;\;
\mathcal{Y}_{2,b}^\pm
\,\equiv\,
\frac{\mathcal{R}_0^\pm}{(4\textrm{i})^2}
      \sum_{j =0}^\infty (-1)^{j}\, \widetilde{\mathcal{D}}_{2,j}^\pm\, \big(\mathcal{R}_0^\pm\big)^{j}
\ee
with
\be
\label{D2-tilde-j-def}
\widetilde{\mathcal{D}}_{2,j}^\pm
\,\equiv\,
\mathcal{P}_2^\pm - \mathcal{D}_1^\pm\, \mathcal{P}_1^\pm
+
 \frac{j}{2}\,\big(\mathcal{P}_1^\pm\big)^2 
\ee
and a similar analysis for $\mathcal{Y}_3^\pm $ in (\ref{B3-from-P}) leads to
\be
\label{B3-from-P-expansion}
\mathcal{Y}_3^\pm 
\,=\,
\mathcal{B}_3^\pm - \mathcal{B}_2^\pm \, \mathcal{B}_1^\pm + \frac{1}{3}\big(\mathcal{B}_1^\pm\big)^3
\,=\,
\mathcal{Y}_{3,a}^\pm + \mathcal{Y}_{3,b}^\pm 
\ee
where
\be
\label{B3-from-P-a-b}
\mathcal{Y}_{3,a}^\pm
\,\equiv\,
\frac{1}{(4\textrm{i})^3}
\left(
\mathcal{D}_3^\pm - \mathcal{D}_2^\pm \, \mathcal{D}_1^\pm + \frac{1}{3}\big(\mathcal{D}_1^\pm\big)^3
\right)
\;\;\qquad\;\;
\mathcal{Y}_{3,b}^\pm
\,\equiv\,
\frac{\mathcal{R}_0^\pm}{(4\textrm{i})^3}
      \sum_{j =0}^\infty (-1)^{j}\, \widetilde{\mathcal{D}}_{3,j}^\pm \,\big(\mathcal{R}_0^\pm\big)^{j}
\ee
with
\be
\label{D3-tilde-j-def}
\widetilde{\mathcal{D}}_{3,j}^\pm
\,\equiv\,
\mathcal{P}_3^\pm + \mathcal{P}_1^\pm \big(\mathcal{D}_1^\pm \big)^2 
- \mathcal{P}_2^\pm \, \mathcal{D}_1^\pm - \mathcal{P}_1^\pm \,\mathcal{D}_2^\pm
+ 
j\,\mathcal{P}_1^\pm
\Big(
\mathcal{P}_2^\pm  - \mathcal{P}_1^\pm \,\mathcal{D}_1^\pm
\Big)
+
\frac{j(j-1)}{6}\, \big(\mathcal{P}_1^\pm\big)^3  \,.
\ee
Notice the similar structure occurring  in (\ref{log-expansion-R0}),
(\ref{B1-from-P-a-b}), (\ref{B2-from-P-expansion})
and (\ref{B3-from-P-expansion}).

By using 
(\ref{Stilde-alpha-app-nustar-step2}), (\ref{log-Ttilde-expansion}) and (\ref{log-expansion-R0}), 
the leading term in (\ref{exp-S-infty-123}) becomes
\be
\label{SA-infty-integral-0}
  \widetilde{S}_{A,\infty,0}^{(\alpha)} 
\,=\,
   \frac{1}{2\pi \textrm{i}}\,
   \sum_{j =1}^\infty \frac{(-1)^{j+1}}{j}
  \int_{\mathbb{R}}
 \big( \partial_y\hat{s}_\alpha(y) \big)\,
  \Big[
\big(\mathcal{R}_0^{-}\big)^j - \big(\mathcal{R}_0^{+}\big)^j
\,  \Big]
 \, \textrm{d}y  \,.
\ee

These integrals can be evaluated through the residue theorem as follows. 
Since, from \eqref{R_0_pm_explicit}, we have that
$(\mathcal{R}_0^{-} )^j $ and $(\mathcal{R}_0^{+} )^j $
contain the factors $ (4\eta)^{+ 4\textrm{i}y j}$ and $ (4\eta)^{- 4\textrm{i}y j}$
respectively and $\eta \to +\infty$, 
for the integrals corresponding to $(\mathcal{R}_0^{-} )^j$ or $(\mathcal{R}_0^{+} )^j$
we chose a contour enclosing the upper half plane or the lower half plane respectively;
hence non-trivial contributions can come from 
the residues of the corresponding integrands 
in the upper half plane or the lower half plane respectively.
Since the function $\Omega(z)$ defined in (\ref{Omega-def})
has poles at $z=-k$ with integer $k \geqslant 1$, 
the factor $\Omega(\mp \nu_\star^{\pm})$ occurring in $\mathcal{R}_0^{\pm}$
has poles at $\nu_\star^{\pm} = \pm k$,
i.e. at $y = \mp \,\textrm{i}\, (1/2 - k)$, with integer $k \geqslant 1$.
Thus, the poles of the factor $\Omega(+ \nu_\star^{-})^j$ in $(\mathcal{R}_0^{-} )^j $
and of the factor $\Omega(- \nu_\star^{+})^j$ in $(\mathcal{R}_0^{+} )^j $
are located  in the lower half plane and in the upper half plane respectively;
hence they do not contribute to the integral in (\ref{SA-infty-integral-0}).
The contributions coming from the poles of $\partial_y\hat{s}_\alpha(y)$ in (\ref{SA-infty-integral-0}),
which have been described in the text below (\ref{der-s-y-explicit}),
can be found as follows. 
Since $\Omega(z)$ in (\ref{Omega-def}) has zeros of second order 
at $z = k$ with integer $k \geqslant 0$,
the factor $\Omega(\mp \nu_\star^{\pm})$ in $\mathcal{R}_0^{\pm}$
 has second order zeros at $y = \mp \, \textrm{i} (k+1/2)$.
 Combining this observation with the structures of the poles described in the text below (\ref{der-s-y-explicit}),
 one finds that the integrand of (\ref{SA-infty-integral-0}) 
 does not have singularities  when $\alpha=1$,
 while it has simple poles at (\ref{y_k-tilde-def}) for finite $\alpha >0$ and $\alpha \neq 1$.

The above analysis tells us that \eqref{SA-infty-integral-0} vanishes when $\alpha =1$
and that for finite $\alpha >1$ it can be written as follows
\bea
\label{SA-infty-integral-0-sum}
  \widetilde{S}_{A,\infty,0}^{(\alpha)} 
&=&
   \frac{1}{1-\alpha }\,
   \sum_{j =1}^\infty \frac{(-1)^{j+1}}{j}\,
   \Bigg\{
\sum_{k=0}^{\infty}
  \Big[
\big(\mathcal{R}_0^{-}\big)^j \big|_{\tilde{y}_k} + \big(\mathcal{R}_0^{+}\big)^j \big|_{\tilde{y}_{-k-1}}
\,  \Big]
\Bigg\}
\\
\label{SA-infty-integral-0-sum-v2}
\rule{0pt}{.7cm}
&=&
   \frac{2}{\alpha - 1}\,
   \sum_{j =1}^\infty \frac{(-1)^{j}}{j}\;\cos(4\eta j)\,
   \sum_{k=0}^{\infty}
 \frac{\Omega\big((k+1/2)/\alpha -1/2\big)^j}{(4\eta)^{2j(2k+1)/\alpha}}
\eea
where the last expression has been found by using (\ref{R_0_pm_explicit}) and 
\be
 \frac{ \Omega( - \textrm{i} y - 1/2)}{  (4\eta)^{- 4\textrm{i}y}} \,\bigg|_{\tilde{y}_k}
 =\,
  \frac{ \Omega( + \textrm{i} y - 1/2)}{  (4\eta)^{+ 4\textrm{i}y}} \,\bigg|_{\tilde{y}_{-k-1}}
 \!\! =\,
     \frac{\Omega\big((k+1/2)/\alpha -1/2\big)}{(4\eta)^{2(2k+1)/\alpha}}  \,.
\ee
Notice that (\ref{SA-infty-integral-0-sum-v2}) is real, as expected.

As for the $O(1/\eta)$ term in (\ref{exp-S-infty-123}),
by employing (\ref{Stilde-alpha-app-nustar-step2}), (\ref{log-Ttilde-expansion}),
and (\ref{B1-from-P-a-b})
it becomes
\bea
\label{SA-infty-integral-1}
  \widetilde{S}_{A,\infty,1}^{(\alpha)} 
&=&
   \frac{1}{2\pi \textrm{i}}
  \int_{\mathbb{R}}
 \big( \partial_y\hat{s}_\alpha(y) \big)\,
  \big[\,
\mathcal{B}_1^{-}  - \mathcal{B}_1^{+}
 \,\big]
 \, \textrm{d}y
 \\
 \rule{0pt}{.8cm}
 \label{SA-infty-integral-1-b}
 &=&
   \frac{1}{8\pi }
  \int_{\mathbb{R}}
  \big[\,
\mathcal{D}_1^{+}  - \mathcal{D}_1^{-}
 \,\big]\, \partial_y\hat{s}_\alpha(y)
 \, \textrm{d}y
  \nonumber
 \\
  \rule{0pt}{.7cm}
 & & 
 +\,
    \frac{1}{8\pi }
     \sum_{j =0}^\infty (-1)^{j}
  \int_{\mathbb{R}}
 \big( \partial_y\hat{s}_\alpha(y) \big)\,
  \Big[\,
\mathcal{P}_1^{+} \, \big(\mathcal{R}_0^{+}\big)^{j+1} - \mathcal{P}_1^{-} \, \big(\mathcal{R}_0^{-}\big)^{j+1}
 \,\Big]
 \, \textrm{d}y  \,.
\eea
From \eqref{D-012-def}, \eqref{nu_star_app_y_def} and \eqref{R0tilde-D-P-def},
we find that $\mathcal{D}_1^{+}  - \mathcal{D}_1^{-} = 1 - 12\, y^2$,
which is an even function. 
Combining this observation with the fact that 
the functions in (\ref{der-s-y-explicit}) are odd,
we have that  the integral in the first line of (\ref{SA-infty-integral-1-b}) vanishes
because its integrand is an odd function. 
The integral in the second line of (\ref{SA-infty-integral-1-b}) 
can be evaluated by adapting the analysis made above to obtain 
(\ref{SA-infty-integral-0-sum}) and (\ref{SA-infty-integral-0-sum-v2}).
In particular, this integral gives a vanishing contribution when $\alpha = 1$.
Instead, for positive $\alpha \neq 1$ we have 
\bea
\label{SA-infty-integral-1-sum}
  \widetilde{S}_{A,\infty,1}^{(\alpha)} 
&=&
   -\, \frac{1}{4\,\textrm{i}\,(\alpha - 1)}
     \sum_{j =0}^\infty (-1)^{j}\,
\sum_{k=0}^{\infty}
  \Big[\,
  \mathcal{P}_1^{+} \, \big(\mathcal{R}_0^{+}\big)^{j+1} \big|_{\tilde{y}_{-k - 1}} 
  \!+
  \mathcal{P}_1^{-} \, \big(\mathcal{R}_0^{-}\big)^{j+1}\big|_{\tilde{y}_k} 
 \,\Big]
 \hspace{1cm}
 \\
 \rule{0pt}{.7cm}
 &=&
     \sum_{j =1}^\infty 
        \frac{(-1)^{j}   \sin(4\eta j)}{2\,(\alpha - 1)}\,
   \sum_{k=0}^{\infty}
   \frac{\Omega\big((k+1/2)/\alpha -1/2\big)^j}{(4\eta)^{2j(2k+1)/\alpha}}\;
    \mathcal{P}_1^{-} \big|_{\tilde{y}_k} 
\eea
where we used that $\mathcal{P}_1^{-} = - \,\mathcal{P}_1^{+} = 1- 12\, y^2$
and  $\mathcal{P}_1^{-}|_{\tilde{y}_k} = 1+ 3(2k+1)^2/\alpha^2$,
which have been obtained from (\ref{R0tilde-D-P-def}), (\ref{nu_star_app_y_def}) and (\ref{y_k-tilde-def})
and do not depend on the index $j$.

The $O(1/\eta^2)$ term in (\ref{exp-S-infty-123}) can be studied
by employing (\ref{Stilde-alpha-app-nustar-step2}), (\ref{log-Ttilde-expansion}), 
(\ref{B2-from-P-expansion}) and (\ref{B2-from-P-a-b}).
The result reads
\bea
\label{SA-infty-integral-2}
  \widetilde{S}_{A,\infty,2}^{(\alpha)} 
 &=&
   \frac{1}{2\pi \textrm{i}}
  \int_{\mathbb{R}}
 \big( \partial_y\hat{s}_\alpha(y) \big)\,
  \big[\,
\mathcal{B}_{2,a}^{-} - \mathcal{B}_{2,a}^{+}
 \,\big]
 \, \textrm{d}y
 +
    \frac{1}{2\pi \textrm{i}}
  \int_{\mathbb{R}}
 \big( \partial_y\hat{s}_\alpha(y) \big)\,
  \big[\,
\mathcal{B}_{2,b}^{-} - \mathcal{B}_{2,b}^{+}
 \,\big]
 \, \textrm{d}y
 \hspace{1cm}
 \\
 \label{SA-infty-integral-2-b}
  \rule{0pt}{.8cm}
 & =& 
    \frac{1}{32\pi \textrm{i}}
  \int_{\mathbb{R}}
 \big( \partial_y\hat{s}_\alpha(y) \big)\,
\left[
\left( \mathcal{D}_2^{+} - \frac{1}{2} \big(\mathcal{D}_1^{+}\big)^2 \right)
-
\left( \mathcal{D}_2^{-} - \frac{1}{2} \big(\mathcal{D}_1^{-}\big)^2 \right)
\right]
\textrm{d}y
 \nonumber
 \\
   \rule{0pt}{.7cm}
 & & 
 +\,
     \frac{1}{32\pi \textrm{i}} 
     \sum_{j =0}^\infty (-1)^{j}
  \int_{\mathbb{R}}
 \big( \partial_y\hat{s}_\alpha(y) \big)\,
  \Big[\,
\widetilde{\mathcal{D}}_{2,j}^{+}\, \big(\mathcal{R}_0^{+}\big)^{j+1} 
- 
\widetilde{\mathcal{D}}_{2,j}^{-}\, \big(\mathcal{R}_0^{-}\big)^{j+1}
 \,\Big]
 \, \textrm{d}y
 \eea
 where $\widetilde{\mathcal{D}}_{2,j}^{\pm}$ is given in (\ref{D2-tilde-j-def}).
From (\ref{D-012-def}), (\ref{nu_star_app_y_def}) and (\ref{R0tilde-D-P-def}),
we find that the integrand in the first integral in (\ref{SA-infty-integral-2-b}) becomes
\be
\left( \mathcal{D}_2^{+} - \frac{1}{2} \big(\mathcal{D}_1^{+}\big)^2 \right)
-
\left( \mathcal{D}_2^{-} - \frac{1}{2} \big(\mathcal{D}_1^{-}\big)^2 \right)
=
10\, \textrm{i} \, y \big( 4 y^2 - 1 \big)  \,.
\ee
Thus, for any finite $\alpha >0$, we obtain
 \be
 \label{S-tilde-inf-2-0int}
   \widetilde{S}_{A,\infty,2}^{(\alpha)} 
 =
\frac{(1+\alpha)\, (3\,\alpha^2 - 7)}{384\, \alpha^3}
 -
      \frac{1}{32\pi \textrm{i}}
     \sum_{j =1}^\infty (-1)^{j}
  \int_{\mathbb{R}}
 \big( \partial_y\hat{s}_\alpha(y) \big)\,
  \Big[\,
\widetilde{\mathcal{D}}_{2,j-1}^{+}\, \big(\mathcal{R}_0^{+}\big)^{j} 
- 
\widetilde{\mathcal{D}}_{2,j-1}^{-}\, \big(\mathcal{R}_0^{-}\big)^{j}
 \,\Big]
 \, \textrm{d}y
\ee
where, from (\ref{D2-tilde-j-def}), we have that 
\be
\label{D-tilde-2-y}
\widetilde{\mathcal{D}}_{2,j-1}^{\pm}
\,=\,
\frac{j}{2} \big( 12y^2 -1 \big)^{2} \mp 10 \,\textrm{i}\,y \,\big(4y^2-1\big)
\ee
which are real when $y$ is purely imaginary. 
The integral in (\ref{S-tilde-inf-2-0int}) can be analysed 
by adapting  to this expression the procedure described above.
In particular, when $\alpha = 1$ 
 the integral in (\ref{S-tilde-inf-2-0int}) gives vanishing contribution, 
while for finite $\alpha \neq 1$ we obtain
 \be
 \label{S-tilde-inf-2-0int-sum}
   \widetilde{S}_{A,\infty,2}^{(\alpha)} 
\,=\,
\frac{(\alpha + 1)\, (3\,\alpha^2 - 7)}{384\, \alpha^3}
 +
     \sum_{j =1}^\infty 
           \frac{(-1)^{j}}{16 \, (1-\alpha)}
     \sum_{k=0}^{\infty}
  \Big[\,
    \widetilde{\mathcal{D}}_{2,j-1}^{-} \, \big(\mathcal{R}_0^{-}\big)^{j}\big|_{\tilde{y}_k} 
  \!+
 \widetilde{\mathcal{D}}_{2,j-1}^{+} \, \big(\mathcal{R}_0^{+}\big)^{j} \big|_{\tilde{y}_{-k - 1}} 
 \,\Big]  \,.
\ee
Since $\widetilde{\mathcal{D}}_{2,j}^{-} |_{\tilde{y}_k} = 
 \widetilde{\mathcal{D}}_{2,j}^{+} |_{\tilde{y}_{-k - 1}} $, this becomes 
\be
 \label{S-tilde-inf-2-0int-sum-final}
   \widetilde{S}_{A,\infty,2}^{(\alpha)} 
\,=\,
\frac{(\alpha + 1)\, (3\,\alpha^2 - 7)}{384\, \alpha^3}
 +
     \sum_{j =1}^\infty 
      \frac{(-1)^{j} \cos(4\eta j)}{8 \, (1-\alpha)}\,
   \sum_{k=0}^{\infty}
   \frac{\Omega\big((k+1/2)/\alpha -1/2\big)^j}{(4\eta)^{2j(2k+1)/\alpha}}\;
    \widetilde{\mathcal{D}}_{2,j-1}^{-}  \big|_{\tilde{y}_k}   \,.
\ee
By using the fact that $\tilde{y}_k$ is purely imaginary in (\ref{D-tilde-2-y}),
it is straightforward to observe that $\widetilde{\mathcal{D}}_{2,j-1}^{-}|_{\tilde{y}_k} $ is real;
hence (\ref{S-tilde-inf-2-0int-sum-final}) is real.

As for the $O(1/\eta^3)$ term in (\ref{exp-S-infty-123}), 
by employing (\ref{Stilde-alpha-app-nustar-step2}), (\ref{log-Ttilde-expansion}), 
(\ref{B3-from-P-expansion}) and (\ref{B3-from-P-a-b}),
we find
\bea
\label{SA-infty-integral-3}
  \widetilde{S}_{A,\infty,3}^{(\alpha)} 
 &=&
   \frac{1}{2\pi \textrm{i}}
  \int_{\mathbb{R}}
 \big( \partial_y\hat{s}_\alpha(y) \big)\,
  \big[\,
\mathcal{B}_{3,a}^{-} - \mathcal{B}_{3,a}^{+}
 \,\big]
 \, \textrm{d}y
 +
    \frac{1}{2\pi \textrm{i}}
  \int_{\mathbb{R}}
 \big( \partial_y\hat{s}_\alpha(y) \big)\,
  \big[\,
\mathcal{B}_{3,b}^{-} - \mathcal{B}_{3,b}^{+}
 \,\big]
 \, \textrm{d}y
 \hspace{1cm}
 \\
 \label{SA-infty-integral-3-b}
  \rule{0pt}{.8cm}
 & =& 
    \frac{1}{128\pi }
  \int_{\mathbb{R}}
 \big( \partial_y\hat{s}_\alpha(y) \big)\,
\left[
\left( \mathcal{D}_3^{-} - \mathcal{D}_2^{-} \, \mathcal{D}_1^{-} + \frac{1}{3}\big(\mathcal{D}_1^{-}\big)^3 \right)
-
\left( \mathcal{D}_3^{+} - \mathcal{D}_2^{+} \, \mathcal{D}_1^{+} + \frac{1}{3}\big(\mathcal{D}_1^{+}\big)^3 \right)
\right]
\textrm{d}y
 \nonumber
 \\
   \rule{0pt}{.7cm}
 & & 
 +\,
     \frac{1}{128\pi}
     \sum_{j =0}^\infty (-1)^{j}
  \int_{\mathbb{R}}
 \big( \partial_y\hat{s}_\alpha(y) \big)\,
  \Big[\,
\widetilde{\mathcal{D}}_{3,j}^{-}\, \big(\mathcal{R}_0^{-}\big)^{j+1} 
- 
\widetilde{\mathcal{D}}_{3,j}^{+}\, \big(\mathcal{R}_0^{+}\big)^{j+1}
 \,\Big]
 \, \textrm{d}y
\eea
where $\widetilde{\mathcal{D}}_{3,j}^{\pm}$ is defined in (\ref{D3-tilde-j-def}).
By using (\ref{D-3-def}), (\ref{nu_star_app_y_def}), (\ref{R0tilde-D-P-def}) 
and (\ref{D3-tilde-j-def}),
for the integrand of the first  integral in (\ref{SA-infty-integral-3-b}) we obtain
\be
\label{Dtilde3-pm-difference}
\left( \mathcal{D}_3^{+} - \mathcal{D}_2^{+} \, \mathcal{D}_1^{+} + \frac{1}{3}\big(\mathcal{D}_1^{+}\big)^3 \right)
-
\left( \mathcal{D}_3^{-} - \mathcal{D}_2^{-} \, \mathcal{D}_1^{-} + \frac{1}{3}\big(\mathcal{D}_1^{-}\big)^3 \right)
=\,
220\, y^4- 114\, y^2 +\frac{37}{12}  \,.
\ee
Since this is an even function and 
the function $ \partial_y\hat{s}_\alpha(y)$ is odd (see (\ref{der-s-y-explicit})),
the first integral of (\ref{SA-infty-integral-3-b}) gives a vanishing contribution. 
The remaining series in (\ref{SA-infty-integral-3-b}) 
can be studied through a slight modification of the analyses made above, 
finding 
\be
\label{SA-infty-integral-3-step2}
  \widetilde{S}_{A,\infty,3}^{(\alpha)} 
 \,=\,
     \frac{1}{64\,\textrm{i}\, (1-\alpha)}\,
      \sum_{j =1}^\infty (-1)^{j}
     \sum_{k=0}^{\infty}
  \Big[\,
    \widetilde{\mathcal{D}}_{3,j-1}^{-} \, \big(\mathcal{R}_0^{-}\big)^{j}\big|_{\tilde{y}_k} 
  \!+
 \widetilde{\mathcal{D}}_{3,j-1}^{+} \, \big(\mathcal{R}_0^{+}\big)^{j} \big|_{\tilde{y}_{-k - 1}} 
 \,\Big]
\ee
where, from (\ref{D3-tilde-j-def}), we have
\be
\label{Dtilde3-j-y}
\widetilde{\mathcal{D}}_{3,j-1}^{\pm}
=
\pm\,\frac{j^2}{6} \big( 12y^2 -1 \big)^{3} 
- 10 \,\textrm{i}\,j\, y \,\big(48 y^4 -16 y^2 +1\big)
\mp \left( 220 y^4 - 114 y^2 + \frac{37}{12}\, \right)
\ee
which is real when $y$ is purely imaginary.
Finally, since from (\ref{Dtilde3-j-y}) and (\ref{y_k-tilde-def})
we have that $\widetilde{\mathcal{D}}_{3,j}^{-} |_{\tilde{y}_k} = -\, \widetilde{\mathcal{D}}_{3,j}^{+} |_{\tilde{y}_{-k - 1}} $,
for (\ref{SA-infty-integral-3-step2}) we obtain
\be
\label{SA-infty-integral-3-step3} 
  \widetilde{S}_{A,\infty,3}^{(\alpha)} 
 \,=\,
      \sum_{j =1}^\infty 
\frac{(-1)^{j}   \sin(4\eta j)}{32\, (1-\alpha)}\,
   \sum_{k=0}^{\infty}
   \frac{\Omega\big((k+1/2)/\alpha -1/2\big)^j}{(4\eta)^{2j(2k+1)/\alpha}}\;
    \widetilde{\mathcal{D}}_{3,j-1}^{-}  \big|_{\tilde{y}_k}   \,.
\ee
This is a real function because (\ref{Dtilde3-j-y}) evaluated along the imaginary axes is real.

\subsubsection{Single copy entanglement}
\label{app-large-eta-single-copy}

The large $\eta$ asymptotic expansion of the single copy entanglement $S^{(\infty)}_A$ 
requires a separate discussion because the function \eqref{s-infty-y-def}
occurring in the integrand of \eqref{Stilde-alpha-app-nustar} is not entire.
This function has cusp singularities along the whole line $\mathrm{Re}(y)=0$;
hence this case cannot be considered a special case of the above analysis,
which employs the residue theorem.

In the limit $\alpha \to \infty$,  the integral (\ref{Stilde-alpha-app-nustar-step2}) becomes
\be
\label{Stilde-alpha-infty-app-nustar}
 \widetilde{S}_{A,\infty}^{(\infty)} 
\,=\,
   \frac{1}{2\pi \textrm{i}}
  \int_{\mathbb{R}}
 \big( \partial_y\hat{s}_\infty(y) \big)\,
  \Big[
  \log \! \big( 1 + \widetilde{\mathcal{T}}_\infty^{-} \big)
  -
  \log \! \big( 1 + \widetilde{\mathcal{T}}_\infty^{+} \big)
  \Big]
 \, \textrm{d}y
\ee
where $\partial_y\hat{s}_\infty(y) $ can be obtained from (\ref{s-infty-y-def}), finding
\be
\label{der-s-infty-y-def}
\partial_y\hat{s}_\infty(y) 
\,=\,
\pi\,\textrm{sign}(y) \, \big( \tanh(\pi |y|) - 1 \big)
\ee
and the expansion (\ref{log-Ttilde-expansion}) of 
$ \log( 1 + \widetilde{\mathcal{T}}_\infty^{\pm} )$ discussed in the  
Appendix\;\ref{app_large_eta_Ttilde} can be employed. 
This leads to an expansion like (\ref{exp-S-infty-123}) 
for (\ref{Stilde-alpha-infty-app-nustar}).
For the sake of simplicity, 
in the following we consider only the leading term.

By using the expansion (\ref{log-expansion-R0})
for the leading term of (\ref{log-Ttilde-expansion}),
one finds that the leading contribution to (\ref{Stilde-alpha-infty-app-nustar}) 
can be written as follows
\bea
\label{Stilde-alpha-infty-0}
  \widetilde{S}_{A,\infty,0}^{(\infty)} 
\,=\,
& &
\\
& & \hspace{-1.6cm}
=  
    \sum_{j =1}^\infty \frac{(-1)^{j+1}}{2\textrm{i}\,j}
  \int_0^{\infty}
\! \big( \tanh(\pi y) - 1 \big)
  \Big[\,
\big(\mathcal{R}_0^{-}(y) \big)^j - \big(\mathcal{R}_0^{+}(y) \big)^j
- \big(\mathcal{R}_0^{-}(-y) \big)^j + \big(\mathcal{R}_0^{+}(-y) \big)^j
  \,\Big]
 \, \textrm{d}y  \,.
 \nonumber
\eea
From (\ref{R_0_pm_explicit}) it is straightforward to obtain
\be
\big(\mathcal{R}_0^{\mp}(y) \big)^j + \big(\mathcal{R}_0^{\pm}(-y) \big)^j
 \,=\,
\frac{ 2\cos(4\eta j)\,\Omega( \mp \textrm{i} y - 1/2)^j}{ (4\eta)^{\mp 4\textrm{i}y\, j}}  
\ee
which naturally leads us to write (\ref{Stilde-alpha-infty-0}) as
\be
\label{tildeS-infty-0}
  \widetilde{S}_{A,\infty,0}^{(\infty)} 
=  \frac{1}{\textrm{i}}
    \sum_{j =1}^\infty \frac{(-1)^{j+1}}{j}\, \cos(4\eta j)\,
    \Big(  \mathcal{I}_{0,j}^{+} - \mathcal{I}_{0,j}^{-} \Big)
\ee
where
\be
\label{I-0-pm-def}
\mathcal{I}_{0,j}^{\pm} \,
\equiv 
  \int_0^{\infty}
\! \big( \tanh(\pi y) - 1 \big)\,
\frac{\Omega( \mp \textrm{i} y - 1/2)^j}{ (4\eta)^{\mp 4\textrm{i}y\, j}}  
 \; \textrm{d}y  \,.
\ee

In the integrand of (\ref{I-0-pm-def}),
the function $\Omega( \mp \textrm{i} y - 1/2)$ 
has second order poles at $y = \mp \,\textrm{i}(k+1/2)$ with integer $k \geqslant 0$,
while the simple poles of $\tanh(\pi y) - 1$
are located at $y = \,\textrm{i}(k+1/2)$ with integer $k \in \mathbb{Z}$.
The singularities of $\tanh(\pi y) - 1$ in the upper (lower) half plane
are canceled by the zeros of $\Omega( \mp \textrm{i} y - 1/2)$.
These observations allow to write the integral (\ref{I-0-pm-def}) as follows
\be
\label{integration-dec-0}
\int_0^\infty \big( \dots \big) \textrm{d}y\;
=
\lim_{\Lambda \to +\infty}
\left[\,
\int_0^{\pm \textrm{i} L} \big( \dots \big) \textrm{d}y
+
\int_{\pm \textrm{i} L}^{\pm \textrm{i} L + \Lambda} \big( \dots \big) \textrm{d}y
+
\int_{\pm \textrm{i} L + \Lambda}^{\Lambda} \big( \dots \big) \textrm{d}y
\,\right]
\ee
for any finite value of $L>0$.
Since the integrand of (\ref{I-0-pm-def}) is infinitesimal as $\Lambda \to +\infty$,
the last integral in the r.h.s. of (\ref{integration-dec-0}) vanishes in this limit.
As for the second integral in the r.h.s. of (\ref{integration-dec-0})
in the limit $\Lambda \to +\infty$, we find
\\
\be
\frac{1}{(4\eta)^{4Lj}}
  \int_0^{\infty}
\! \big( \tanh(\pi [y \pm \textrm{i}L] ) - 1 \big)\,
\frac{\Omega( \mp \textrm{i} [y \pm \textrm{i}L] - 1/2)^j}{ (4\eta)^{\mp 4\textrm{i} j \,y}}  
 \; \textrm{d}y
 \,=\,
 O\big(1/\eta^{4Lj}\big)
\ee
because the absolute value of the integrand is independent of $\eta$.

By introducing the integration variable $w = \mp \textrm{i} y$ 
for the first integral in the r.h.s. of (\ref{integration-dec-0}), 
for (\ref{I-0-pm-def}) we obtain
\bea
\mathcal{I}_{0,j}^{\pm} \,
&=&
\int_0^{\pm \textrm{i} L}
\! \big( \tanh(\pi y) - 1 \big)\,
\frac{\Omega( \mp \textrm{i} y - 1/2)^j}{ (4\eta)^{\mp 4\textrm{i}y\, j}}  
 \; \textrm{d}y
 +
  O\big(1/\eta^{4Lj}\big)
  \\
      \label{Int-0-j-pm}
&=&
- \int_0^{L}
\! \big( \tan(\pi  w) \pm \textrm{i} \big)\,
\frac{\Omega( w - 1/2)^j}{ (4\eta)^{4 w\, j}}  
 \; \textrm{d}w
 +
  O\big(1/\eta^{4Lj}\big)  \,.
\eea
By employing this result in (\ref{tildeS-infty-0}), we find
\be
\label{tildeS-infty-0-def}
  \widetilde{S}_{A,\infty,0}^{(\infty)} 
\,=  \,
2
    \sum_{j =1}^\infty \frac{(-1)^{j}}{j}\, \cos(4\eta j)
    \left[\;
\int_0^{L}
\frac{\Omega( w - 1/2)^j}{ (4\eta)^{4 w\, j}}  
 \; \textrm{d}w
 +
  O\big(1/\eta^{4Lj}\big)\,
  \right]  .
\ee

The Taylor expansion of $\Omega( w - 1/2)^j$ with $j \geqslant 1$ about $w=0$ reads
\be
\label{exp-Omega-0}
\Omega( w - 1/2)^j
\,=\,
\sum_{p=0}^{\infty} \frac{\widetilde{\Omega}_{j,p} }{p!} \; w^p
\;\;\;\qquad\;\;\;
\widetilde{\Omega}_{j,p} \equiv \big[\partial^p_w \,\Omega( w - 1/2)^j \big] \!\big|_{w=0}
\ee
where the generic coefficient can be expressed through 
the Fa\'a di Bruno's formula\footnote{The Fa\'a di Bruno's formula
generalises the chain rule to higher derivatives $\partial_w^n f(g(w))$.
It reads
\be
\partial_w^n f(g(w)) = 
\sum_{k=1}^n
f^{(k)}(g(w))\,
B_{n,k}\big( g'(w), g''(w), \dots , g^{(n-k+1)}(w) \big)
\ee
in terms of the Bell polynomials $B_{n,k}(x_1, x_2, \dots, x_{n-k+1})$.}
for $f(g(w))$ 
with $f(z)=z^j$ and $g(w)=\Omega( w - 1/2)$,
finding
\be
\label{Faa-di-Bruno-omega}
\widetilde{\Omega}_{j,p} 
=\,
\sum_{k=0}^p
\Bigg[\, \prod_{l=0}^{k-1} \big(j-l\big)\Bigg]
B_{p,k}\big( \Omega'(-1/2), \Omega''(-1/2), \dots , \Omega^{(p-k+1)}(-1/2) \big)
\ee
where $p \geqslant 1$, $p - k + 1 >0$,
we used that $\Omega(-1/2) = 1$ and assumed $\prod_{l=0}^{-1} (\cdots) = 1$.
Notice that
\be
\label{a-series-gamma}
\prod_{l=0}^{k-1} \big(j-l\big) = \frac{\Gamma(j+1)}{\Gamma(j+1-k)} 
\equiv \sum_{l=0}^k a_{k,l}\, j^l  \,.
\ee
The radius of convergence of the power series (\ref{exp-Omega-0})
 for $\Omega( w - 1/2)^j$ centered at $w=0$ 
is equal to $1/2$
because of the singularity occurring at $w = -1/2$.
This implies that $L<1/2$ in (\ref{Int-0-j-pm}). 
We choose $L=1/4$.

By using (\ref{exp-Omega-0}) and (\ref{Faa-di-Bruno-omega}) into
(\ref{tildeS-infty-0-def}) with $L=1/4$, we obtain
\bea
  \widetilde{S}_{A,\infty,0}^{(\infty)} 
&=&
2
    \sum_{j =1}^\infty \frac{(-1)^{j}}{j}\, \cos(4\eta j)\;
    \sum_{p=0}^{\infty} 
    \frac{\widetilde{\Omega}_{j,p}}{p!}
\int_0^{1/4} \!\!
\frac{w^p}{ (4\eta)^{4 w\, j}} \; \textrm{d}w
 +
  O(1/\eta)
  \\
  \label{S-infty-Omega-series}
  \rule{0pt}{.8cm}
  &=&
  2
    \sum_{j =1}^\infty \frac{(-1)^{j}}{j}\, \cos(4\eta j)\;
    \sum_{p=0}^{\infty} 
    \frac{\widetilde{\Omega}_{j,p}}{[ \,4j\, \log(4\eta) \,]^{p+1}}
 +
  O(1/\eta)
\eea
where we used that
\be
\int_0^{\tfrac{1}{4}} \frac{w^p}{ (4\eta)^{4 w\, j}}  \,\textrm{d}w
\,=\,
\frac{\Gamma(p+1)- \Gamma(p+1, j \log(4\eta))}{[ \,4j\, \log(4\eta) \, ]^{p+1}}
\,=\,
\frac{p!}{[ \,4j\, \log(4\eta) \,]^{p+1}} + O(1/\eta^j)  \,.
\ee
The last expression has been obtained by exploiting the fact that
the incomplete Gamma function $\Gamma(a,z)$
vanishes like $e^{-z} \, z^{a-1}$  as $z \to +\infty$
and that $\Gamma(p+1) = p!$ for integer $p$.

Plugging (\ref{Faa-di-Bruno-omega}) and (\ref{a-series-gamma}) into (\ref{S-infty-Omega-series}), 
one encounters the following double sum
\bea
  2
    \sum_{j =1}^\infty 
    \frac{(-1)^{j}}{j^{p+2}}\;\cos(4\eta j)
    \sum_{l=0}^k a_{k,l}\, j^l
    &=&
    2 \sum_{l=0}^k a_{k,l}
    \sum_{j =1}^\infty \frac{(-1)^{j}}{j^{p+2-l}}\;
    \cos(4\eta j)
    \\
    &=&
            \sum_{l=0}^k a_{k,l}
            \Big[
            \textrm{Li}_{2+p-l}(-e^{4\textrm{i} \eta}) + \textrm{Li}_{2+p-l}(-e^{-4\textrm{i} \eta})
            \Big]
            \hspace{1cm}
\eea
hence (\ref{S-infty-Omega-series}) can be written as follows
\bea
  \widetilde{S}_{A,\infty,0}^{(\infty)} 
&=&
      \sum_{p=0}^{\infty} 
      \frac{1}{[ \,4 \log(4\eta) \, ]^{p+1}}
        \sum_{k=0}^p
        \sum_{l=0}^k 
        a_{k,l}
            \Big[
            \textrm{Li}_{2+p-l}(-e^{4\textrm{i} \eta}) + \textrm{Li}_{2+p-l}(-e^{-4\textrm{i} \eta})
            \Big]
            \\
            & & \hspace{4.5cm}
\times \, B_{p,k}\big( \Omega'(-1/2), \Omega''(-1/2), \dots , \Omega^{(p-k+1)}(-1/2) \big)
\nonumber
\eea
up to $O(1/\eta)$ term, that has been neglected.
Since $ \sum_{k=0}^p \sum_{l=0}^k (\dots)  =  \sum_{l=0}^p \sum_{k=l}^p (\dots) $,
this expression becomes
\be
\label{S-single-copy-app-final}
  \widetilde{S}_{A,\infty,0}^{(\infty)} 
\,=\,
      \sum_{p=0}^{\infty} 
      \frac{1}{[ \,4 \log(4\eta) \, ]^{p+1}}
        \sum_{l=0}^p
        \tilde{a}_{p,l}
            \Big[
            \textrm{Li}_{2+p-l}(-e^{4\textrm{i} \eta}) + \textrm{Li}_{2+p-l}(-e^{-4\textrm{i} \eta})
            \Big]
            +O(1/\eta)
\ee
where 
\be
\tilde{a}_{p,l}
\,\equiv\,
        \sum_{k=l}^p      a_{k,l}\,
        B_{p,k}\big( \Omega'(-1/2), \Omega''(-1/2), \dots , \Omega^{(p-k+1)}(-1/2) \big)  \,.
\ee

The result (\ref{expansion-infty-app-34-order}) in the main text
is obtained by considering the first five terms of the series occurring in (\ref{S-single-copy-app-final}),
corresponding to $p\in \{0,1,2,3,4\}$.

\section{A double scaling limit of the lattice results}
\label{sec_large_distance_lattice}

In this appendix we briefly mention some lattice results related 
to the quantities that have been studied in the main text through a particular double scaling limit.

The Hamiltonian of the free fermionic chain on the infinite line is
\cite{EislerPeschel:2009review}
\be
\label{hamiltonian-tight-binding}
H \,=\,
- \!\! \sum_{i=-\infty}^{+\infty}
\left[\,
\hat{c}_i^\dagger\, \hat{c}_{i+1}
+
 \hat{c}_{i+1}^\dagger\, \hat{c}_{i}
 -2h \left(\hat{c}_i^\dagger\, \hat{c}_{i} -\frac{1}{2}\right)
\,\right]
\ee
where $ \hat{c}_{i}$ describe spinless fermionic degrees of freedom
satisfying the anticommutation relation $\{\hat{c}_{i}, \hat{c}_{j}\} = \delta_{i,j}$
and $h$ is the chemical potential. 
The ground state of \eqref{hamiltonian-tight-binding} is a Fermi sea 
with Fermi momentum $\kappa_{\textrm{\tiny F}}  = \textrm{arccos}( |h|) \in[0,\pi]$.
A Jordan Wigner transformation maps
the Hamiltonian \eqref{hamiltonian-tight-binding} 
into the Hamiltonian of the spin-$\tfrac{1}{2}$ Heisenberg XX chain in a magnetic field $h$.
The two-point correlator of this lattice model reads
\be
\label{correlator-fermionic-chain}
\mathcal{C}_{i,i} \equiv \frac{\kappa_{\textrm{\tiny F}}}{\pi}
\;\;\;\qquad\;\;\;
\mathcal{C}_{i,j} \equiv
\frac{\sin [\kappa_{\textrm{\tiny F}} (i-j)  ]}{\pi (i - j)}
\qquad
i \neq j
\qquad
i, j \,\in\, \mathbb{Z}  \,.
\ee

Considering the bipartition of the infinite chain given by 
a block $A$ made by $L$ consecutive sites and its complement, 
many numerical analyses can be performed by considering 
the $L \times L$ reduced correlation matrix $\boldsymbol{\mathcal{C}}_A$,
whose generic element is \eqref{correlator-fermionic-chain} with $i,j \in A$,
which turns out to be a  Toeplitz matrix.

It is insightful to study the lattice results in the following double scaling limit \cite{Krasovsky-11, Abanov-2011, EislerPeschelProlate}
\be
\label{dsl}
L \rightarrow +\infty 
\;\;\qquad \;\;
\kappa_{\textrm{\tiny F}} \rightarrow 0 
\;\;\;\;\qquad \;\;\;\;
\kappa_{\textrm{\tiny F}} L  \equiv 2\eta \,. 
\ee

The full counting statistic generating function $\chi(\zeta) $ 
allows us to study the cumulants of the particle number operator 
$N_A$ in the block $A$ 
%
and its Toeplitz determinant representation reads
\cite{Lieb-61, Abanov-2011}
\be
\label{toeplitz}
\chi(\zeta) 
\,\equiv\, \langle \e^{\textrm{i} \zeta N_A} \rangle
\,=\,
\textrm{det} \big( \boldsymbol{1} + (\e^{\textrm{i} \zeta} - 1) \, \boldsymbol{\mathcal{C}}_A \big)
\,=\,
\textrm{det} \big( \boldsymbol{1} - z^{-1}\,\boldsymbol{\mathcal{C}}_A \big)
\ee
where $ \boldsymbol{1} $ is the $L\times L$ identity matrix and $\zeta = 2\pi \nu_\star$,
with $\nu_\star=\nu_\star(z)$ being defined in (\ref{nu-star-main}).
%
The logarithm of (\ref{toeplitz}) gives
\be
\label{log-toeplitz}
\log\! \big[ \chi(\zeta) \big]
=
\log\! \big[  \langle \e^{\textrm{i} \zeta N_A} \rangle \big]
=
\textrm{Tr} \big[ \log  \big( \boldsymbol{1} + (\e^{\textrm{i} \zeta} - 1) \, \boldsymbol{\mathcal{C}}_A \big) \,\big]
\ee
which generates the cumulants.
The first cumulants (mean value, variance and skewness) read respectively
\bea
\label{mean-NA-app}
\langle N_A \rangle
&=&
\big[\partial_{\ri \zeta} \log(\chi)\big]\big|_{\zeta=0}
=\,
\textrm{Tr}(\boldsymbol{\mathcal{C}}_A )
\\
\label{var-NA-app}
\rule{0pt}{.6cm}
\langle \big(N_A - \langle N_A \rangle \big)^2  \rangle
&=&
\langle N_A^2 \rangle - \langle N_A \rangle^2
\,=\,
\big[\partial^2_{\ri \zeta} \log(\chi)\big]\big|_{\zeta=0}
=\,
\textrm{Tr}\big(\boldsymbol{\mathcal{C}}_A - \boldsymbol{\mathcal{C}}_A^2\big)
\\
\label{3rd-cum-NA-app}
\rule{0pt}{.6cm}
\langle \big(N_A - \langle N_A \rangle \big)^3  \rangle
&=&
\langle N_A^3 \rangle - 3\, \langle N_A^2 \rangle \, \langle N_A \rangle + 2\, \langle N_A \rangle^3
\nonumber
\\
&=&
\big[\partial^3_{\ri \zeta} \log(\chi)\big]\big|_{\zeta=0}
=\,
\textrm{Tr}\big(\boldsymbol{\mathcal{C}}_A - 3\,\boldsymbol{\mathcal{C}}_A^2 + 2\,\boldsymbol{\mathcal{C}}_A^3\big)  \,.
\eea

Since $\boldsymbol{\mathcal{C}}_A $ is a Toeplitz matrix, 
the Fisher-Hartwig conjecture \cite{FHc} (proved in \cite{Basor-91})
and its generalisation \cite{Deift-11, Basor-91, Calabrese-Essler-10} 
can be employed to study the leading and the subleading terms respectively 
of the expansion of the determinant \eqref{toeplitz} as $L \rightarrow \infty$.
In particular, the double scaling limit \eqref{dsl} of the Fisher-Hartwig conjecture 
gives (\ref{tau-tilde-infty-def}).
Its generalised version include also some subleading corrections:
for instance, the double scaling limit \eqref{dsl} of the expansion reported in Eq.\,(84) of \cite{Calabrese-Essler-10} 
gives \eqref{cal-T-infty-split-101-bis} with the infinite sums in $k$ 
truncated to the finite sums including only the terms corresponding to $k \in \{0,1,2\}$.
We emphasise that all order corrections occur in
(\ref{tau-expansion-large-eta}), which holds in the double scaling limit.

The cumulants generated by \eqref{log-toeplitz} have been studied 
by employing the Fisher-Hartwig conjecture and its generalisation 
e.g. in \cite{Abanov-2011, Ivanov-2013}.

The entanglement entropies of the block $A$ are  obtained 
through the eigenvalues of the reduced correlation matrix $\boldsymbol{\mathcal{C}}_A $
\cite{Peschel:2003rdm,EislerPeschel:2009review}.
Analytic results for these entanglement entropies for large $L$ have been obtained 
by employing the Fisher-Hartwig conjecture  for the leading terms \cite{Jin_2004, Keating_04}
and the generalised Fisher-Hartwig conjecture 
combined with further computations for the subleading terms \cite{Calabrese-Essler-10}.
Taking  the double scaling limit (\ref{dsl}) in the result for 
the leading terms of $S^{(\alpha)}_A$ reported in \cite{Jin_2004} for large $\eta$,
one obtains \eqref{S_alpha_main_large_eta_0}.
Similarly, for the subleading terms of $S^{(\alpha)}_A$,
we have that the double scaling limit \eqref{dsl} 
of the lattice result in Eq.\,(10) of \cite{Calabrese-Essler-10}
gives \eqref{S-infty-tilde-ab-dec} for $N \in \{ 0,1,2\}$. 
In order to obtain the $N=3$ term in \eqref{S-infty-tilde-ab-dec} 
through the double scaling limit \eqref{dsl},
higher order terms must be computed in the lattice analysis,
along the lines discussed in \cite{Calabrese-Essler-10}.

Beside the analysis involving the Fisher-Hartwig conjecture, 
in \cite{Jin_2004} lattice results have been reported that can be compared 
with the small $\eta$ expansions discussed in this manuscript. 
As for the entanglement entropy,
considering the double scaling limit (\ref{dsl}) of the results of \cite{Jin_2004} for small $\eta$,
we have that the first term in the last expression of (\ref{EE-small-c-regime}) 
agrees with Eq.\,(5) of \cite{Jin_2004} for $\alpha = 1$ up to missing factors of 2, 
whose absence there seems just a typo.
As for the R\'enyi entropies,
the expansion (\ref{renyi_small_leading}) agrees with Eq.\,(5) of \cite{Jin_2004} for $\alpha \neq 1$.
We find it worth mentioning also that the sine kernel tau function \eqref{tau-function-sine} evaluated at $z=1$
provides the emptiness formation probability \cite{korepin-book,korepin-1994,essler-1994} of the XX chain and in the double scaling limit
\cite{Claeys_11,Stephan-14,Kozlowska_2019,Ares:2019rad}.
More recently, the approach of \cite{Jin_2004} has been extended to compute the EE of two disjoint intervals on the XX chain separated by a single site \cite{Brightmore_2020}. This problem has not exactly a well defined continuum limit, since in such a limit, the gap between the two intervals goes to zero giving a trivial subsystem. However, in the same lines, it would be very interesting to study the EE for two (or more) intervals separated by a finite distance in the continuum QFT. This problem has been addressed in \cite{Spitzer-14} for the leading term. Subleading corrections to that result are still unknown.

\bibliographystyle{nb}

\bibliography{refsSchrod}

\end{document}